\documentclass[lettersize,journal]{IEEEtran}
\usepackage{macro}
\usepackage{amsmath,amsfonts}
\usepackage{amssymb}
\usepackage{amsthm}
\usepackage{algorithm}
\usepackage[noend]{algpseudocode}
\usepackage{array}
\usepackage[caption=false,font=normalsize,labelfont=sf,textfont=sf]{subfig}
\usepackage{textcomp}
\usepackage{stfloats}
\usepackage{url}
\usepackage{verbatim}
\usepackage{graphicx}
\usepackage{cite}
\usepackage{hyperref}
\usepackage{color}
\usepackage{multirow}
\newtheorem{Thm}{Theorem}
\newtheorem{Def}{Definition}
\newtheorem{Lem}{Lemma}
\newtheorem{Ass}{Assumption}
\newtheorem{Rem}{Remark}
\newtheorem{Cor}{Corollary}
\newtheorem{Pro}{Proposition}

\begin{document}


\title{R-FAST: Robust Fully-Asynchronous Stochastic Gradient Tracking over General Topology}

\author{Zehan~Zhu, Ye~Tian, Yan~Huang, Jinming~Xu$^\dagger$, Shibo~He
\thanks{Z. Zhu, Y. Huang, J. Xu and S. He are with the College of Control Science and Engineering, Zhejiang University, China,
Hangzhou, 310027. Y. Tian is with Amazon, and his work of this paper is done prior to him joining Amazon.}
\thanks{$^\dagger$Correspondence to jimmyxu@zju.edu.cn (Jinming Xu).}}
\markboth{Journal of \LaTeX\ Class Files,~Vol.~14, No.~8, August~2021}%
{Shell \MakeLowercase{\textit{et al.}}: A Sample Article Using IEEEtran.cls for IEEE Journals}

\maketitle
\begin{abstract} 
We propose a Robust Fully-Asynchronous Stochastic Gradient Tracking method (R-FAST) for distributed machine learning problems over a network of nodes, where each node performs local computation and communication at its own pace without any form of synchronization. Different from existing asynchronous distributed algorithms, R-FAST can eliminate the impact of data heterogeneity across nodes on convergence performance and allow for packet losses by employing a robust gradient tracking strategy that relies on properly designed auxiliary variables for tracking and buffering the overall gradient vector. Moreover, the proposed method utilizes two spanning-tree graphs for communication so long as both share at least one common root, enabling flexible designs in communication topologies. We show that R-FAST converges in expectation to a neighborhood of the optimum with a geometric rate for smooth and strongly convex objectives; and to a stationary point with a sublinear rate for general non-convex problems. Extensive experiments demonstrate that R-FAST runs 1.5-2 times faster than synchronous benchmark algorithms, such as Ring-AllReduce and D-PSGD, while still achieving comparable accuracy, and outperforms the existing well-known asynchronous algorithms, such as AD-PSGD and OSGP, especially in the presence of stragglers.
\end{abstract}


\section{Introduction}
\label{sec:introduction}
\IEEEPARstart{I}{n} the past decade, deep learning\cite{lecun2015deep} has shown great success in various fields, such as computer vision\cite{voulodimos2018deep}, natural language processing\cite{hirschberg2015advances}, autonomous driving\cite{muhammad2020deep}. Training modern deep neural networks to desirable accuracy usually requires an enormous number of training samples, which results in significant consumption of computing resources. Multiple computing devices can thus be employed to accelerate such large-scale training tasks~\cite{langer2020distributed}. In particular, one consider solving the following distributed stochastic optimization problem via a group of workers:
\begin{equation}
\label{global_loss_function}
\underset{x\in \mathbb{R}^p}{\min}F\left( x \right) \triangleq \sum_{i=1}^n{\underset{\triangleq f_i\left( x \right)}{\underbrace{\mathbb{E}_{\zeta _i\sim \mathcal{D}_i}\left[ l\left( x;\zeta _i \right) \right] }}}
\end{equation}
where $\mathcal{D}_i$ denotes the distribution of sample $\zeta _{i}$ locally stored at worker $i$ and $l\left(x;\zeta_i\right)$ denotes the loss function with respect to the model parameter $x\in \mathbb{R}^p$ and sample $\zeta_i$; $f_i\left( x \right)$ and $F\left( x \right)$ denote the local objective function at worker $i$ and the global objective function, respectively.
This setting also covers the empirical risk minimization (ERM) problems with $\mathcal{D}_i$ being the local training dataset. The workers are connected over a communication network for information exchange.  Collaboratively, all workers seek the global optimal parameter $x$ minimizing the global objective function $F$. 

Stochastic gradient descent (SGD) is a commonly used technique for solving the above large-scale training problem~\cite{nemirovski2009robust,moulines2011non,ghadimi2013stochastic}. However, standard SGD does not scale well to large data sets due to its inherently sequential way of updating~\cite{dean2012large}. 
In particular, parallel and distributed methods are employed for large-scale training using SGD, e.g., Parallel SGD~\cite{zinkevich2010parallelized,mcmahan2017communication} which has a parameter server responsible for aggregating gradients from clients and updating model parameters;  Ring-AllReduce SGD \cite{sergeev2018horovod, goyal2017accurate} which computes the exact average of gradients via a ring network in a decentralized manner; and Distributed SGD \cite{lian2017can}, D$^2$ \cite{tang2018d}, DSGT \cite{zhang2019decentralized}, S-AB \cite{xin2019distributed} that rely on approximate averaging of model parameters via certain gossip protocols over a peer-to-peer network. These abovementioned algorithms need to rely on perfect synchronization during communications, which limits their application to real scenarios where one usually observes the presence of stragglers and high latency of the network communication channel due to limited bandwidth~\cite{luo2020prague, nadiradze2021asynchronous}.

Numerous asynchronous parallel and distributed algorithms have thus been proposed to overcome the above drawback~\cite{verbraeken2020survey}. 
However, most of these methods still rely on certain form of synchronization, such as specific node activating rules~\cite{ram2009asynchronous}, real-time information mixing among a subset of nodes~\cite{lian2018asynchronous}, and computing at the same pace for all nodes~\cite{assran2019stochastic}, not being fully asynchronous.
Besides, they usually suffer from the deteriorating effect on the convergence performance caused by data heterogeneity across nodes~\cite{assran2019stochastic,lian2018asynchronous} and can not effectively deal with unpredictable packet losses~\cite{lian2018asynchronous,assran2019stochastic,zhang2019fully}; and rely on certain topology for communication, such as undirected graphs~\cite{lian2018asynchronous} or fixed strongly connected digraphs~\cite{zhang2019fully,tian2020achieving}, which restricts the flexibility of network topologies.

In this paper, we propose a Robust Fully-Asynchronous Stochastic Gradient Tracking method (termed R-FAST) to address all the above identified issues.  Our main contributions are summarized as follows:
\begin{itemize}
\item [1)] \textbf{New robust fully asynchronous algorithms over general topology.} Different from the existing works, the proposed R-FAST method allows each node to perform the local communication and computation at its own pace, without any form of synchronization and imposing no restriction on the arriving order of communication messages. Besides, R-FAST avoids the deteriorating effect from data heterogeneity on the convergence rate, and allows for packet losses thanks to the introduced robust gradient tracking scheme. Moreover, R-FAST works on general communication graphs containing spanning trees that share at least one common root, which enables flexible design of the underlying network topology, including PS, Ring and Gossip structure as special cases. 
\item [2)] \textbf{Provable convergence guarantee.} 
We prove that R-FAST converges linearly in expectation to a neighborhood of the optimal solution, for strongly convex objectives and finds a stationary point at a rate of $\mathcal{O}(1/\sqrt{K})$ for non-convex objectives.
Our proof relies on a properly designed augmented system to account for unpredictable delays and packet losses. Different from~\cite{pu2020push, xin2019distributed, zhang2019fully, tian2020achieving, kungurtsev2021decentralized}, leveraging \emph{two-time-scale techniques}, we are able to carry out the analysis over a time period in which common root nodes of two spanning-tree graphs are activated at least once, ensuring a valid contraction of optimality gap for strongly convex cases and a valid descent towards the stationary point for non-convex cases.
\item [3)] \textbf{Extensive experimental evaluations.} 
Experiments of training logistic regression model on MNIST dataset show that the proposed R-FAST algorithm can work over general network topologies such as binary tree, line, and directed ring graphs.
In addition, large-scale image classification tasks of training ResNet-50 on ImageNet dataset show that R-FAST converges 1.5-2 times faster than the synchronous algorithms while enjoys higher testing accuracy than the well-known asynchronous algorithms such as AD-PSGD~\cite{lian2018asynchronous} and OSGP~\cite{assran2019stochastic}, especially in the presence of a straggler.
Moreover, we show the good scalability of R-FAST in the number of nodes.
\end{itemize}

\section{RELATED WORK}
\label{related_work}
\subsection{On Asynchronous Model}
Parallel schemes where there is a center (e.g., parameter server) responsible for aggregating gradients and updating model parameters have been widely employed for the parallel update of SGD, and corresponding asynchronous versions have been proposed recently.
For example, Recht \textit{et al}.~\cite{recht2011hogwild} designed an asynchronous algorithm called HOGWILD!, where different processors have access to a shared memory that can be overwritten at any time. 
Duchi \textit{et al}.~\cite{agarwal2011distributed} proposed an A-PSGD algorithm that allows the central parameter server to use stale gradients received from clients to update model parameters. 
Dean \textit{et al}.~\cite{dean2012large} developed a DistBelief framework based on which they designed Downpour SGD where the model replicas run independently of each other and the parameter server maintains the current model parameters.  Another line of research is focused on seeking a balance between optimizing the global problem \eqref{global_loss_function} and optimizing the local problem $\min_{x\in \mathbb{R}^p} f_i(x)$, motivated by applications where local optimality and global cooperation are both respected, such as language models on smart phone and robotic exploration. For instance, \cite{bedi2022fedbc} considers a star-topology in the context of federated learning, while \cite{bedi2019asynchronous} considers a general mesh network with asynchronous update (delayed gradients).  Instead of \textit{enforcing} consensus among agents, these works only impose constraint requirements that the differences among agents' local estimates are reasonably upper bounded.

Building on A-PSGD, Lian \textit{et al}.~\cite{lian2015asynchronous} studied two asynchronous parallel implementations of SGD 
for non-convex problems and established a sublinear rate of $\mathcal{O}\left( 1/\sqrt{K} \right)$. 
To obtain better performance, Zhang \textit{et al}.~\cite{zhang2015staleness} proposed staleness-aware async-SGD, where the central server keeps track of the staleness associated with each gradient computation and adjusts the learning rate according to the staleness value.
To deal with both the issues of stale gradients and non-IID data, Zhou \textit{et al}.~\cite{zhou2022towards} proposed an asynchronous algorithm termed WKAFL, where the central server 
assigns each of the first $K$ received stale gradients with a weight based on its cosine similarity with respect to the estimated global unbiased gradient.
The above asynchronous parallel schemes, however, rely on a center to aggregate gradients from clients, which may become the potential bottleneck due to high communication burden and suffer from the single point failure~\cite{lian2018asynchronous}.

To address the above issues, numerous decentralized optimization algorithms are proposed for training deep neural networks, 
and some form of asynchrony are allowed.
For instance, Ram \textit{et al}.~\cite{ram2009asynchronous} proposed an asynchronous decentralized algorithm based on gossip protocols where each worker wakes up randomly according to the local Poisson clock and selects a neighbor to exchange parameters, while no any form of communication delay is allowed which calls for certain coordination among the workers. 
Lian \textit{et al}.~\cite{lian2018asynchronous} developed an asynchronous decentralized algorithm, termed AD-PSGD, which allows for using stale gradients for update. However, they consider only undirected graph and require real-time communication and information mixing among a group of workers at each iteration assuming the frequency of updating for each worker is known a prior, which imposes a key challenge for synchronization.
To make it applicable to general digraphs, Assran \textit{et al}.~\cite{assran2019stochastic} proposed SGP based on push-sum protocol~\cite{kempe2003gossip} and developed an asynchronous variant OSGP which allows for communication delays but requires all workers perform computation at the same pace.
Building on SGP, Spiridonoff \textit{et al}.~\cite{spiridonoff2020robust} developed RASGP which enables each worker updating at its own pace, but requires all messages arriving in the order they were sent.
There has been also some other efforts attempting to decompose the operation of exact average of all workers (c.f., AllReduce~\cite{sergeev2018horovod}) into a series of exact average of a subset of workers~\cite{luo2020prague, miao2021heterogeneity}. In particular, they use Partial-Reduce primitive instead of AllReduce primitive to compute the exact average whenever a bunch of workers finish updating, thus accounting for computing heterogeneity among workers but still require blocking for synchronization among that bunch of workers.
To sum up, these abovementioned asynchronous distributed algorithms usually assume stringent conditions on their asynchronous models and thus can not be implemented in a \emph{fully asynchronous} manner.

\subsection{On Robustness}
Existing asynchronous methods usually suffer from certain robustness issues due to data heterogeneity across computing nodes, packet losses and strict requirement on the underlying network topology. 
For instance, the algorithms in ~\cite{assran2019stochastic,lian2018asynchronous} suffer from the deteriorating effect on their convergence rates due to data heterogeneity across computing nodes. To cope with such issue, Zhang \textit{et al}.~\cite{zhang2019fully} proposed an asynchronous APPG algorithm which can alleviate the effect of data heterogeneity leveraging gradient tracking scheme~\cite{di2016next, xu2015augmented,pu2021distributed, zhang2019decentralized, xin2019distributed} that corrects the gradient descent direction over time, but works only for deterministic problems and does not allow for packet losses.
The aforementioned RASGP~\cite{spiridonoff2020robust} can handle packet losses but the result therein is established only for strongly convex objectives. The recently proposed ASY-SONATA algorithm~\cite{tian2020achieving} can deal with both the issues of data heterogeneity and packet losses, but works only for deterministic optimization problems. Kungurtsev \textit{et al}.~\cite{kungurtsev2021decentralized} extended the ASY-SONATA algorithm to consider stochastic gradient update and established a sublinear rate merely for non-convex objective functions.

To the best of our knowledge, most decentralized optimization algorithms are only applicable to doubly stochastic weight matrices which are, indeed, not easy to design or even impossible for certain graphs~\cite{gharesifard2010does}.
The aforementioned algorithms~\cite{zhang2019fully, tian2020achieving, kungurtsev2021decentralized}  based on gradient tracking strategy can work on two separate weight matrices (i.e., one being row stochastic and the other column stochastic) but require the two corresponding subgraphs to be both strongly-connected.
Push-Pull methods~\cite{pu2020push} relax the above requirements and can work on two separate subgraphs as long as both contain a spanning tree and share a common root.
This property admits great flexibility in designing the underlying network topology, including popular structures such as Parameter Server, Ring and Gossip. However, similar to APPG, Push-Pull methods work only for deterministic optimization problems.
Recently, a robust variant of Push-Pull methods termed R-Push-Pull, was proposed in~\cite{pu2020robust}, which is robust against noisy information exchange over communication links. However, no any form of packet/message delay or loss is allowed in R-Push-Pull, due to its synchronous update mechanism.

\section{PRELIMINARY}
\label{preliminary}
\subsection{Network Model}
We consider that all workers cooperate to solve Problem~\eqref{global_loss_function} over a network modeled as a directed graph $\mathcal{G}=(\mathcal{V},\mathcal{E})$ where $\mathcal{V} = \left\{ 1,2,...,n \right\} $ denotes the set of nodes and $\mathcal{E} \subset \mathcal{V} \times \mathcal{V}$ denotes the set of edges/communication links. Let $\mathcal{G}(M)$ denote the graph induced by a non-negative matrix $M\in \mathbb{R}^{n \times n}$ such that  $(j,i)\in \mathcal{E}(M)$ if and only if $M_{ij} > 0$. We use $\mathcal{N}^{\text{in}}_i (M) \triangleq \left\{ j\,\vert\, M_{ij} >0,\,\, j\in \mathcal{V} \right\}$ and $\mathcal{N}^{\text{out}}_i (M) \triangleq \left\{ j\,\vert\, M_{ji} >0,\,\, j\in \mathcal{V} \right\}$ to denote the sets of in-neighbors and out-neighbors that node $i$ can communicate with.

\subsection{Push-Pull Protocol}
 The proposed algorithm is developed based on synchronous push-pull algorithm proposed in~\cite{cai2012average}, which has been recently employed to solve the ERM problem~\eqref{global_loss_function} over general digraphs~\cite{pu2020push}. The update of each node $i$ is as follows
\begin{subequations} \label{Alg_syn}
 \begin{alignat}{2}
&x_{i}^{t+1}=\sum_{j=1}^n{w_{ij}\left( x_{j\!\:}^{t}-\gamma ^tz_{j}^{t} \right)}
\\
&z_{i}^{t+1}=\sum_{j=1}^n{a_{ij}z_{j\!\:}^{t}}+\nabla f_i\left( x_{i}^{t+1} \right) -\nabla f_i\left( x_{i}^{t} \right) 
\end{alignat}
\end{subequations}
with $z_i^0=\nabla f_i(x_i^0)$ for $\forall i\in \mathcal{V}$.
In~\eqref{Alg_syn}, $x_i^t$ and $z_i^t$ are two estimates maintained by node $i$ for the global optimal decision variable and the global gradient $\frac{1}{n} \nabla F$, respectively. It relies on two properly designed weight matrices for information exchange among nodes, that is, each node $i$ uses $W \triangleq [w_{ij}]_{n\times n}$ to pull information of the decision variable from its in-neighbors $j\in \mathcal{N}_{i}^{\mathrm{in}}(W)$ for reaching consensus, while using $A \triangleq [a_{ij}]_{n\times n}$ to push gradient information to its out-neighbors $j\in \mathcal{N}_{i}^{\mathrm{out}}(A)$ for tracking the global gradient. To ensure the convergence of the update \eqref{Alg_syn}, the following assumptions are imposed on the two weight matrices as well as their induced graphs~\cite{pu2020push}. 

\begin{Ass} [On the weight matrices]\label{Ass_weight_matrix}
Let ${M}\in \mathbb{R}^{n\times n}$ denote either ${A}$ or ${W}$. We have: \\
\romannumeral1) $M_{ii}> 0$, $\forall i\in \mathcal{V}$; and there exists $\bar{m} > 0$ such that $\min \left\{M_{ij} \,\vert \, M_{ij} > 0\right\} \geq \bar{m}$.\\
\romannumeral2)
$W$ is row-stochastic and $A$ is column-stochastic, i.e., $W\mathbf{1}=\mathbf{1}$ and $\mathbf{1}^\top A=\mathbf{1}^\top$. 
\end{Ass}

\begin{Ass}[On the induced graphs] \label{Ass_graph}
The graph $\mathcal{G}(W)$ and $\mathcal{G}({A}^{\top})$ each contains at least one spanning tree, and at least one pair of the spanning trees of $\mathcal{G}(W)$ and $\mathcal{G}({A}^\top)$ share a common root.  That is, $\mathcal{R}\triangleq \mathcal{R}_{{W}}\,\,\cap \,\,\mathcal{R}_{{A}^\top}\,\,\ne \,\,\varnothing $, where $\mathcal{R}_{{W}}$ (resp. $\mathcal{R}_{{A}^{\top}}$) denotes the set of roots of all possible spanning trees in the graph $\mathcal{G}({{W})}$ (resp. $\mathcal{G}({{A}^{\top}})$). Define $r\triangleq \left| \mathcal{R} \right|$.
\end{Ass}

\begin{figure}[tb]
    \centering
    {
        \begin{minipage}[t]{0.48\textwidth}
        \centering          
         \includegraphics[width=\textwidth]{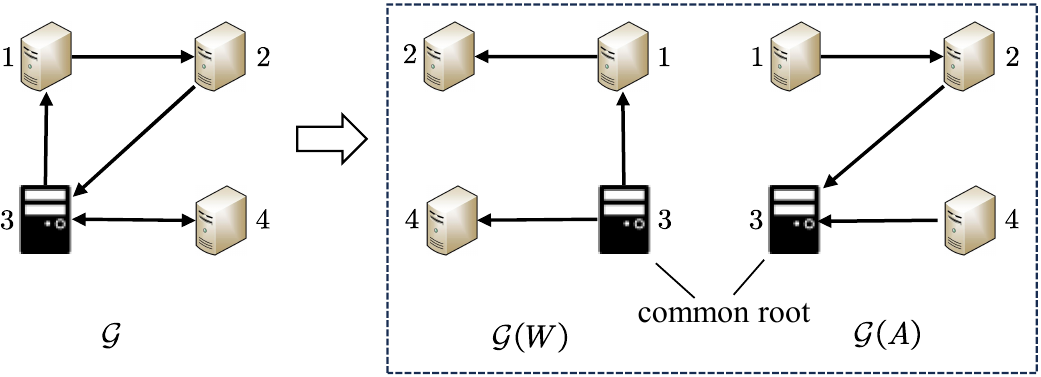}
        \end{minipage}%
    }
    \caption{An illustration of constructing two communication graphs over a strongly connected topology $\mathcal{G}$ wiht 4 nodes. $\mathcal{G}\left( W \right)$ (resp., $\mathcal{G}\left( A \right)$) is a (resp., reversed) spanning tree with node $3$ being the common root.}
    \label{spanning tree}
\end{figure}

\begin{Rem}
Assumption~\ref{Ass_graph} imposes the minimum requirement to guarantee an unobstructed information flow among nodes, which is much weaker than requiring the two communication graphs $\mathcal{G}(W)$ and $\mathcal{G}({A})$ be strongly connected. The assumption of being strongly connected has been used in most existing decentralized algorithms~\cite{xin2018linear, xin2019distributed} which limits the flexibility of graph design. Suppose we have an underlying topology $\mathcal{G}$, which is minimal strongly connected, i.e., removing any edge breaks strong connectivity. The approach to construct communication graphs $\mathcal{G}\left( W \right)$ and $\mathcal{G}\left( A \right)$ required by \cite{xin2018linear, xin2019distributed} is to set $\mathcal{G}\left( W \right) = \mathcal{G}\left( A \right) = \mathcal{G}$. In contrast, our Assumption~2 allows more flexible ways to construct graphs $\mathcal{G}\left( W \right)$ and $\mathcal{G}\left( A \right)$. For instance, we can pick any node $i \in \mathcal{V}$, and let $\mathcal{G}\left( W \right)$ (resp., $\mathcal{G}\left( A \right)$) be a spanning tree (resp., reversed spanning tree) of $\mathcal{G}$ with root $i$, as illustrated in Figure~\ref{spanning tree}. More examples about the flexibility of graph design can be found in Appendix~\ref{weghit_matrix_used} in the supplemental material.
\end{Rem}

\section{The Proposed R-FAST Algorithm}
\label{algorithm_development}
In this section, we present the proposed R-FAST method which employs three key schemes to allow it to be implemented in a fully asynchronous manner and enhance its robustness and efficiency, which are depicted as follows:
\begin{itemize}
\setlength{\itemsep}{0.02cm}
\item [i)] \textbf{Fully-Asynchronous mechanism.} Each node performs autonomous computation and communication at its own pace without any synchronization; 
\item [ii)] \textbf{Communication over spanning trees.} Nodes exchange messages over two subgraphs $\mathcal{G}(W)$ and $\mathcal{G}(A)$ induced by the weight matrices $W$ and $A$ respectively, so long as each contains a spanning tree and both have at least one common root;
\item [iii)] \textbf{Robust gradient tracking.} We employ the standard gradient tracking scheme to track the global gradient. Additionally, we introduce the auxiliary running-sum variable $\rho_{ji}$ accumulating the tracking variable $z_i$ (up to some scaling factor) and the buffer variable $\tilde{\rho}_{ij}$ to tackle packet losses.
\end{itemize}
Now, we proceed to the specific implementation of the algorithm, as summarized in Algorithm~\ref{Myalgorithm_R-FAST}. It depicts the updates of any node, from a local perspective. Being \textit{blind} to the progress of the entire system, every node $i$ maintains a local iteration counter $t_i$ (the subscript $i$ is omitted from $i$'s local view to simplify notation), and updates at its own pace in a fully asynchronous way.

\begin{figure}[tb]        
\centering          
 \includegraphics[width=0.4\textwidth]{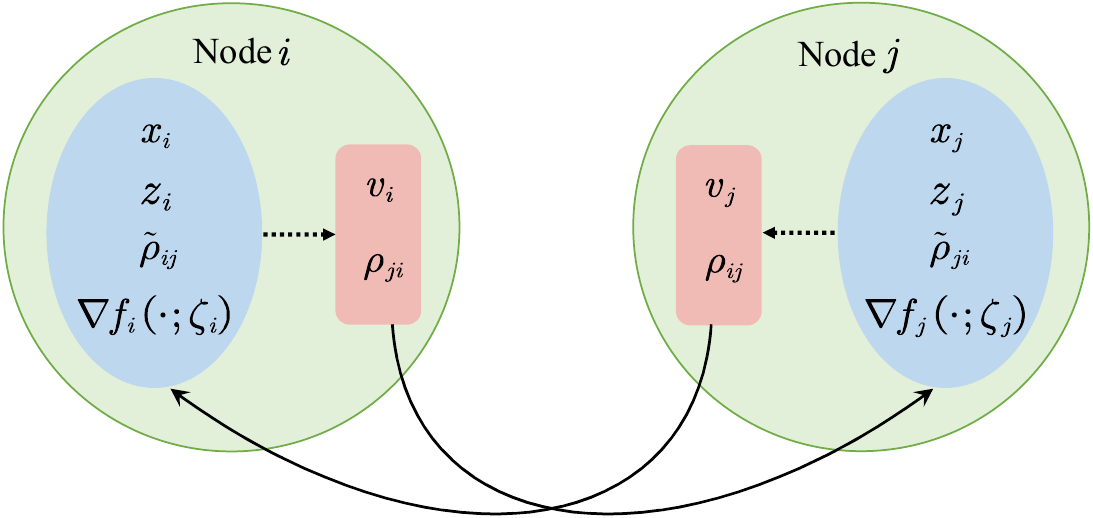}
\caption{An illustration of the storage and communications of variables within each node. The blue ellipse contains the \textit{private} variables meant for internal computations, and the red rectangle contains the \textit{communicating} variables to be sent to out-neighbors.  Solid lines indicate actual communications in the network while dotted lines indicate internal computational dependencies. }
\label{communication pattern}
\end{figure}

\begin{center}
    \begin{algorithm}
      \caption{\textbf{R}obust \textbf{F}ully \textbf{A}synchronous \textbf{S}tochastic Gradient \textbf{T}racking (R-FAST)} 
      \label{Myalgorithm_R-FAST} 
      \begin{algorithmic}[1]
        \State \textbf{Initialization:}  $x_{i}^{0}\in \mathbb{R}^p$, $z_{i}^{0}=\nabla f_i(x_{i}^{0};\zeta _{i}^{0})$,  $\tau _{v,ij}^0 = \tau _{\rho, ij}^0 =0$,  $v_{i}^0 = \rho _{ij}^0 =  \tilde{\rho}_{ij}^{0}=0$, $t=0$.
        \While{each node $i$ asynchronously}
        \State (S1) \textbf{Perform local descent:}\quad 
        $ v_{i}^{t+1}=x_{i}^{t}-\gamma^{t} z_{i}^{t}  $
        \State (S2) \textbf{Process received messages:}
        \State \qquad a)
          $x_i^{t+1}=w_{ii}v_{i}^{t+1}+\sum_{j\in \mathcal{N}_{i}^{\text{in}}\left( {W} \right)}{w_{i j}v_{j}^{\tau _{v,ij}^{t}}}$ 
        \State 
          with  $\tau _{v,ij}^{t}$ being the largest local iteration stamp received from $j \in \mathcal{N}_{i}^{\text{in}}\left( {W} \right) $
        \State \qquad b)
        $z_{i}^{t+\frac{1}{2}}=z_{i}^{t}+\sum_{j\in \mathcal{N}_{i}^{\text{in}}\left( {A} \right)}{\left( \rho _{ij}^{\tau _{\rho, ij}^{t}}-\tilde{\rho}_{ij}^{t} \right)}+\nabla f_i (x_i^{t+1};\zeta _i^{t+1})-\nabla f_{i}(x_{i}^{t};\zeta _{i}^{t})$
        \State 
          with  $\tau _{\rho, ij}^{t}$ being the largest local iteration stamp received from $j \in \mathcal{N}_{i}^{\text{in}}\left( {A} \right)  $ 
        \State \qquad c) $z_{i}^{t+1}=a_{i i}z_{i}^{t+\frac{1}{2}}$; \quad $\rho _{ji}^{t+1}=\rho _{ji}^{t}+a_{ji}z_{i}^{t+\frac{1}{2}}$
        \State (S3) \textbf{Send information to its out-neighbors:}
        \State \qquad a) Send $(t+1, \,\, v_{i}^{t+1})$ to every $ j\in \mathcal{N}_{i}^{\text{out}}\left( {W} \right)$
        \State \qquad b) Send $(t+1,\,\,  \rho _{ji}^{t+1})$ to every $ j \in \mathcal{N}_{i}^{\text{out}}\left( {A} \right)$
        \State (S4) \textbf{Update buffer:}  \quad $\tilde{\rho}_{ij}^{t+1}=\rho _{ij}^{\tau _{\rho,ij}^{t}},\quad \forall j\in \mathcal{N}_{i}^{\text{in}}\left( {A} \right)$
        \State (S5) \textbf{Increase local iteration counter:} \quad $t = t+1$
        \EndWhile
        \State \textbf{end}
        \end{algorithmic}
    \end{algorithm}
\end{center}
The algorithm consists of three main steps: \textbf{i) Local descent.} With $z_i^t$ being a proxy to the stochastic $\nabla F(\cdot)$, each node $i$ firstly performs an approximate stochastic gradient descent on $x_i^t$ using step size $\gamma^t$ at (S1), generating the intermediate result $v_i^{t+1}$.  The node will then communicate with its neighbors over the two directed sub-graphs $\mathcal{G}(W)$ and $\mathcal{G}(A)$ in order to: 1) force consensus on $v_i^{t+1}$ to form the next estimate $x_i^{t+1}$, and 2) update the gradient tracking variable $z_i^t$. 
\textbf{ii) Local communication for consensus and gradient-tracking.} At step (S2, a), the node forms $x_i^{t+1}$ as a weighted average of its most recent  $v_i^{t+1}$ and the $v$ variables received from its in-neighbors, using the weights $w_{ij}.$  For any $v_j$ variable from $j \in \mathcal{N}_{i}^{\text{in}}(W)$, node $i$ will not be able to understand the extent of how outdated $v_j$ currently is.  However, among all $v_j$ variables being available to $i$, node $i$ is able to pick up the most updated one, because each node  attaches its local iteration counter for any information packet it sends out.  Thus we use $\tau^t_{v,ij}$ to denote $j$'s largest local iteration counter attached to the $v_j$ variable, which is picked up by node $i$ when forming $x_i^{t+1}.$  Regarding the update of $z$ variables, we adopt the robust tracking method from \cite{tian2020achieving}.  Node $i$ maintains variables $\rho_{ji}$  as the running sum of $z_i$ (up to some scaling factor) over time for each out-neighbor $j\in \mathcal{N}_{i}^{\text{out}}(A),$ and also a buffer variable $\tilde{\rho}_{ij}$ for each in-neighbor  $j\in \mathcal{N}_{i}^{\text{in}}(A).$  It picks up the most updated $\rho_{ij} $ for each $j\in \mathcal{N}_{i}^{\text{in}}(A)$ among all the available.  The index notation $\tau_{\rho, ij}^t$ possesses the same logic as $\tau_{v, ij}^t$, but for the $\rho$ variable now.  The difference $\rho_{ij}^{\tau_{\rho,ij}^t} - \tilde{\rho}_{ij}^t$ captures the sum of the scaled $ z_j$ variables generated by $j$ but unseen by $i$ until $i$'s $t$-th local iteration.  In addition, it samples a random local gradient $\nabla f_i(x_i^{t+1};\zeta_i^{t+1})$ at $x_i^{t+1}$.  With all the information ready, node $i$ generates $z_i^{t+1/2}$ at (S2, b), through adding the new gradient sample $\nabla f_i(x_i^{t+1};\zeta_i^{t+1})$ while clearing out the old $\nabla f_i(x_i^{t};\zeta_i^{t})$.
 Afterwards, $z_i^{t+1/2}$ is used to update $z_i^{t+1}$ and $\rho_{ji}^{t+1}$ respectively at step (S2, c).  \textbf{iii) Update the buffer variables.} After sending the information packets to its out-neighbors at (S3), node $i$ set the buffer variable $\tilde{\rho}$ as most recently consumed $\rho$ variable at step (S4). Lastly, it increase the local iteration counter at (S5).
Thanks to the introduction of robust gradient tracking scheme, the proposed algorithm is able to deal with unpredictable packet losses due to the fact the running-sum variable $\rho$ maintains all the historical information regarding the gradient $z$.  To further illustrate the above process, Fig.~\ref{communication pattern} is introduced to show the role of variables maintained by each node in Algorithm~\ref{Myalgorithm_R-FAST}.

\begin{center}
    \begin{algorithm}
      \caption{\textbf{R}obust \textbf{F}ully \textbf{A}synchronous \textbf{S}tochastic
gradient \textbf{T}racking (R-FAST, Global View)} 
      \label{Myalgorithm_RASGT_GV} 
      \begin{algorithmic}
        \State \textbf{Initialization:} 
        $x_{i}^{0}\in \mathbb{R}^p$, $z_{i}^{0}=\nabla f_i(x_{i}^{0};\zeta _{i}^{0})$,  $v_{i}^0 = \rho _{ij}^0 =  \tilde{\rho}_{ij}^{0}=0$;\,\, $\rho _{ij}^{\ell}={v}_{i}^{\ell}=0$,\,\, $\forall \, \ell\in \{-D,-D+1,...,0\}$;\,\, $t_i^0=k=0$ 
        \While{a termination criterion is not met}
        \State  (S.0) \textbf{Pick:} $i^k$, $d_{v,j}^k$ for $\forall \, j\in \mathcal{N}_{i^k}^{\text{in}}(W)$, $d_{\rho,j}^k$ for $\forall \, j\in \mathcal{N}_{i^k}^{\text{in}}(A)$  
        \State (S.1) \textbf{Perform local descent:} \quad 
        $
            {v}_{i^k}^{k+1}={x}_{i^k}^k-\gamma^k{z}_{i^k}^k
        $
        \State (S.2) \textbf{Process received messages:}
        \State \qquad a)
          ${x}_{i^k}^{k+1}=w_{i^ki^k}{v}_{i^k}^{k+1}+\sum_{j\in \mathcal{N}_{i^k}^{\text{in}}(W)}w_{i^kj}v_{j}^{k-d_{v,j}^k}$ 
        \State \qquad b)
        ${z}_{i^k}^{k+\frac{1}{2}}={z}_{i^k}^{k}+\sum_{j\in \mathcal{N}_{i^k}^{\text{in}}(A)}{\left( \rho _{i^kj}^{k - d_{\rho,j}^k}-\tilde{\rho}_{i^kj}^{k} \right)}
        +\nabla f_{i^k}({x}_{i^k}^{k+1};\zeta_{i^k}^{k+1})-\nabla f_{i^k}({x}_{i^k}^k;\zeta_{i^k}^{k})$
        \State \qquad c) ${z}_{i^k}^{k+1}=a_{i^ki^k}{z}_{i^k}^{k+\frac{1}{2}}, \qquad \rho _{ji^k}^{k+1}=\rho_{ji^k}^k+a_{ji^k}{z}_{i^k}^{k+\frac{1}{2}},  \quad \forall j\in \mathcal{N}_{i^k}^{\text{out}}(A)$
        \State  (S.3) \textbf{Send information to out-neighbors:}
        \State \qquad a) Send $(t_{i^k}^k+1,\,\,  v_{i^k}^{k+1})$ to every $j\in \mathcal{N}_{i^k}^{\text{out}}(W)$
        \State \qquad b) Send $(t_{i^k}^k+1,\,\,\rho_{ji^k}^{k+1})$ to every $j\in \mathcal{N}_{i^k}^{\text{out}}(A)$
        \State  (S.4) \textbf{Update buffer:} \quad 
            $\tilde{\rho}_{i^kj}^{k+1}=\rho _{i^kj}^{k - d_{\rho,j}^k}, \quad \forall j\in \mathcal{N}_{i^k}^{\text{in}}(A)$
        \State (S.5) \textbf{Increase iteration counter:}
        \State \qquad a) $t_{i^k}^k =  t_{i^k}^k+1 $
        \State \qquad b) Untouched state variables shift to state $k+1$ while keeping the same value
        \State \qquad c) $ k\gets k+1$
        \EndWhile
        \end{algorithmic}
    \end{algorithm}
\end{center}

\textbf{From local view to global view.} While Algorithm~\ref{Myalgorithm_R-FAST} and the above description are from the local view of node $i$ in order to facilitate the understanding of the training process of R-FAST, we need a global iteration counter $k$ to measure the progress of the entire system and raise theoretical convergence analysis.  Algorithm~\ref{Myalgorithm_RASGT_GV} depicts the global view.  The increase on the global iteration counter $k \leftarrow k+1$ is triggered whenever any node finishes one local iteration, and we use $i^k$ to indicate such node. $t_{i^k}^k$ is $i^k$'s local iteration counter.  Posterior to the asynchronous process, we define: i) $l:\,k \to (i, t) $, mapping the global iteration counter $k$ to the associated active node $i = i^k$ and its local iteration counter $t$ as of the $k$-th global iteration; and ii) the inverse of $l$ as $g: (i, t) \to k.$  With these tools, we can measure the delayed extent of any received packet 
$v_{j}^{\tau _{v,ij}^{t}}$ in the global view: since $v_{j}^{\tau _{v,ij}^{t}}$ is generated at node $j$'s  local iteration $ \tau _{v,ij}^{t}-1$ and equivalently the global iteration $g(j, \,\tau _{v,ij}^{t}-1)$, the delay of the packet in global view is thus $d_{v,j}^k = k - g(j, \,\tau _{v,ij}^{t}-1)-1$, with $k = g(i, t)$. We define $d_{\rho,j}^k$ similarly for the $\rho$ variables. We consider the following asynchronous model in this work.

\begin{Ass}[On the asynchronous model] \label{Ass_asyn} \text{}\\
\romannumeral1) $\exists \, T\in \mathbb{N}$ and $T\geqslant n$ such that $\cup _{t=k}^{k+T-1}i^t=\mathcal{V}$,\,\,  for all $k\in \mathbb{N}$;\\
\romannumeral2) $\exists \, D\in \mathbb{N}$ such that $0\leqslant d_{v,j}^{k},\, d_{\rho, j}^{k} \leqslant D$,\,\,  for all $k\in \mathbb{N}$.
\end{Ass}
The above assumption rules out scenarios where some nodes have infinitely long inactive time and/or some transmission links fail for infinitely long time.  
The following remark shows that the synchronous counterpart of R-FAST, indeed, falls into our asynchronous model.

\begin{Rem}
\label{rem:sync}
In Algorithm~\ref{Myalgorithm_R-FAST}, if we have $\tau_{v,ij}^t =\tau_{\rho,ij}^t = t$ hold for $\forall\, i,\,j,\, t$, the algorithm reduces to a synchronous one.  
For example, to compute $x_i^{t+1}$, agent $i$ needs to wait until all $v_j^{t}$ from $\forall \, j \in  \mathcal{N}_{i}^{\text{in}}\left( {W} \right)$ become available.  The same coordination applies to the update of the $z$ variable.  To cast the synchronous scenario into the global view introduced above, one can set $i^k = (k\, \text{mod} \,n) + 1$, i.e., $\{i^k\}_{k=0}^{\infty}\,=\, \{1,2,3,\ldots,n,1,2,\ldots\}.$  Therefore, we have $d_{v,j}^k = d_{\rho,j}^k = k - g(j,t-1)-1 = k - g\left(j,\lfloor\frac{k}{n}\rfloor-1 \right)-1 = k-(\lfloor\frac{k}{n}\rfloor-1)n - j < 2n-2.$  Clearly, synchronous updates 
satisfies Assumption~\ref{Ass_asyn} with $T = 
n$ and $D = 2n-2.$
\end{Rem}

\section{Convergence Analysis}
\label{Convergence Analysis}
In this section, we present the theoretical convergence results of the proposed R-FAST (Algorithm~\ref{Myalgorithm_RASGT_GV}). Beforehand,
we denote $x^k \triangleq \left[ x_{1}^{k},x_{2}^{k},\ldots, x_{n}^{k} \right]^\top \in \mathbb{R}^{n\times p}$  as the concatenation of all local variables, and $\bar{x}^k\triangleq \frac{1}{n}\sum_{i=1}^n{x_{i}^{k}}\in \mathbb{R}^{p} $;
we use $x^{\star}\in \mathbb{R}^{p}$ to denote the optimal solution of problem \eqref{global_loss_function} when $F$ is strongly convex; $\left\| \cdot \right\| $ denotes the Frobenius norm and $I$ denotes the identity matrix.  
Given a matrix $M \triangleq (M_{ij})_{i,j=1}^n$, $M_{i,:}$ and $M_{:,j}$ denote its $i$-th row vector and $j$-th column vector. Given the sequence $\left\{ M^t \right\} _{t=s}^k$ with $k \geqslant  s$, $M^{k:s}\triangleq M^kM^{k-1}\cdot \cdot \cdot M^{s+1}M^s$ if $k>s$ and $M^{k:s}\triangleq M^s$ otherwise. 
The dimensions of the all-one vector $\mathbf{1}$ and the $i$-th canonical vector $e_i$ will be clear from the context.
$\left\| M \right\|_2 $ denotes the spectral norm of matrix $M$ and $[\cdot]_i$ represents the $i$-th component of a vector.
To facilitate reading, we also include Table~\ref{table_constants_definition} summarizing all the definitions of several frequently used problem-related constants.

\begin{table}[htbp!]
\centering
\caption{Definitions of proof-related constants}
\begin{tabular}{|c|c|}
    \hline
    \textbf{Constant} & \textbf{Definition} \\
    \hline
    $n$ & the number of nodes \\
    \hline
    $D$ & the maximum of message delays \\
    \hline
    $T$ & time period \\
    \hline
    $\bar{m}$ & lower bound of all non-zero mixing weights \\
    \hline
    $L_i$ & Lipschitz constant of function $f_i \left( \cdot \right)$ \\
    \hline
    $C_L$ & $\max \left\{ L_i \right\} _{i=1}^{n}$ \\
    \hline
    $L$ & $\sum_{i=1}^n{L_i}$ \\
    \hline
    $S$ & $n+\left( D+1 \right) \left| \mathcal{E}(A)\right|$ \\
    \hline
    $K_1$ & $\left( 2n-1 \right) \cdot T+n\cdot D$ \\
    \hline
    $\eta$ & $\bar{m}^{K_1}$ \\
    \hline
    $\rho$ & $(1-\eta)^\frac{1}{K_1}$ \\
    \hline
    $C$ & $2\frac{1+\bar{m}^{-K_1}}{1-\bar{m}^{K_1}}$ \\
    \hline
    $C_1$ & $\frac{2\sqrt{2S}\left( 1+\bar{m}^{-K_1} \right)}{\rho \left( 1-\bar{m}^{K_1} \right)}$ \\
    \hline
    $C_2$ & $\frac{2\sqrt{\left( D+2 \right) n}\left( 1+\bar{m}^{-K_1} \right)}{1-\bar{m}^{K_1}}$ \\
    \hline
\end{tabular}
\label{table_constants_definition}
\end{table}

\textbf{Sketch of Proofs.} To facilitate the reading of the proofs, we first provide the sketch of the proofs as follows:
\begin{itemize}
    \item[i)] We first develop an augmented system by adding virtual nodes into the graph $\mathcal{G}(W)$ and $\mathcal{G}(A)$ to virtually store the value of delayed variables that are transmitted on the way.  The augmented system transforms the asynchronous updates and communications into equivalent synchronous operations indexed by global iterations (c.f., Section~\ref{sec:PAC} and~\ref{graph_augmented_A}).
    \item[ii)] To overcome the challenge stemming from stochastic gradients, we use an virtual auxiliary sequence that resembles the recursion of tracking variables but with local full gradients.
    The auxiliary sequence is instrumental in quantifying the stochastic errors induced by random gradient sampling (c.f., Section~\ref{vitual_sequence_deal_with_sampling_error}).
    \item[iii)] We are then ready to analyze the convergence rate for the case of strongly convex objectives.    
    We start by connecting the dynamics of 
    the consensus error, gradient tracking error and optimality gap on the augmented graph, aiming to establish the evolving contraction of each  term up to perturbations (c.f., Proposition~\ref{proposition 1}).  However, the new challenge beyond the asynchronous model is our more relaxed Assumption~\ref{Ass_graph} on communication graphs.
    Specifically, in any global iteration triggered by a non-root node, the optimality gap of the overall networked system does not have a sufficient contraction, which renders the existing analysis techniques~\cite{zhang2019fully, tian2020achieving} invalid.
    To address this challenge, we employ a \emph{two-time-scale} technique to analyze over a sequence of iteration windows (c.f., proof of Proposition~\ref{proposition 1}).  Within each iteration window, all root nodes are activated at least once, ensuring a sufficient contraction of the optimality gap
    (c.f.,~\eqref{eq_4_in_pro1}).
    Finally, we derive the linear convergence result utilizing the generalized small-gain theorem~\cite{tian2020achieving}, with the additional consideration of gradient noises.
    \item[iv)] For general non-convex objectives, due to our more relaxed Assumption~\ref{Ass_graph} again, the vanilla descent lemma~\cite{tian2020achieving, kungurtsev2021decentralized} based on the smooth property no longer can be established at every global iteration. To overcome the challenge, we develop a new descent lemma (c.f., Lemma~\ref{new_lemma7}) leveraging \emph{two-time-scale} techniques that ensure a valid descent towards the stationary point up to perturbations constituted by consensus error and gradient tracking error. Building on this lemma, we manage to arrive at a sublinear convergence rate by bounding the aforementioned two errors with the norm of gradient vectors which vanishes over time.
\end{itemize}

In addition, we make the following assumptions on the objectives, which are commonly used in the decentralized stochastic optimization methods~\cite{xin2019distributed}. 

\begin{Ass}[Smoothness]\label{Ass_smoo}
\label{assumption1}
Each $f_i\left( \cdot \right), \,\,i=1,...,n $ is
$L_i$-Lipschitz differentiable. Define $C_L\triangleq \max \left\{ L_i \right\} _{i=1}^{n}$, and $L\triangleq \sum_{i=1}^n{L_i}$.
\end{Ass}

\begin{Ass}[Bounded variance]\label{Ass_bounded_var}
The stochastic gradient $\nabla f_i\left( x;\zeta _i \right)$, $\zeta _i\sim \mathcal{D}_i$ generated by each node $i$ is unbiased with bounded variance, i,e.
$\mathbb{E}\left[ \nabla f_i\left( x;\zeta _i \right) \right] =\nabla f_i\left( x \right)$ and there exists a constant $\sigma \geqslant 0$ such that $\mathbb{E}\left[ \left\| \nabla f_i\left( x;\zeta _i \right) -\nabla f_i\left( x \right) \right\| ^2 \right] \leqslant \sigma ^2$.
\end{Ass}

\subsection{Augmented System}\label{sec:aug_system}
In this section, we first recast the proposed R-FAST method (Algorithm~\ref{Myalgorithm_RASGT_GV}) as an \textit{\textbf{augmented system}} to deal with the delays. In particular, leveraging \textit{augmented} graphs, we can account for the delay by adding virtual nodes into the graph $\mathcal{G}(W)$ and $\mathcal{G}(A)$ to represent the value of delayed variables that are transmitted on the way, yielding an equivalent synchronous algorithm with augmented weight matrices $\hat{W}^k$ (resp. $\hat{A}^k $), corresponding to $W$ (resp. $A $), that governs the information exchange over the augmented graphs.

\subsubsection{\textbf{An augmented system for consensus scheme (augmented graph of $\mathcal{G}(W)$)}}
\label{sec:PAC}
We first add $D+1$ virtual nodes for each node $i$, denoted by $i[0], i[1],...,i[D]$, to store delayed information $v_i^{k}, v_i^{k-1},...,v_i^{k-D}$. It should be noted that any virtual node $i[d], d=D,D-1,...1$ can only receive information from virtual node $i[d-1]$; besides, $i[0]$ can only receive information from the real node $i$ or keep the value unchanged. Define ${v}^k \triangleq [{v}_1^k, \cdots, {v}_n^k]^{\top}\in \mathbb{R}^{n\times p}$. 
Then, the consensus scheme can be rewritten as an augmented system  as follows
\begin{align}
\label{iterate of h^k}
{h}^{k+1} = \hat{W}^k({h}^{k} - \gamma^k e_{i^k} \left( z_{i^k}^k \right )^{\top}),
\end{align}
where \begin{equation}
\label{def_of_h_vector}
h^{k} \triangleq [{{x}^k}; {{v}^k}; {{v}^{k-1}}; \cdots; {{v}^{k-D}}]\in \mathbb{R}^{\left( D+2 \right) n\times p},
\end{equation}
and $\hat{W}^k \in \mathbb{R}^{\left( D+2 \right) n\times \left( D+2 \right) n}$  denotes the \textit{row stochastic} augmented matrix, whose definition can be found at~\eqref{def_of_hat_W} in the supplemental material. For more details about the augmented system for consensus, the readers are referred to Appendix~\ref{appendix_augmented_graph_of_W} available in the supplemental material.

The following lemma captures the asymptotic behavior of  $\hat{W}^k$ over spanning-tree graphs, whose proofs can be adapted from the proof of Lemma 17 in~\cite{tian2020achieving}.
 \begin{Lem}
\label{lemma_contraction_W^k}
Suppose Assumptions~\ref{Ass_weight_matrix}-\ref{Ass_asyn} hold.
Define $K_1 \triangleq \left( 2n-1 \right) \cdot T+n\cdot D$, $C_2 \triangleq \frac{2\sqrt{\left( D+2 \right) n}\left( 1+\bar{m}^{-K_1} \right)}{1-\bar{m}^{K_1}}$, $\eta \triangleq \bar{m}^{K_1}$ and $\rho \triangleq (1-\eta)^\frac{1}{K_1}$. We have for any $k \geqslant t \geqslant 0$: i) $\hat{W}^k$ is row stochastic; ii) there exists a sequence
of stochastic vectors $\left\{ \psi ^k \right\} _{k \geqslant 0}$ such that $\psi _{i}^{k}\geqslant \eta$ for all  $i\in \mathcal{R}_{{W}} $, and
\begin{equation}
    \left\| {\hat{W}}^{k:t}-\mathbf{1}\left( \psi ^t \right) ^{\top} \right\|_2 \leqslant C_2\rho ^{k-t}.
\end{equation}
\end{Lem}

Next, we introduce an auxiliary variable 
\begin{equation}
\label{new_def_of_x_psi_0}
x_{\psi}^{k} \triangleq \left( \psi ^k \right) ^{\top}h^k
\end{equation}
to construct a weighted average system corresponding to ~\eqref{iterate of h^k}, whose evolution will be crucial to our subsequent analysis in upper-bounding the consensus error and optimality gap. 
In particular, using~\eqref{new_def_of_x_psi_0}, with~\eqref{iterate of h^k} and the fact $\left( \psi ^{k+1} \right) ^{\top}\hat{W}^k=\left( \psi ^k \right) ^{\top}$~\cite{tian2019asynchronous}, we get
\begin{equation}
\label{each_iterate_of_x_psi_k}
\CV^{k+1} = \CV^{k} - \gamma^k (\psi^k)^{\top} e_{i^k} (z_{i^k}^k)^{\top}
=\CV^{k} - \gamma^k \psi_{i^k}^k (z_{i^k}^k)^{\top}.
\end{equation}

\subsubsection{\textbf{An augmented system for gradient tracking scheme (augmented graph of $\mathcal{G}(A)$)}}
\label{graph_augmented_A}
For the augmented system of gradient tracking scheme, different from that of the consensus scheme, we add $D+1$ virtual nodes for each edge $\left( j,i \right) \in \mathcal{E}\left( A \right)$, denoted by $(j,i)^0, (j,i)^1,...,(j,i)^D$, to store the delayed information $z_{(j,i)^0}^k,z_{(j,i)^1}^k,...,z_{(j,i)^D}^k$
that has been generated by node $j \in \mathcal{N}_i^{\mathrm{in}}(A)$ for node $i$ but not received by node $i$ yet.
Then, we define the set of real and virtual nodes  as $\hat{\mathcal{V}}\triangleq \mathcal{V}\cup \left\{ \left( j,i \right) ^d\left| \left( j,i \right) \in \mathcal{E}(A) , \right. d=0,1,...,D \right\} 
$
and its cardinality as $S\triangleq \left| \hat{\mathcal{V}} \right|=n+\left( D+1 \right) \left| \mathcal{E}(A)\right| $.  
We use $z_{s^d}^{k}$ to denote $z_{\left( j,i \right) ^d}^{k}$ if $(j,i)$ is the $s$-th edge of $\mathcal{E}(A) $.
Define $z^k\triangleq \left[ z_{1}^{k},z_{2}^{k},...,z_{n}^{k} \right] ^{\top}\in \mathbb{R}^{n\times p}$ and $z_{\mathcal{E}\left( A \right) ^d}^{k}\triangleq \left[ z_{1^d}^{k},z_{2^d}^{k},...,z_{\left| \mathcal{E}\left( A \right) \right|^d}^{k} \right] ^{\top}\in \mathbb{R}^{\left| \mathcal{E}\left( A \right) \right|\times p}$. 
Then, the gradient tracking scheme can be rewritten as an augmented system as follows:
\begin{equation}
\label{iterate_of_z_k}
{\hat{z}}^{k+1}={\hat{A}}^k{\hat{z}}^k+{P}^k{e}_{i^k}(\epsilon ^k)^{\top},
\end{equation}
where 
\begin{equation}
\label{def_hat_z^k}
\hat{z}^k\triangleq \left[ z_{i}^{k} \right] _{i=1}^{S}=\left[ z^k;z_{\mathcal{E}\left( A \right) ^0}^{k};z_{\mathcal{E}\left( A \right) ^1}^{k}...;z_{\mathcal{E}\left( A \right) ^D}^{k} \right] \in \mathbb{R}^{S\times p}
\end{equation}
denotes the concatenated tracking variables 
with initialization $z_{i}^{0}=\nabla f_i(x_{i}^{0};\zeta_i^0)$ for $i \in  \mathcal{V}$  and $z^{0}_i = 0$ for $i \in  \widehat{\mathcal{V}} \setminus \mathcal{V}$; 
$P^k\in \mathbb{R}^{S \times S}$ and $\hat{A}^k\in \mathbb{R}^{S \times S}$ are two \textit{column stochastic} augmented matrices defined at~\eqref{def_of_P_k} and~\eqref{def_of_hat_A_k} respectively in the supplemental material,
and 
\begin{equation}
\label{pertubtion_stochastic_gradient}
    \epsilon ^k=\nabla f_{i^k}\left( x_{i^k}^{k+1};\zeta _{i^k}^{k+1} \right) -\nabla f_{i^k}\left( x_{i^k}^{k};\zeta _{i^k}^{k} \right).
\end{equation}
More details about the augmented system for gradient tracking scheme can be found in Appendix~\ref{appendix_augmented_graph_of_A} available in the supplemental material. 

Similar to Lemma~\ref{lemma_contraction_W^k}, the asymptotic behavior of $\hat{A}^k$ is depicted in the following lemma.

\begin{Lem}
\label{lemma_contraction_A^k}
Suppose Assumptions~\ref{Ass_weight_matrix}-\ref{Ass_asyn} hold and define $C \triangleq 2\frac{1+\bar{m}^{-K_1}}{1-\bar{m}^{K_1}}$. We have for any $k \geqslant t \geqslant 0$: i) $\hat{A}^k$ is column stochastic; ii) there exists a sequence
of stochastic vectors $\left\{ {\xi }^k \right\} _{k \geqslant 0}$ such that $\xi _{i}^{k}\geqslant \eta$ for all $i\in \mathcal{R}_{{A}^{\top}} $, and
\begin{equation}
\left| {\hat{A}}_{ij}^{k:t}-\xi _{i}^{k} \right|\leqslant C\rho ^{k-t}
\end{equation}
for all $i,j\in \left\{ 1,...,S \right\}$.
\end{Lem}
The following Lemma provides the conservation property of the above gradient-tracking scheme.
\begin{Lem}
\label{lemma tracking_property}
Suppose Assumptions~\ref{Ass_weight_matrix}-\ref{Ass_asyn} hold. Let $\left \{ \hat{z}^k \right \}_{k=0}^{\infty}$ be the sequence generated by the system~\eqref{iterate_of_z_k}, we have for all $k\geqslant 0$,
\begin{equation}
\label{eq:tracking_property}
\mathbf{1}^{\top}\hat{z}^k=\left(\sum_{i=1}^n{\nabla f_i\left( x_{i}^{k};\zeta _{i}^{k} \right)}\right)^{\top}.
\end{equation}
\end{Lem}

\begin{proof}
Applying \eqref{iterate_of_z_k} telescopically, yields:
$\hat{z}^k=\hat{A}^{k-1:0}\hat{z}^0+\sum_{l=1}^{k-1}{\hat{A}^{k-1:l}P^{l-1}e_{i^{l-1}}\left(\epsilon ^{l-1}\right)^\top}+P^{k-1}e_{i^{k-1}}\left(\epsilon ^{k-1}\right)^{\top}$.
Left multiplying $\mathbf{1}^{\top}$ from both sides of the above equation, and using the column  stochasticity of $\hat{A}^k$ and $P^k$, we can get
\begin{equation*}
\begin{aligned}
&\mathbf{1}^{\top}\hat{z}^k=\mathbf{1}^{\top}\hat{z}^0+\sum_{l=0}^{k-1}{\left(\epsilon ^l\right)^\top}
=\left(\sum_{i=1}^n{\nabla f_i\left( x_{i}^{0};\zeta _{i}^{0} \right)}\right)^\top \\
&+\left(\sum_{l=0}^{k-1}{\left( \nabla f_{i^l}\left( x_{i^l}^{l+1};\zeta _{i^l}^{l+1} \right) -\nabla f_{i^l}\left( x_{i^l}^{l};\zeta _{i^l}^{l} \right) \right)}\right)^\top
\\
&\overset{\left( a \right)}{=}\left(\sum_{i=1}^n{\nabla f_i\left( x_{i}^{k};\zeta _{i}^{k} \right)}\right)^{\top},
\end{aligned}
\end{equation*}
where in $(a)$ we used $x_j^{l+1}=x_j^{l}$ and $\zeta_{j}^{l+1}=\zeta_{j}^{l}$ for $j\ne i^l$.
\end{proof}

\subsubsection{\textbf{Auxiliary sequence for tackling stochastic errors}}
\label{vitual_sequence_deal_with_sampling_error}
To tackle the stochastic error induced by random gradient sampling, we introduce an \textit{{auxiliary sequence}} $\left\{ \bar{z}_{i}^{k} \right\} _{i\in \mathcal{V}}$ (initialized as 
$\bar{z}_{i}^{0}=\nabla f_i\left( x_{i}^{0} \right)$ for $i\in \mathcal{V}$), which resembles the recursion of tracking variable $\left\{ z_{i}^{k} \right\} _{i\in \mathcal{V}}$ but uses the local full gradients for updating~\cite{kungurtsev2021decentralized}.
Let ${\bar{z}}^{k}$ denote the augmented auxiliary variable corresponding to the tracking variables $\hat{z}^k$. Then, in view of \eqref{iterate_of_z_k},
the recursion of the auxiliary variable $\bar{z}^k$ can be rewritten in a compact form as follows:
\begin{equation}
\label{compact_form_virtual}
    {\bar{z}}^{k+1}={\hat{A}}^k{\bar{z}}^k+{P}^k{e}_{i^k}(\bar{\epsilon}^k)^\top,
\end{equation}
where 
\begin{equation}
\label{virtual_sequence}
\bar{\epsilon}^k=\nabla f_{i^k}\left( x_{i^k}^{k+1} \right) -\nabla f_{i^k}\left( x_{i^k}^{k} \right),
\end{equation}
with $x_{i^k}^k$ being generated from Algorithm \ref{Myalgorithm_RASGT_GV}.

\begin{Rem}
The auxiliary variable $\left\{ \bar{z}_{i}^{k} \right\} _{i\in \mathcal{V}}$ indeed acts the same as the real tracking variable $\left\{ z_{i}^{k} \right\} _{i\in \mathcal{V}}$ except for employing local full gradients for updating. As we will show shortly, in a such design of the auxiliary sequence, the sum of gradients will be preserved over time, making it possible to quantify the difference between the auxiliary sequence and the real one. Clearly, this auxiliary sequence has no effect on the algorithm and, in fact, does not come into play when the algorithm runs, whose purpose is, instead, merely for convergence analysis.
\end{Rem}

Next we bound the variance of the tracking variable $z_{i^k}^k$ with respect
to its deterministic counterpart $\bar{z}_{i^k}^k$ (auxiliary sequence) using the following Lemma.

\begin{Lem}
\label{variance of the tracking and auxiliary}
Suppose Assumptions~\ref{Ass_weight_matrix}-\ref{Ass_asyn} and \ref{Ass_bounded_var} hold. We have for $k\geqslant 0$,
\begin{equation}
\label{new_byproduct}
\mathbb{E}\left[ \left\| z_{i^k}^{k}-\bar{z}_{i^k}^{k} \right\| ^2 \right] \leqslant n\sigma ^2.
\end{equation}
\end{Lem}

\begin{proof}
Similar to the proof of Lemma \ref{lemma tracking_property}, 
the auxiliary variables has the following property:
\begin{equation}
\label{auxiliary tracking_property}
\begin{aligned}
&\mathbf{1}^{\top}\bar{z}^k 
\\
&=\left(\sum_{i=1}^n{\nabla f_i\left( x_{i}^{0} \right)}+\sum_{l=0}^{k-1}{\left( \nabla f_{i^l}\left( x_{i^l}^{l+1} \right) -\nabla f_{i^l}\left( x_{i^l}^{l} \right) \right)}\right)^\top  
\\
& =\left(\sum_{i=1}^n\nabla f_i\left( x_{i}^{k} \right)\right)^\top .
\end{aligned}
\end{equation}

Therefore, we have
\begin{equation*}
\begin{aligned}
&\mathbb{E}\left[ \left\| \mathbf{1}^{\top}\hat{z}^k-\mathbf{1}^{\top}\bar{z}^k \right\| ^2 \right] 
\\
\overset{\eqref{eq:tracking_property}}{=} &\mathbb{E}\left[ \left\| \sum_{i=1}^n{\nabla f_i\left( x_{i}^{k};\zeta _{i}^{k} \right)}-\sum_{i=1}^n{\nabla f_i\left( x_{i}^{k} \right)} \right\| ^2 \right] 
\\
=&\sum_{i=1}^n{\mathbb{E}\left[ \left\| \nabla f_i\left( x_{i}^{k};\zeta _{i}^{k} \right) -\nabla f_i\left( x_{i}^{k} \right) \right\| ^2 \right]}
\\
&+\sum_{i\ne i^{\prime}} {\mathbb{E}\Bigg< \nabla f_i\left( x_{i}^{k};\zeta _{i}^{k} \right) -\nabla f_i\left( x_{i}^{k} \right) ,}
\\
&\nabla f_{i^{\prime}}\left( x_{i^{\prime}}^{k};\zeta _{i^{\prime}}^{k} \right) -\nabla f_{i^{\prime}}\left( x_{i^{\prime}}^{k} \right) \Bigg> \overset{\left( a \right)}{\leqslant} n\sigma ^2,
\end{aligned}
\end{equation*}
where in $(a)$ we have used Assumption \ref{Ass_bounded_var} and the fact that the inner product equals to zero since the two terms of the inner product are independent of each other.

Additionally,  knowing that
\begin{equation*}
\mathbb{E}\left[ {\hat{z}}^k-{\bar{z}}^k \right] =0,
\end{equation*}
we then obtain
\begin{equation*}
\mathbb{E}\left[ \left\| z_{i^k}^{k}-\bar{z}_{i^k}^{k} \right\| ^2 \right] \leqslant \mathbb{E}\left[ \left\| {\mathbf{1}}^{\top}{\hat{z}}^k-{\mathbf{1}}^{\top}{\bar{z}}^k \right\| ^2 \right] \leqslant n\sigma ^2,
\end{equation*}
which completes the proof.
\end{proof}

\subsection{Strongly Convex: Geometric convergence} \label{analysis_of_strongly_convex}
We first define the consensus error, gradient tracking error and optimality gap on the augmented graph and our objective is to prove that these errors will converge linearly to a ball. To this end, we derive recursions of these above error terms (c.f., Proposition \ref{proposition 1}) and, with the help of Lemma \ref{supporting lemma}, build an equivalent linear system of inequalities for the obtained error dynamics. By the generalized small gain theorem~\cite{tian2020achieving}, the linear convergence of the above error terms can be obtained, leading to the main results as given in Theorem \ref{Thm_Geo_conv}.

Let $\left\{ x_{i}^{k},v_{i}^{k},z_{i}^{k},\bar{z}_{i}^{k} \right\} _{k \geqslant 0,i\in \left[ n \right]}
$ be the sequence generated by R-FAST (Algorithm~\ref{Myalgorithm_RASGT_GV}). Let $E_{c}^{k}\triangleq \mathbb{E}\left[ \left\| {h}^k-\mathbf{1}x_{\psi}^{k} \right\|^2 \right] $, $E_{t}^{k}\triangleq \mathbb{E}\left[ \left\| \bar{z}_{i^k}^{k}-\xi _{i^k}^{k-1}({\bar{z}}^k)^{{\top}} \mathbf{1} \right\|^2 \right] $, $E_{z}^{k}\triangleq \mathbb{E}\left[ \left\| \bar{z}_{i^k}^{k} \right\|^2 \right] $ and $E_{o}^{k}\triangleq \mathbb{E}\left[ \left\| x_{\psi}^{k}-(x^{\star})^{\top} \right\|^2 \right] 
$ be the consensus error, gradient tracking error, the
magnitude of the tracking variable and optimality gap, respectively. The recursions of these above error terms are established in the following proposition.

\begin{Pro}
\label{proposition 1}
Consider the R-FAST (Algorithm~\ref{Myalgorithm_RASGT_GV}) with a constant step size $\gamma$. Suppose Assumptions \ref{Ass_weight_matrix}-\ref{Ass_bounded_var} hold. Then, we have for all $k\geqslant 0$
\begin{equation}
\label{eq_1_in_pro1}
\hspace{-2.6cm}
E_{c}^{k+1}\leqslant 2C_{2}^{2}E_{c}^{0}\cdot \rho ^{2k}+\frac{4\gamma ^2C_{2}^{2}}{1-\rho}\sum_{l=0}^k{\rho ^{k-l}E_{z}^{l}}
\end{equation}
\vspace{-0.3cm}
\begin{equation*}
\hspace{-5.4cm}
+\frac{4\gamma ^2C_{2}^{2}n}{1-\rho}\sum_{l=0}^k{\rho ^{k-l}\cdot \sigma ^2},
\end{equation*}

\begin{equation}
\hspace{-5.2cm}
E_{t}^{k+1}\leqslant 3C_{1}^{2}\left\| \bar{z}^0 \right\| ^2\cdot \rho ^{2k}
\end{equation}
\begin{equation*}
\hspace{-0.1cm}
+\sum_{l=0}^k{\rho ^{k-l}\left( \frac{27C_{1}^{2}C_{L}^{2}}{1-\rho}E_{c}^{l}+\frac{54\gamma ^2C_{1}^{2}C_{L}^{2}}{1-\rho}E_{z}^{l}+\frac{54\gamma ^2C_{1}^{2}C_{L}^{2}n}{1-\rho}\sigma ^2 \right)},
\end{equation*}
\begin{equation}
\hspace{-3.9cm}
E_{z}^{k}\leqslant 3E_{t}^{k}+3C_{L}^{2}nE_{c}^{k}+3L^2E_{o}^{k}.
\end{equation}
Further, assuming that $F$ is $\tau$-strongly convex and $\gamma \leqslant \frac{1}{L}$, we have for all $k\geqslant 0$
\begin{equation}
\label{eq_4_in_pro1}
\hspace{-3.3cm}
E_{o}^{k+1}\leqslant 4\mathcal{L}\left( \gamma \right) ^{-2r}E_{o}^{0}\cdot \mathcal{L}\left( \gamma \right) ^{\frac{2r}{T}\cdot \left( k+1 \right)}
\end{equation}
\begin{equation*}
\hspace{-0.7cm}
+\frac{4\mathcal{L}\left( \gamma \right) ^{-2r}\gamma ^2}{1-\mathcal{L}\left( \gamma \right) ^{\frac{r}{T}}}\sum_{l=0}^k{\mathcal{L}\left( \gamma \right) ^{\frac{r}{T}\left( k-l \right)}\left( C_{L}^{2}nE_{c}^{l}+E_{t}^{l}+n\sigma ^2 \right)}
,
\end{equation*}
where 
\begin{equation}
\label{def_C_1}
\mathcal{L}\left( \gamma \right) \triangleq 1-\tau \eta ^2\gamma 
, \quad C_1\triangleq \frac{2\sqrt{2S}\left( 1+\bar{m}^{-K_1} \right)}{\rho \left( 1-\bar{m}^{K_1} \right)}.
\end{equation}
\end{Pro}

\begin{proof}
See Appendix~\ref{appendix Proof of Proposition 1} available in the supplemental material for the detailed proofs.
\end{proof}

\begin{Rem}\label{remark_strong_cvx}
Since R-FAST works on more relaxed spanning-tree based communication graphs and in a more general fully asynchronous manner, the analysis is more involved and challenging compared to existing results. Specifically, the activation of non-root node cannot guarantee a sufficient contraction of optimality gap $E_o^k$, which renders the analysis techniques in~\cite{zhang2019fully, tian2020achieving} (which requires that a sufficient contraction of the optimality gap holds every global iteration) invalid and thus fails to guarantee the convergence of the optimality gap (c.f., \eqref{eq_4_in_pro1}). Instead, we employ a \textbf{\textit{two-time-scale}} technique to carry out the analysis over a period of length $T$, within which each of $r$ root nodes is activated at least once, enabling a sufficient contraction of optimality gap within such a window, and eventually allowing us to establish the convergence of the optimality gap in~\eqref{eq_4_in_pro1}
(c.f., the proof from~\eqref{optimal_gap_expectation} to~\eqref{response_eq_2} in the supplemental material). 
\end{Rem}

In what follows,  we show that each of the aforementioned error terms $E_c^k$, $E_t^k$, $E_z^k$ and $E_o^k$ will linearly converge to a neighborhood of zero with respect to the global iteration $k$. To this end, we introduce the following definition and the corresponding lemma which will be crucial to our subsequent analysis based on the generalized small gain theorem~\cite{tian2020achieving}.

\begin{Def}
For given sequence $\left\{ u^k \right\} _{k=0}^{\infty}$, constant $\lambda \in \left( 0,1 \right)$ and $N\in \mathbb{N}$, we define 
\begin{equation}
\left| u \right|^{\lambda ,N}=\underset{k=0,1,2...,N}{\max}\frac{\left| u^k \right|}{\lambda ^k}.
\end{equation}
\end{Def}

\begin{Lem}
\label{supporting lemma}
Consider non-negative sequences $\left\{ u^k \right\} _{k=0}^{\infty}$, and $\left\{ v_{i}^{k} \right\} _{k=0}^{\infty}$ for $i=1,...,m$. If there exist constants $\lambda _0,\lambda _1,...\lambda _m\in \left( 0,1 \right)$, $R_0,R_1,...R_m\in \mathbb{R}_+$, $\sigma >0$, $\lambda _h\in \left( 0,1 \right)$ and $R_h\in \mathbb{R}_+$ such that
\begin{equation*}
u^{k+1}\leqslant R_0\left( \lambda _0 \right) ^k+\sum_{i=1}^m{R_i\sum_{l=0}^k{\left( \lambda _i \right) ^{k-l}v_{i}^{l}}}+R_h\sum_{l=0}^k{\left( \lambda _h \right) ^{k-l}\sigma^2},
\end{equation*}
then, there holds
\begin{equation*}
\label{lamada_sequnce}
\left| u \right|^{\lambda ,N}\leqslant u^0+\frac{R_0}{\lambda}+\sum_{i=1}^m{\left( \frac{R_i}{\lambda -\lambda _i}\cdot \left| v_i \right|^{\lambda ,N} \right)}+\frac{R_h}{1 -\lambda _h}\cdot \frac{\sigma^2}{\lambda ^N},
\end{equation*}
for any $\lambda \in \left( \underset{i=0,1,...,m}{\max}\lambda _i,1 \right)$ and $N\in \mathbb{N}$.
\end{Lem}
With these above supporting results, we are ready to derive the convergence result of the proposed algorithm (c.f., Algorithm~\ref{Myalgorithm_RASGT_GV}) for strongly convex objective function.

\begin{Thm}[Geometric convergence]\label{Thm_Geo_conv}
Suppose Assumptions \ref{Ass_weight_matrix}-\ref{Ass_bounded_var} hold. Let $F$ be $\tau$-strongly convex and the constant step size $\gamma$ be sufficiently small. There exists $\lambda \in \left( \max \left\{ \rho ,(1-\tau \eta ^2\gamma )^{\frac{r}{T}} \right\} ,1 \right)$ such that for $\forall k \geqslant 0$, we have 
\begin{equation}
\begin{aligned}
\mathbb{E}\left[ \left\| x^k-\mathbf{1}\left( x^{\star} \right) ^{\top} \right\|^2 \right] \leqslant \mathcal{O}\left( \lambda ^k \right) +\varXi(\gamma) \sigma^2 
,
\end{aligned}
\end{equation}
where $\rho =\left( 1-\bar{m}^{\left( 2n-1 \right) \cdot T+n\cdot D} \right) ^{\frac{1}{\left( 2n-1 \right) \cdot T+n\cdot D}}\in \left( 0,1 \right) 
$, $\eta = \bar{m}^{\left( 2n-1 \right)\cdot T+n\cdot D}\in \left( 0,1 \right)$ and $\varXi(\gamma)$ is defined at~\eqref{definition_of_steady_error}. 
\end{Thm}

\begin{proof}
We reformulate the error dynamic system as established in Proposition \ref{proposition 1} into a linear system using Lemma \ref{supporting lemma}. Specifically, choosing $\lambda \in \left( \max \left\{ \rho ,(1-\tau \eta ^2\gamma )^{\frac{r}{T}} \right\} ,1 \right)$, and applying Lemma~\ref{supporting lemma} to the inequalities~\eqref{eq_1_in_pro1}-\eqref{eq_4_in_pro1} in Proposition~\ref{proposition 1} for any $N\in \mathbb{N}$, yields that 
\begin{equation*}
\left[ \begin{array}{c}
	\left| E_z \right|^{\lambda ,N}\\
	\left| E_c \right|^{\lambda ,N}\\
	\left| E_t \right|^{\lambda ,N}\\
	\left| E_o \right|^{\lambda ,N}\\
\end{array} \right] \leqslant P\left[ \begin{array}{c}
	\left| E_z \right|^{\lambda ,N}\\
	\left| E_c \right|^{\lambda ,N}\\
	\left| E_t \right|^{\lambda ,N}\\
	\left| E_o \right|^{\lambda ,N}\\
\end{array} \right] +\underset{\triangleq \alpha}{\underbrace{\left[ \begin{array}{c}
	0\\
	E_{c}^{0}+\frac{2C_{2}^{2}E_{c}^{0}}{\lambda}\\
	E_{t}^{0}+\frac{3C_{1}^{2}\left\| \bar{z}^0 \right\| ^2}{\lambda}\\
	E_{o}^{0}+\frac{4\left( 1-\tau \eta ^2\gamma \right) ^{-2r}E_{o}^{0}}{\lambda}\\
\end{array} \right] }}
\end{equation*}
\vspace{-0.25cm}
\begin{equation}
\label{new_small_gain}
+\frac{\sigma ^2}{\lambda ^N}\underset{\triangleq \beta}{\underbrace{\left[ \begin{array}{c}
	0\\
	\frac{4\gamma ^2C_{2}^{2}n}{\left( 1-\rho \right) ^2}\\
	\frac{54\gamma ^2C_{1}^{2}C_{L}^{2}n}{\left( 1-\rho \right) ^2}\\
	\frac{4\left( 1-\tau \eta ^2\gamma \right) ^{-2r}\gamma ^2n}{\left( 1-\left( 1-\tau \eta ^2\gamma \right) ^{\frac{r}{T}} \right) ^2}\\
\end{array} \right] }},
\end{equation}
where 
\begin{equation}\label{def_of_P}
\hspace{-8.3cm}
P\triangleq
\end{equation}
\begin{small}
\begin{equation*}
\hspace{-0.3cm}
\left[ \begin{matrix}
	0&		b_1&		3&		3L^2\\
	\frac{b_3\gamma ^2}{\left( 1-\rho \right) \left( \lambda -\rho \right)}&		0&		0&		0\\
	\frac{b_2\gamma ^2}{\left( 1-\rho \right) \left( \lambda -\rho \right)}&		\frac{b_2}{\left( 1-\rho \right) \left( \lambda -\rho \right)}&		0&		0\\
	0&		\frac{b_2\mathcal{L}\left( \gamma \right) ^{-2r}\gamma ^2}{\left( 1-\mathcal{L}\left( \gamma \right) ^{\frac{r}{T}} \right) \left( \lambda -\mathcal{L}\left( \gamma \right) ^{\frac{r}{T}} \right)}&		\frac{4\mathcal{L}\left( \gamma \right) ^{-2r}\gamma ^2}{\left( 1-\mathcal{L}\left( \gamma \right) ^{\frac{r}{T}} \right) \left( \lambda -\mathcal{L}\left( \gamma \right) ^{\frac{r}{T}} \right)}&		0\\
\end{matrix} \right]
\end{equation*}
\end{small}
$b_1\triangleq 3C_{L}^{2}n$, $b_2\triangleq 54C_{1}^{2}C_{L}^{2}$, $b_3\triangleq 4C_{2}^{2}$ and $\mathcal{L}\left( \gamma \right) = 1-\tau \eta ^2\gamma $.
If the spectral radius $\rho({P})<1$, then using (\ref{new_small_gain}) recursively yields
\begin{equation*}
\left[ \begin{array}{c}
	\frac{E_{z}^{N}}{\lambda ^N}\\
	\frac{E_{c}^{N}}{\lambda ^N}\\
	\frac{E_{t}^{N}}{\lambda ^N}\\
	\frac{E_{o}^{N}}{\lambda ^N}\\
\end{array} \right] \leqslant \left[ \begin{array}{c}
	\left| E_z \right|^{\lambda ,N}\\
	\left| E_c \right|^{\lambda ,N}\\
	\left| E_t \right|^{\lambda ,N}\\
	\left| E_o \right|^{\lambda ,N}\\
\end{array} \right] \leqslant \left( I-P \right) ^{-1}\alpha +\frac{\sigma ^2}{\lambda ^N}\left( I-P \right) ^{-1}\beta .
\end{equation*}
Multiplying by $\lambda^N$ on both sides of the above inequality, we have
\begin{equation}
\label{small_gain_final_inequality}
    \left[ \begin{array}{c}
	E_{z}^{N}\\
	E_{c}^{N}\\
	E_{t}^{N}\\
	E_{o}^{N}\\
\end{array} \right] \leqslant \underset{\text{linearly converge to 0}}{\underbrace{\lambda ^N\cdot \left( I-P \right) ^{-1}\alpha }}+\underset{\text{steady error}}{\underbrace{\left( I-P \right) ^{-1}\beta \sigma ^2}}.
\end{equation}
We note that \eqref{small_gain_final_inequality} holds for any $N\in \mathbb{N}$. 
In what follows we show that the spectral radius $\rho({P})<1$.
The characteristic
polynomial $p_P\left( \vartheta \right)$ of ${P}$ satisfies the conditions of Lemma 24 in \cite{tian2020achieving}, therefore $\rho \left( {P} \right) <1  $ if and only if $p_{{P}}\left( 1 \right) >0$, that is
\begin{equation*}
\begin{aligned} 
& \frac{3b_2\gamma ^2}{\left( 1-\rho \right) \left( \lambda -\rho \right)}+\frac{b_1b_3\gamma ^2}{\left( 1-\rho \right) \left( \lambda -\rho \right)}+\frac{3b_2b_3\gamma ^2}{\left( 1-\rho \right) ^2\left( \lambda -\rho \right) ^2}
\\
&+\frac{12L^2b_2\mathcal{L}\left( \gamma \right) ^{-2r}\gamma ^4+3L^2b_2b_3\mathcal{L}\left( \gamma \right) ^{-2r}\gamma ^4}{\left( 1-\rho \right) \left( \lambda -\rho \right) \left( 1-\mathcal{L}\left( \gamma \right) ^{\frac{r}{T}} \right) \left( \lambda -\mathcal{L}\left( \gamma \right) ^{\frac{r}{T}} \right)}
\\
&+\frac{12L^2b_2b_3\mathcal{L}\left( \gamma \right) ^{-2r}\gamma ^4}{\left( 1-\rho \right) ^2\left( \lambda -\rho \right) ^2\left( 1-\mathcal{L}\left( \gamma \right) ^{\frac{r}{T}} \right) \left( \lambda -\mathcal{L}\left( \gamma \right) ^{\frac{r}{T}} \right)}
\\
& \triangleq \mathcal{B}\left( \lambda; \gamma \right) <1.
\end{aligned}
\end{equation*}
It can be verified that $\mathcal{B}\left( \lambda ;\gamma \right)$ is continuous at $\lambda=1$ for fixed $\gamma$. Therefore, as long as
\begin{align*}
\mathcal{B}\left( 1;\gamma \right) &= \frac{3 b_2\gamma ^2}{\left( 1-\rho \right) ^2}+\frac{b_1b_3\gamma ^2}{\left( 1-\rho \right) ^2}+\frac{3b_2b_3\gamma ^2}{\left( 1-\rho \right) ^4}
\\
&+\frac{12L^2b_2\mathcal{L}\left( \gamma \right) ^{-2r}\gamma ^4+3L^2b_2b_3\mathcal{L}\left( \gamma \right) ^{-2r}\gamma ^4}{\left( 1-\rho \right) ^2\left( 1-\mathcal{L}\left( \gamma \right) ^{\frac{r}{T}} \right) ^2}
\\
&+\frac{12L^2b_2b_3\mathcal{L}\left( \gamma \right) ^{-2r}\gamma ^4}{\left( 1-\rho \right) ^4\left( 1-\mathcal{L}\left( \gamma \right) ^{\frac{r}{T}} \right) ^2}<1,
\end{align*}
it is sufficient to claim the existence of some $\lambda \in \left( \max \left\{ \rho ,\mathcal{L}\left( \gamma \right) ^{\frac{r}{T}} \right\} ,1 \right) $ such that $\mathcal{B}\left( \lambda ;\gamma \right) <1$. 
We now show that $\mathcal{B}\left( 1;\gamma \right) <1$ for sufficiently small $\gamma$. 
We only need to prove boundedness of the $\frac{\gamma}{1-\mathcal{L}\left( \gamma \right) ^{\frac{r}{T}}}$ when $\gamma \downarrow 0$.
According to L'H${\hat{o}}$pital's rule, we get
\begin{align*}
    &\underset{\gamma \rightarrow  0}{\lim}\frac{\gamma}{1-\mathcal{L}\left( \gamma \right) ^{\frac{r}{T}}}=\underset{\gamma \rightarrow  0}{\lim}\frac{\gamma}{1-\left( 1-\tau \eta ^2\gamma \right) ^{\frac{r}{T}}}
    \\
    &=\frac{1}{\frac{r}{T}\left( 1-\tau \eta ^2\gamma \right) ^{\frac{r}{T}-1}\cdot \tau \eta ^2}\left| \begin{array}{c}
	\\
	\gamma =0\\
\end{array}=\frac{T}{\tau \eta ^2r} \right. <\infty,
\end{align*}
we thus proved that for a sufficiently small $\gamma$, there exists $\lambda \in \left( \max \left\{ \rho ,\left( 1-\tau \eta ^2\gamma \right) ^{\frac{r}{T}} \right\} ,1 \right)$ such that~\eqref{small_gain_final_inequality} holds for any $N\in \mathbb{N}$.

Now, we are ready to prove the main results in Theorem \ref{Thm_Geo_conv}. It follows from~\eqref{small_gain_final_inequality} that
\begin{align*}
&\mathbb{E}\left[ \left\| x_{\psi}^{k}-(x^{\star})^{\top} \right\|^2 \right] \leqslant \mathcal{O}\left( \lambda ^k \right) +\left[ \left( I-P \right) ^{-1}\beta \right] _4 \sigma^2 
,
\\
&\mathbb{E}\left[ \left\| h^k-\mathbf{1}x_{\psi}^{k} \right\|^2 \right] \leqslant \mathcal{O}\left( \lambda ^k \right) +\left[ \left( I-P \right) ^{-1}\beta \right] _2 \sigma^2 
.
\end{align*}
Then, we further have
\begin{equation*}
\hspace{-0.2cm}
\begin{aligned}
&\mathbb{E}\left[ \left\| x^k-\mathbf{1}\left( x^{\star} \right) ^{\top} \right\| ^2 \right] =\mathbb{E}\left[ \left\| x^k-\mathbf{1}x_{\psi}^{k}+\mathbf{1}x_{\psi}^{k}-\mathbf{1}\left( x^{\star} \right) ^{\top} \right\| ^2 \right] 
\\
&\leqslant 2\mathbb{E}\left[ \left\| x^k-\mathbf{1}x_{\psi}^{k} \right\| ^2 \right] +2n\mathbb{E}\left[ \left\| x_{\psi}^{k}-\left( x^{\star} \right) ^{\top} \right\| ^2 \right] 
\end{aligned}
\end{equation*}
\begin{equation}
\label{definition_of_steady_error}
\hspace{-2.3cm}
\leqslant 2\mathbb{E}\left[ \left\| h^k-\mathbf{1}x_{\psi}^{k} \right\| ^2 \right] +2n\mathbb{E}\left[ \left\| x_{\psi}^{k}-\left( x^{\star} \right) ^{\top} \right\|^2 \right] 
\end{equation}
\begin{equation*}
\leqslant \mathcal{O}\left( \lambda ^k \right) +\underset{\triangleq \varXi (\gamma )}{\underbrace{\left( 2\left[ \left( I-P \right) ^{-1}\beta \right] _2+2n\left[ \left( I-P \right) ^{-1}\beta \right] _4 \right) }}\cdot \sigma ^2
,
\end{equation*}
which completes the proof.
\end{proof}

\begin{Rem}
In Theorem \ref{Thm_Geo_conv}, we show that in expectation, the proposed R-FAST (Algorithm~\ref{Myalgorithm_RASGT_GV}) converges to a neighborhood of the optimal solution with linear rate $\lambda$, which aligns with the convergence result of the synchronous stochastic distributed optimization algorithm S-AB~\cite{xin2019distributed} over directed network. 
In addition, according to Remark~\ref{rem:sync} and setting $T=n$ and $D=2n-2$, we obtain the convergence result for the synchronous R-FAST, which is, to the best of our knowledge, the first convergence result for the stochastic synchronous algorithms using the push-pull protocol. The geometric exact convergence result of the deterministic synchronous push-pull algorithm~\cite{pu2020push} is readily recovered by further setting $\sigma^2=0$.
\end{Rem}

\subsection{Non-Convex: Sublinear convergence} \label{analysis_of_non_convex}
In this section, we consider the scenario that the overall objective function $F$ in (\ref{global_loss_function}) is only smooth (possibly non-convex).
We use the following merit function to measure the distance to the stationary point\footnote{For brevity, in what follows we use  $\mathbb{E}\left\| \cdot \right\| ^2$ to denote $\mathbb{E}\left[ \left\| \cdot \right\| ^2 \right]$}:
\begin{equation}
\label{merit_non_convex}
M_F\left( x^k \right) \triangleq \left\| \nabla F\left( \bar{x}^k \right) \right\| ^2+\left\| x^k-\mathbf{1} \left(\bar{x}^k\right)^\top  \right\| ^2.
\end{equation}

The following lemma provides an upper bound for the accumulative consensus error $\sum_{l=0}^k{\mathbb{E} \left\| {h}^l-\mathbf{1}x_{\psi}^{l} \right\| ^2 }$ and  tracking error $\sum_{l=0}^k{\mathbb{E} \left\| \bar{z}_{i^l}^{l}-\xi _{i^l}^{l-1}({\bar{z}}^l)^{\top}\mathbf{1} \right\| ^2 }$ , using the accumulative magnitude of gradient $\sum_{l=0}^k{\mathbb{E} \left\| \nabla F\left( x_{\psi}^{l} \right) \right\| ^2 }$.
\begin{Lem}
\label{lemma8}
Suppose Assumptions \ref{Ass_weight_matrix}-\ref{Ass_bounded_var} hold. Let the constant step size $\gamma$ satisfy
\begin{equation}
\label{the_first_gamma_condition}
    \gamma \,\,< \frac{1}{\sqrt{3\varrho _cC_{L}^{2}n+3\varrho _t}}.
\end{equation}
Then, for all $k\geqslant 0$, we have
\begin{small}
\begin{equation*}
\sum_{l=0}^k{\mathbb{E}\left\| h^l-\mathbf{1}x_{\psi}^{l} \right\| ^2}\leqslant \frac{3\varrho _c\gamma ^2}{1-3\left( \varrho _cC_{L}^{2}n+\varrho _t \right) \gamma ^2}\sum_{l=0}^k{\mathbb{E}\left\| \nabla F\left( x_{\psi}^{l} \right) \right\| ^2}
\end{equation*}
\end{small}
\begin{small}
\begin{equation}
\label{upper_bound_running_sum_consensus_error}
\hspace{-0.5cm}
+\frac{\left( 1-3\varrho _t\gamma ^2 \right) c_c+3c_t\varrho _c\gamma ^2}{1-3\left( \varrho _cC_{L}^{2}n+\varrho _t \right) \gamma ^2}+\frac{\varrho _cn\gamma ^2\left( k+1 \right)}{1-3\left( \varrho _cC_{L}^{2}n+\varrho _t \right) \gamma ^2}\sigma ^2
,
\end{equation}
\end{small}
and

\begin{small}
\begin{equation}
\label{upper_bound_running_sum_tracking_error}
\hspace{-1.4cm}
\begin{aligned}
&\sum_{l=0}^k{\mathbb{E}\left\| \bar{z}_{i^l}^{l}-\xi _{i^l}^{l-1}(\bar{z}^l)^{\top}\mathbf{1} \right\| ^2}\leqslant \frac{\varrho _tn\gamma ^2\left( k+1 \right)}{1-3\left( \varrho _cC_{L}^{2}n+\varrho _t \right) \gamma ^2}\sigma ^2
\\
&+\frac{3\varrho _t\gamma ^2}{1-3\left( \varrho _cC_{L}^{2}n+\varrho _t \right) \gamma ^2}\sum_{l=0}^k{\mathbb{E}\left\| \nabla F\left( x_{\psi}^{l} \right) \right\| ^2}
\\
&+\frac{\left( 1-3\varrho _cC_{L}^{2}n\gamma ^2 \right) c_t+3c_c\varrho _tC_{L}^{2}n\gamma ^2}{1-3\left( \varrho _cC_{L}^{2}n+\varrho _t \right) \gamma ^2},
\end{aligned}
\end{equation}
\end{small}
where $c_c$, $\varrho _c$, $c_t$ and $\varrho _t$ are constants given as below:
\begin{equation}
\label{def_of_c_c_and_varrho_c}
\hspace{-0.3cm}
c_c\triangleq\left( 1+\frac{2C_{2}^{2}}{1-\rho ^2} \right)\left\| h^0-\mathbf{1}x_{\psi}^{0} \right\| ^2, \quad \varrho _c\triangleq\frac{4C_{2}^{2}}{\left( 1-\rho \right) ^2},
\end{equation}
\vspace{-0.18cm}
\begin{equation*}
\hspace{-0.37cm}
c_t\triangleq \frac{3C_{1}^{2}\left\| \bar{z}^0 \right\| ^2}{1-\rho ^2}+\frac{27C_{1}^{2}C_{L}^{2}\left( 2C_{2}^{2}+1-\rho ^2 \right)}{\left( 1-\rho \right) ^4}\left\| h^0-\mathbf{1}x_{\psi}^{0} \right\| ^2
\end{equation*}
\begin{equation}
\label{def_of_c_t and def_of_varrho_t}
\hspace{-0.37cm}
+\left\| \bar{z}_{i^0}^{0}-\xi _{i^0}^{-1}(\bar{z}^0)^{\top}\mathbf{1} \right\| ^2,  \varrho _t\triangleq \frac{54C_{1}^{2}C_{L}^{2}\left[ 4C_{2}^{2}+\left( 1-\rho \right) ^2 \right]}{\left( 1-\rho \right) ^4}.
\end{equation}
\end{Lem}

\begin{proof}
See Appendix~\ref{appendix_proof_of_lemma8} available in the supplemental material for the detailed proofs.
\end{proof}

The following lemma based on \emph{two-time-scale techniques} provides an upper bound of $\sum_{k=0}^{\bar{k}T-1}\mathbb{E}\left\| \nabla F\left( x_{\psi}^{k} \right) \right\| ^2$.

\begin{Lem}
\label{new_lemma7}
Suppose Assumptions \ref{Ass_weight_matrix}-\ref{Ass_bounded_var} hold. Let the constant step size $\gamma$ satisfy
\begin{equation}
\label{step_size_condition_1}
\gamma \leqslant \min \left\{ \frac{2}{\left( 2T^2+r\eta ^2T \right) L},\frac{1}{8L} \right\}.
\end{equation}
Then, for all $\bar{k} \geqslant 1$, we have
\begin{equation}
\label{complex_descent_lemma}
\hspace{-0.6cm}
\left( \frac{\gamma r\eta ^2}{8T}-\frac{\gamma ^2}{2} \right) \sum_{k=0}^{\bar{k}T-1}{\mathbb{E}\left\| \nabla F\left( x_{\psi}^{k} \right) \right\| ^2}\leqslant F\left( x_{\psi}^{0} \right) -F^{\star}
\end{equation}
\vspace{-0.2cm}
\begin{equation*}
\begin{aligned}
&+\frac{1+\gamma}{2}\sum_{k=0}^{\bar{k}T-1}{\mathbb{E}\left\| \bar{z}_{i^k}^{k}-\xi _{i^k}^{k-1}(\bar{z}^k)^{\top}\mathbf{1} \right\| ^2}+\frac{1}{2}r\eta ^2L^2\bar{k}T^2n\gamma ^3\sigma ^2
\\
&+\frac{\gamma C_{L}^{2}n}{2}\sum_{k=0}^{\bar{k}T-1}{\mathbb{E}\left\| h^k-\mathbf{1}x_{\psi}^{k} \right\| ^2}+Ln\bar{k}T\gamma ^2\sigma ^2+L^2\bar{k}T^3n\gamma ^3\sigma ^2 ,
\end{aligned}
\end{equation*}
where $x_\psi^0$ is defined in~\eqref{new_def_of_x_psi_0} and
$F^{\star}\triangleq  \min_{x\in \mathbb{R}^p} F(x)$.
\end{Lem}
\begin{proof}
See Appendix~\ref{appecdix_proof_of_new_lemma_7} available in the supplemental material for the detailed proofs.
\end{proof}

\begin{Rem}
Similar to Remark~\ref{remark_strong_cvx} for strongly-convex cases, the activation of non-root node cannot guarantee a sufficient descent towards stationary point for general non-convex objective, which renders the vanilla descent lemma~\cite{tian2020achieving, kungurtsev2021decentralized} based on the smooth property invalid. 
We thus resort to the two-time-scale technique again to warrant a sufficient descent towards stationary point over a window of length $T$ (where each of common root nodes is activated at least once), which is fundamental for establishing the above lemma.
\end{Rem}

With the above two supporting lemmas, we provide the sub-linear convergence result of the proposed algorithm.
\begin{Thm} [Sub-linear convergence]
\label{Thm_sublinear_conv}
Suppose Assumptions \ref{Ass_weight_matrix}-\ref{Ass_bounded_var} hold. For any $K > 0 $ being a multiple of $T$, and any constant step size satisfying $\gamma \leqslant \bar{\gamma}$, 
we have for Algorithm \ref{Myalgorithm_RASGT_GV}
\begin{equation}
\label{eq_main_thm_non_convex}
\begin{aligned}
\frac{1}{K}\sum_{k=0}^{K-1}&{\mathbb{E}\left[ M_F\left( x^k \right) \right]}\leqslant \frac{32T\left( F\left( x_{\psi}^{0} \right) -F^{\star}+c_t \right)}{\gamma \eta^2  Kr}
\\
&+\frac{TE_1}{K}+\frac{32T\left( n\varrho _t+nL \right)}{\eta^2}\gamma \sigma ^2+E_2\gamma ^2\sigma ^2,
\end{aligned}
\end{equation}
where constants $c_t$ and $\varrho_t$ are defined in~\eqref{def_of_c_t and def_of_varrho_t};
constants $\bar{\gamma}$, $E_1$ and $E_2$ are defined in~\eqref{total_condion_of_stepsize},~\eqref{def_of_E1} and~\eqref{def_of_E2}, respectively, in the supplemental material.
\end{Thm}

\begin{proof}
The proof involves constructing a loop for the aforementioned three key inequalities (c.f.,~\eqref{upper_bound_running_sum_consensus_error},~\eqref{upper_bound_running_sum_tracking_error} and~\eqref{complex_descent_lemma}) as presented in Lemma~\ref{lemma8} and~\ref{new_lemma7}. See Appendix~\ref{appendix_proof_thm_non_convex} in the supplemental material for the detailed proof.
\end{proof}

In order to show the ability of R-FAST in dealing with data heterogeneity, we first quantify the heterogeneity among data distributions at different nodes in the following definition which is widely used in the existing literature~\cite{assran2019stochastic}. 

\begin{Def}[$\varsigma$-heterogeneity]\label{data_heterogeneity}
There exists a constant $\varsigma \geqslant 0$ such that $\frac{1}{n}\sum_{i=1}^n{\left\lVert \nabla f_i\left( x \right) -\frac{1}{n}\nabla F\left( x \right) \right\rVert ^2}\leqslant \varsigma ^2$.
\end{Def}



\begin{Rem}
\label{Rem_data_heteg}
Most of the existing works under asynchronous settings, such as AD-PSGD~\cite{lian2018asynchronous} and OSGP~\cite{assran2019stochastic}, suffer from the issue of data heterogeneity across nodes, leading to a dependency of $\varsigma$ in the complexity (See Theorem 1 in~\cite{lian2018asynchronous} and $(29)$ in Appendix 
E.3 in~\cite{assran2019stochastic}). In contrast, R-FAST can effectively remove the impact of the data heterogeneity on the convergence rate (i.e., $\varsigma$-free in convergence rates), yielding the robustness to the data heterogeneity across nodes.
\end{Rem}

By carefully choosing the step size in Theorem \ref{Thm_sublinear_conv}, we further obtain the following corollary.
\begin{Cor}  \label{Col_linear_speedup}
Under the conditions in Theorem \ref{Thm_sublinear_conv}, 
if we further set the constant step size
$\gamma \,=\frac{1}{\sigma \sqrt{Kr}+\bar{\gamma}^{-1}}$,
we have for Algorithm~\ref{Myalgorithm_RASGT_GV}
\begin{equation*}
\begin{aligned}
&\frac{1}{K}\sum_{k=0}^{K-1}{\mathbb{E}\left[ M_F\left( x^k \right) \right]}
\\
&\leqslant \frac{32T\left( F\left( x_{\psi}^{0} \right) -F^{\star}+c_t+n\varrho _t+nL \right) \sigma}{\eta ^2\sqrt{Kr}\,}
\\
&+\frac{32r^{-1}\bar{\gamma}^{-1}T\left( F\left( x_{\psi}^{0} \right) -F^{\star}+c_t \right) +\eta ^2\,T\,E_1+r^{-1}\,\eta ^2\,E_2}{\eta ^2K\,},
\end{aligned}
\end{equation*}
where constants $c_t$ and $\varrho_t$ are given in~\eqref{def_of_c_t and def_of_varrho_t};
constants $\bar{\gamma}$, $E_1$ and $E_2$ can be found in~\eqref{total_condion_of_stepsize},~\eqref{def_of_E1} and~\eqref{def_of_E2}, respectively, in the supplemental material.
\end{Cor}

\begin{proof}
The proof is straightforward by substituting $\gamma \,=\frac{1}{\sigma \sqrt{Kr}+\bar{\gamma}^{-1}}$ into~\eqref{eq_main_thm_non_convex}.
\end{proof}

\begin{Rem}
The above result suggests that the convergence rate for R-FAST is of $\mathcal{O}\left( \frac{\sigma}{\sqrt{K}} +\frac{1}{K} \right) $, matching the rate of centralized SGD for non-convex problems~\cite{ghadimi2013stochastic}. Additionally, when $K$ is large enough, the $\frac{1}{K}$ term will be dominated by $\frac{1}{\sqrt{K}}$, leading to a rate of $\mathcal{O}\left( \frac{1}{\sqrt{K}} \right)$.
\end{Rem}

\section{Experiments}
\label{Experiments}
In this section, we conduct several experiments to validate our theoretical results and verify the performance of the proposed R-FAST with comparison to the baseline algorithms. 
We implement all experiments by launching multiple processes within a server, where a process serves as a node, and communication among nodes is achieved by inter-process communication based on distributed communication package \textit{torch.distributed} in PyTorch~\cite{paszke2017pytorch}.
For \textit{asynchronous setup} in our implementation, each process (node) runs its own code independently and the messages are transmitted between processes in a fully-asynchronous way without any blocking. By doing so, each process runs at its own pace without waiting to receive the messages from its in-neighbour process. Therefore, each process may receive the delayed message from its in-neighbours and delays are determined by different updating frequencies among nodes (i.e., degree of asynchrony) and the transmitting time of messages. For \textit{packet losses}, we emulate it in such a way that each node can decide whether to send out updated information for each out-neighbor at each iteration. An node will not send information packet again to its out-neighbors  until it receives the out-neighbor's receipt confirmation for the message it sent last time.  Note that a) all communications including the receipt confirmation are carried out in an asynchronous non-blocking manner; and b) each node, during its waking time, either sends out the newly-updated information packet, or simply discards it.
For all experiments, the total training samples are evenly distributed over $n$ nodes and, as a result, each node has only a partial view of the total training dataset.

\begin{Rem}
In our implementation, we use inter-process communication to mimic communication among nodes, which is a widely used setup in single-server\footnote{Here, by the term `server' we mean a high-performance computer.} distributed training. 
Moreover, our algorithm can be easily adapted to a multi-server network scenario, where each server serves as an individual node. In this case, communication among these nodes (servers) is established through Ethernet using the TCP/IP protocol. Each server (node) performs its execution independently and messages are transmitted between servers (nodes) in a fully-asynchronous non-blocking way. Besides, we do not need to simulate packet/message loss because it naturally exists in the communication channels in real computer networks.
\end{Rem}

\subsection{Regularized logistic regression}\label{sec:sim_sc}
We consider a decentralized training problem over a network composed of $n\in \left\{ 3,7,15,31 \right\} $ nodes, respectively, and train a regularized logistic regression model on 12,000 images of two handwritten digits ($0$ and $1$) from the MNIST dataset~\cite{deng2012mnist}.  Note that the objective function is \textit{smooth and strongly convex}.
We deploy R-FAST on multiple CPU cores, each of which is used by a process (node) to compute.
The hyperparameter setup is: i) mini-batch size: 32 per node; ii) learning rate: 0.001.
\begin{figure}[htbp]
    \centering
    {
        \begin{minipage}[t]{0.47\textwidth}
        \centering          
         \includegraphics[width=\textwidth]{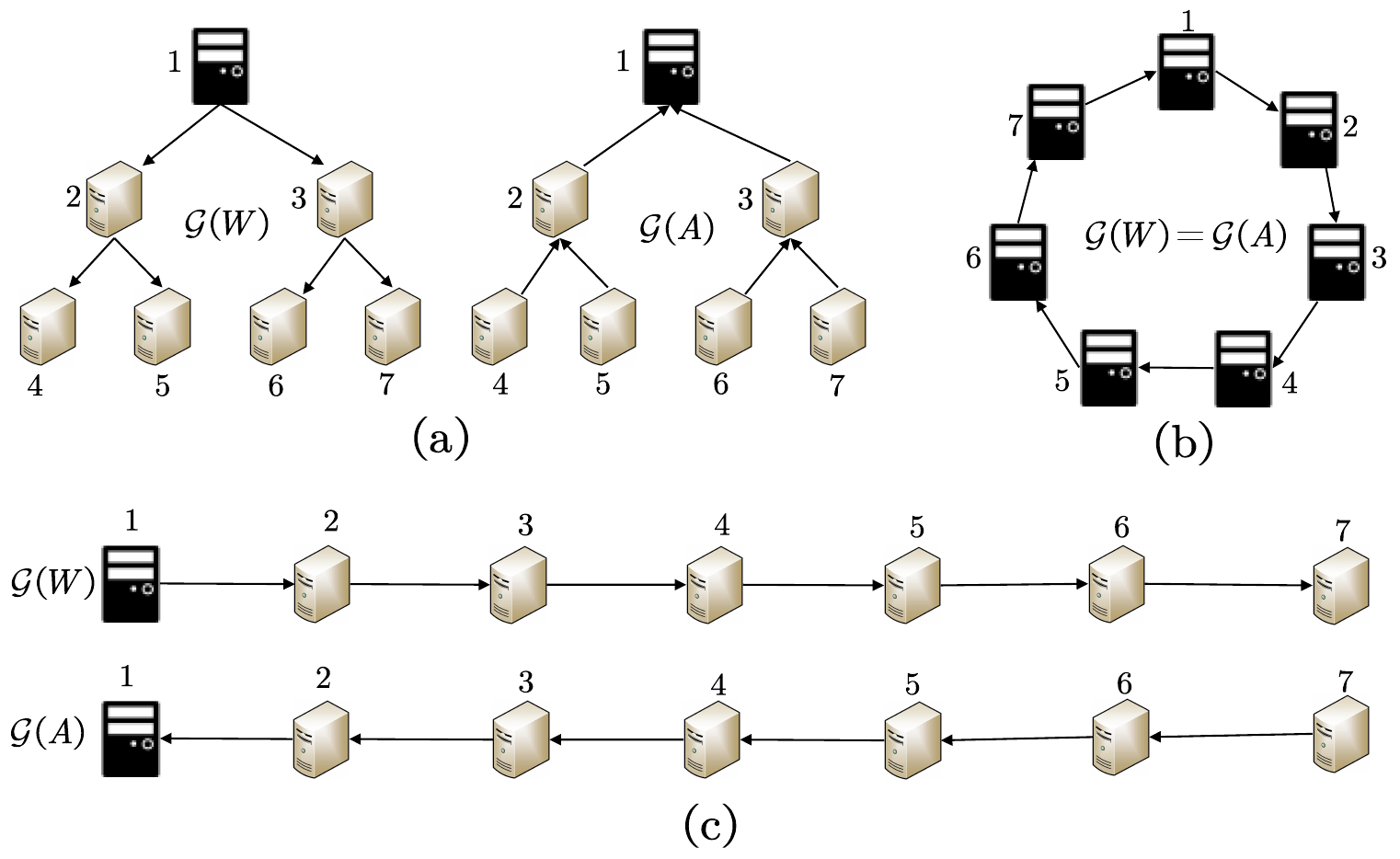}
        \end{minipage}%
    }
    \caption{The network topologies. (a):  binary tree graph; (b): directed ring graph; (c): line graph.}
    \label{topo_fig2}
\end{figure}

\begin{figure}[htbp]
\centering
\subfloat[]{\includegraphics[width=1.6in]{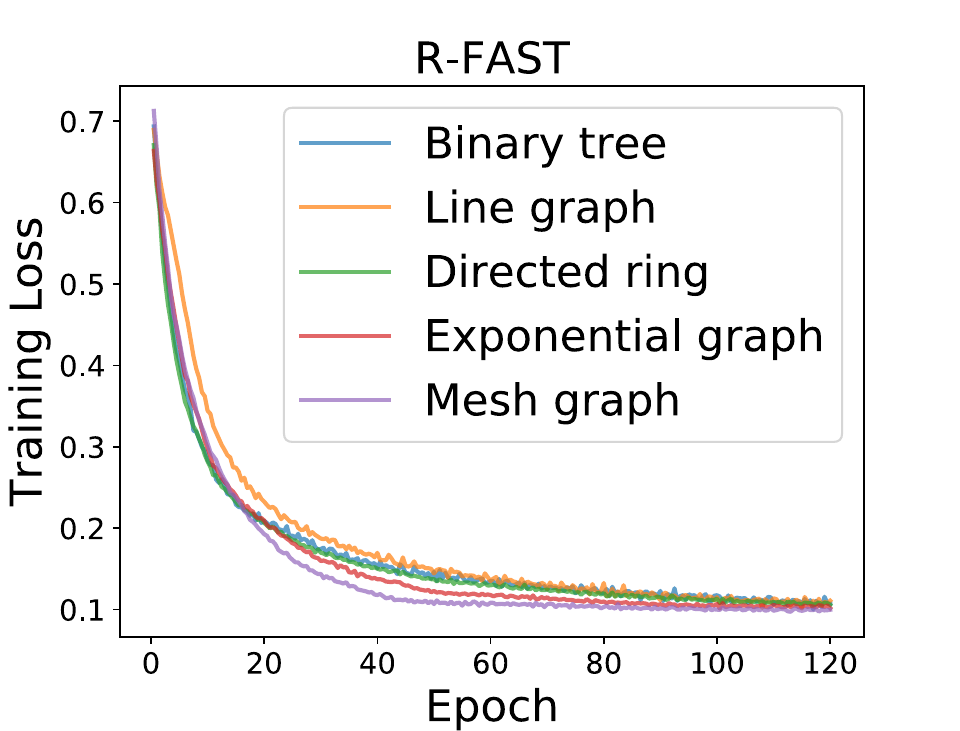}%
\label{fig0a}}
\hfil
\subfloat[]{\includegraphics[width=1.6in]{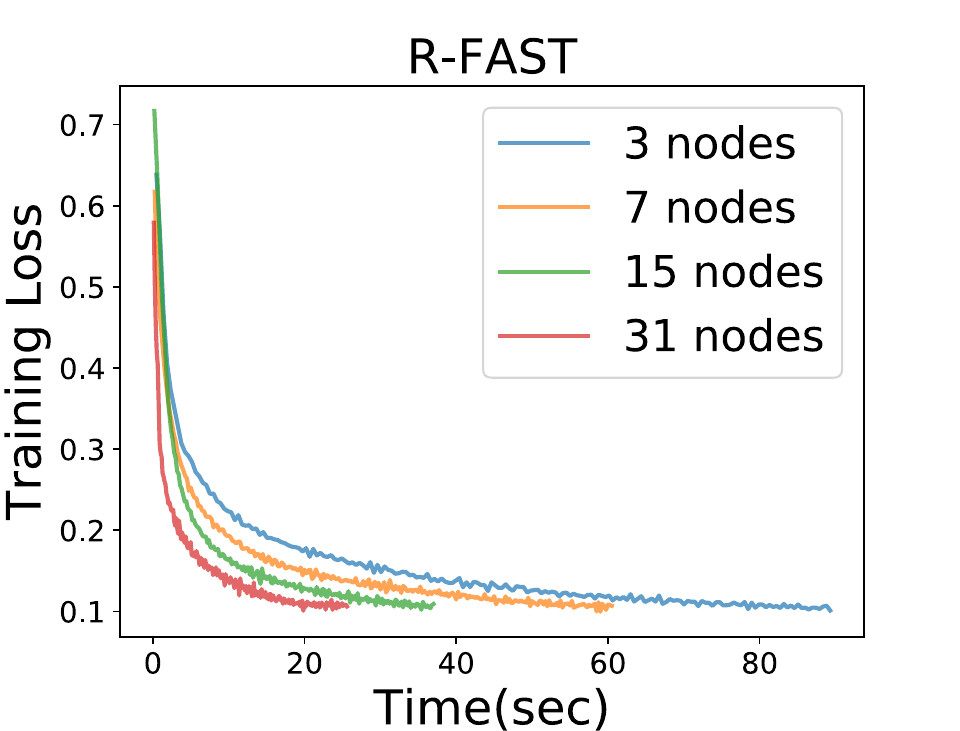}%
\label{fig0c}}
\caption{Performance of R-FAST in training logistic regression model in terms of training loss versus epoch over (a) five different topologies (composed of 7 nodes)  and (b) binary tree topology with different number of nodes.}
\end{figure}

\textbf{Training over general network topologies.}
To verify that the proposed R-FAST algorithm can work over general network topologies satisfying Assumption~\ref{Ass_graph}, we consider five different graphs, i.e., binary tree (c.f., Fig.~\ref{topo_fig2}(a)), line (c.f., Fig.~\ref{topo_fig2}(c)), directed ring (c.f., Fig.~\ref{topo_fig2}(b)), exponential graph (c.f., Fig.~\ref{matrix_exponential} in the supplemental material) and mesh graph (c.f., Fig.~\ref{matrix_mesh} in the supplemental material). For binary tree or line graph, we design the specific two non-strongly-connected sub-graphs according to Assumption~\ref{Ass_graph}, that is, an oriented acyclic tree with one root as sub-graph $\mathcal{G}(W)$, and its inverse graph as sub-graph $\mathcal{G}(A)$.
The corresponding row (resp. column) stochastic weight matrix $W$ (resp. $A$) used for these five network topologies, along with the flexibility of topology design, can be found in Appendix~\ref{weghit_matrix_used} available in the supplemental material.
It follows from Fig.~\ref{fig0a} that,  the proposed R-FAST is able to converge for all the above five topologies. To the best of our knowledge, R-FAST is such an algorithm in the existing literature that can work on two spanning-tree graphs.
Furthermore, we run R-FAST over the binary tree graph for different number of nodes. Fig.~\ref{fig0c} shows that the time consumed for achieving a certain training loss (i.e., $0.1$) is decreasing almost linearly with respect to the number of nodes, indicating that R-FAST can scale well with respect to the size of the network.

\begin{figure*}[htbp]
\centering
\subfloat[]{\includegraphics[width=2.35in]{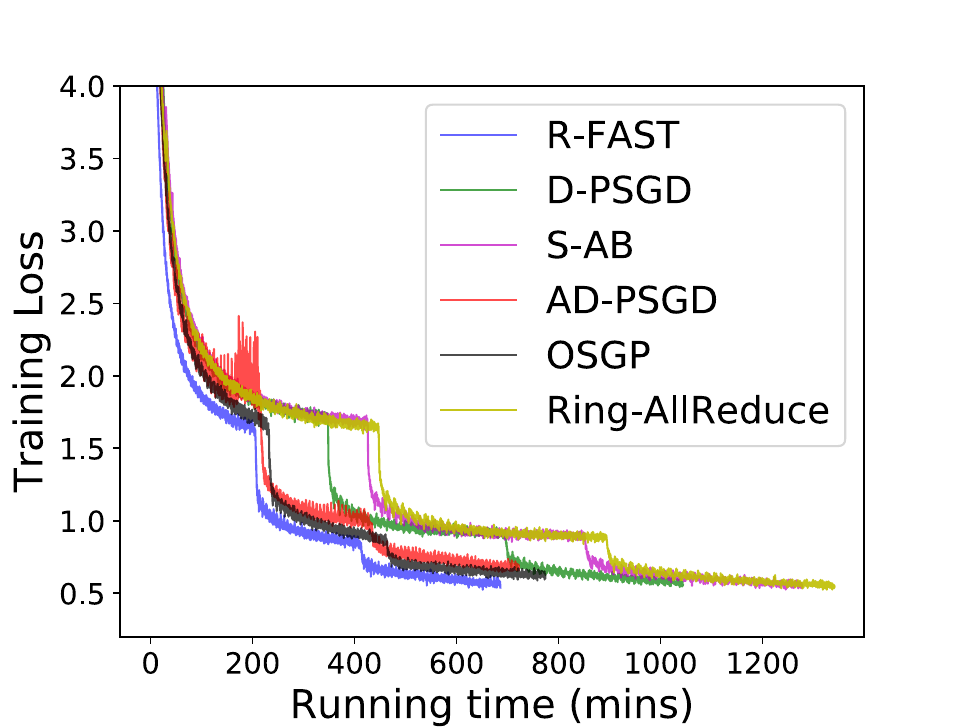}%
\label{fig2a}}
\hfil
\subfloat[]{\includegraphics[width=2.35in]{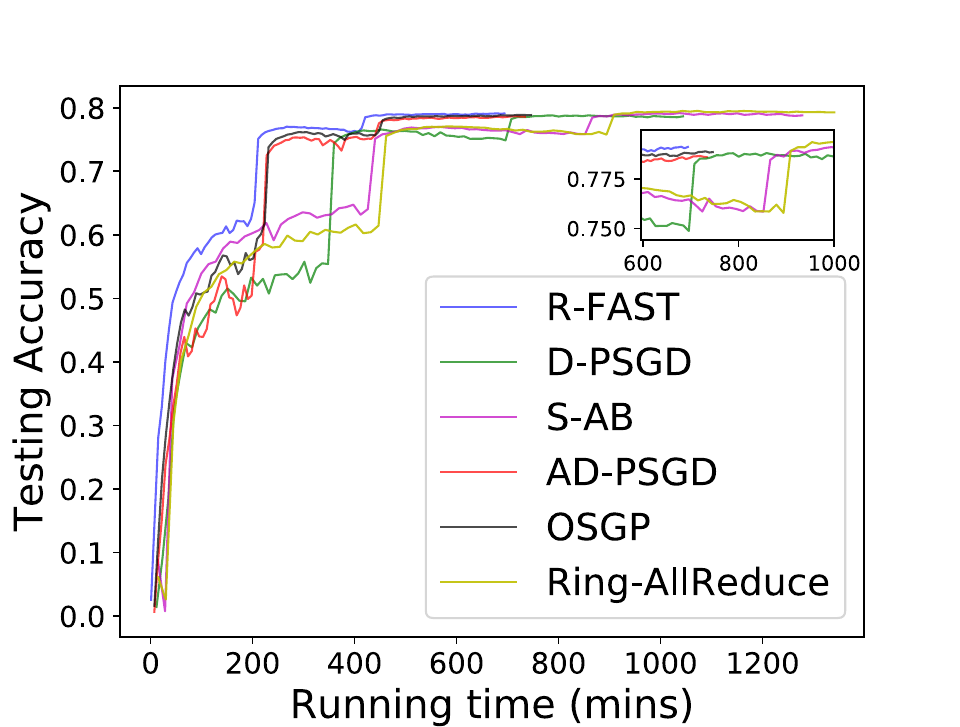}%
\label{fig2b}}
\hfil
\subfloat[]{\includegraphics[width=2.35in]{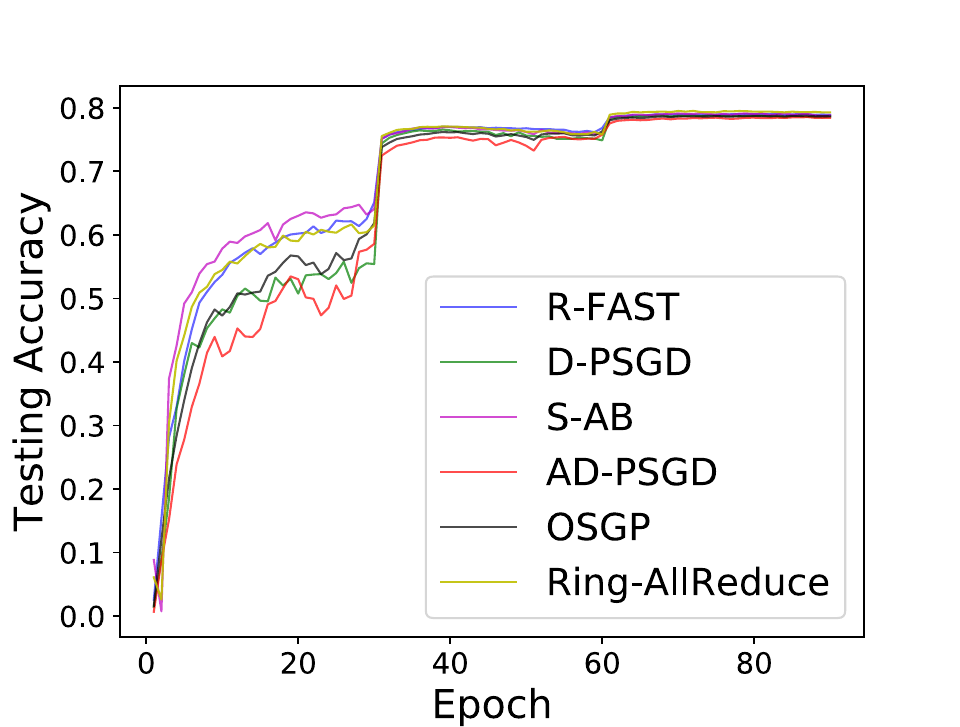}%
\label{fig2c}}
\caption{Performance comparison of R-FAST with D-PSGD, S-AB, AD-PSGD, OSGP and Ring-AllReduce in training ResNet-50 when there is no straggler.
}
\end{figure*}

\begin{figure*}[htbp]
\centering
\subfloat[]{\includegraphics[width=2.35in]{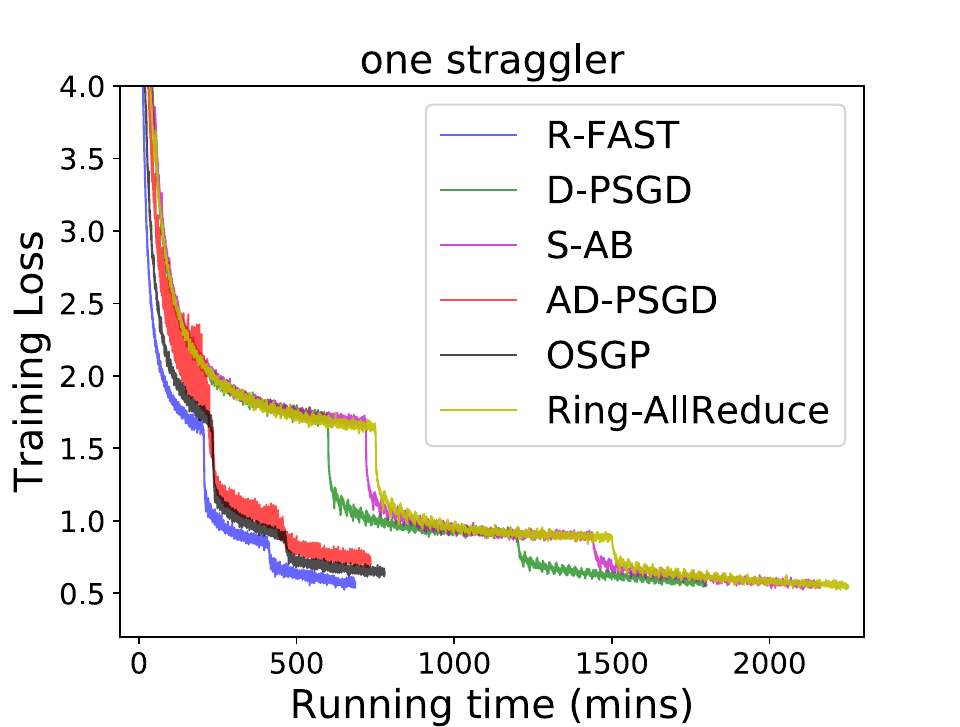}%
\label{fig3a}}
\hfil
\subfloat[]{\includegraphics[width=2.35in]{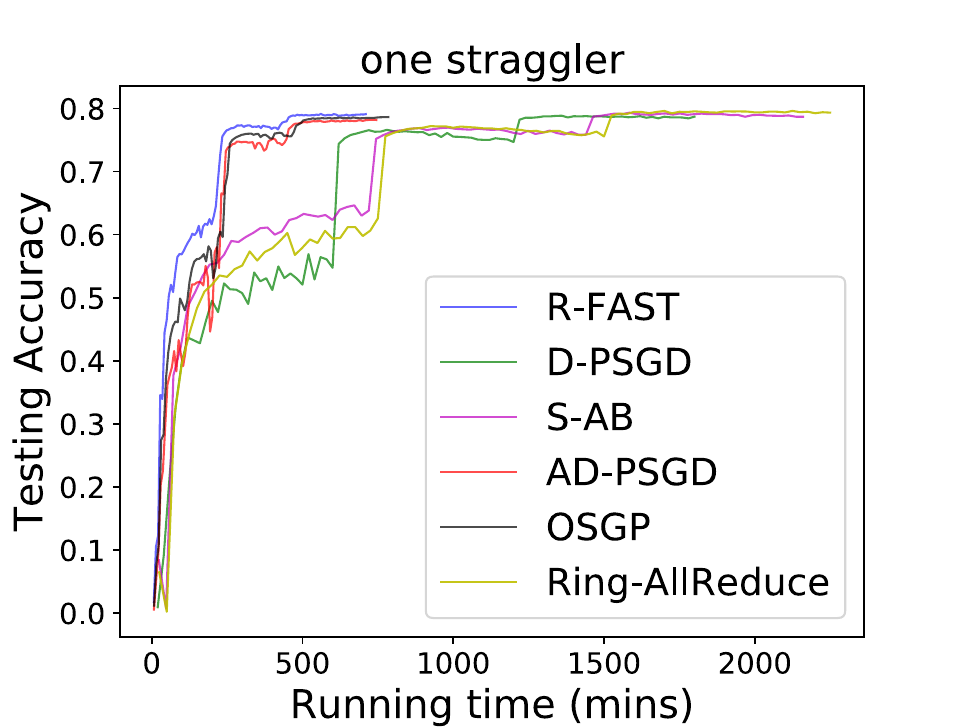}%
\label{fig3b}}
\hfil
\subfloat[]{\includegraphics[width=2.35in]{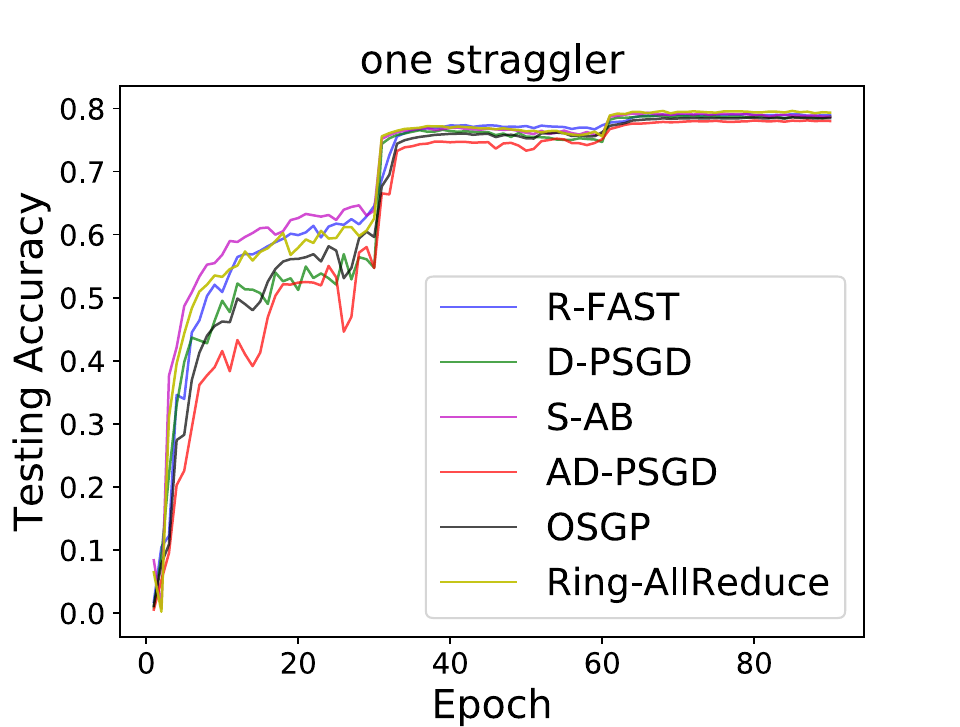}%
\label{fig3c}}
\caption{Performance comparison of R-FAST with D-PSGD, S-AB, AD-PSGD, OSGP and Ring-AllReduce in the presence of \textbf{one slow node/straggler}.}
\end{figure*}

\subsection{Large-scale image classification}\label{sec:sim_nc}
We now report another experiment by training ResNet-50~\cite{he2016deep} 
on ImageNet dataset, to verify the training efficiency and performance of our proposed R-FAST compared with the baselines. 
We randomly select 500 classes from the original ImageNet-1K~\cite{russakovsky2015imagenet} training set as our training set (thus containing 640,000 images), and select the same 500 classes from the original ImageNet-1K testing set as our testing set (thus containing 25,000 images). 
We compare the proposed R-FAST with the baselines: D-PSGD~\cite{lian2017can}, S-AB~\cite{xin2019distributed}, AD-PSGD~\cite{lian2018asynchronous}, OSGP~\cite{assran2019stochastic} and Ring-AllReduce~\cite{sergeev2018horovod}.
All these algorithms are deployed on a high-performance server with 8 Nvidia RTX2080 Ti GPUs, i.e., each process (node) uses a GPU to compute.
We run D-PSGD and AD-PSGD over an undirected ring graph as they require undirected communication topology; and run R-FAST, S-AB and OSGP over a directed ring graph. In addition, for asynchronous algorithm R-FAST, AD-PSGD and OSGP, packet losses exist under our artificial settings as we have presented at the beginning of Section~\ref{Experiments}. 
All experiments are run for 90 epochs, and we use the following commonly used hyperparameter setup for the training process: i) mini-batch size: 32 per node; ii) learning rate: the initial value being $0.1$ and decaying by a factor of $10$ per 30 epochs; iii) momentum: 0.9; and iv) weight decay: $10^{-4}$.

\textbf{Training efficiency.} When there is no slow node (straggler) in the system, i.e., each node has the same computing speed, it follows from Fig.~\ref{fig2a}, \ref{fig2b}, \ref{fig2c} and Table~\ref{table} (column 2, 3) that R-FAST converges $1.5$ times faster in running time than synchronous D-PSGD, S-AB and Ring-AllReduce while maintaining comparable testing accuracy (acc), thanks to the fully asynchronous training mechanism and robust gradient tracking strategy. Furthermore, R-FAST enjoys higher testing accuracy (c.f., Table \ref{table}) comparing to asynchronous AD-PSGD and OSGP, mainly due to the robust gradient tracking strategy which is able to tackle the packet losses.

\begin{table}[htbp]
\caption{Comparison of convergence performance for R-FAST, D-PSGD, S-AB, AD-PSGD, OSGP and Ring-AllReduce when training ResNet-50 under two different settings: i) Each node has the same computing power (no straggler); ii) Presence of a straggler.}
  \label{table}
  \centering
  \begin{tabular}{|c|c|c|c|c|}
    \hline
    \rule{0pt}{9pt}
    \multirow{2}{*}{Algorithm} & \multicolumn{2}{c|}{No straggler} & \multicolumn{2}{c|}{Presence of a straggler}    \\
    \cline{2-5}
    \rule{0pt}{9pt}
     & time(mins)  & acc($\%$) & time(mins) & acc($\%$)    \\
    \hline
    \rule{0pt}{9pt}
    \textbf{R-FAST}  & \textbf{703} & \textbf{79.12} & \textbf{712} & \textbf{79.11}\\
    \rule{0pt}{9pt}
    D-PSGD  & 1044  &78.77 & 1800 & 78.71\\
    \rule{0pt}{9pt}
    S-AB  & 1278  &79.14 & 2160 & 79.12 \\
    \rule{0pt}{9pt}
    AD-PSGD  & 720  &78.63 & 729 & 78.47\\
    \rule{0pt}{9pt}
    OSGP  & 753  &78.86 & 768 & 78.68\\
    \rule{0pt}{9pt}
    Ring-AllReduce  & 1341  &79.43 & 2250 & 79.43\\
    \hline
  \end{tabular}
\end{table}

\textbf{Robustness against straggler.}
To verify the robustness of R-FAST against heterogeneous computing power among nodes, we
randomly select a GPU and allocate extra computing burden to slow down its computation to mimic the straggler. It follows from Fig.~\ref{fig3a}, \ref{fig3b}, \ref{fig3c} and Table \ref{table} (column 4, 5) that the training efficiency of the proposed R-FAST is barely affected and runs much faster than synchronous algorithms (e.g., 3 times faster than Ring-AllReduce) while maintaining comparable accuracy. Moreover, R-FAST can achieve much higher testing accuracy than asynchronous AD-PSGD and OSGP, meaning that R-FAST is more robust against stragglers.

\begin{table}[htbp]
\caption{Performance comparison in training ResNet-50 for R-FAST over 4, 8, and 16 nodes.}
  \label{table_node_scale}
  \centering
  \begin{tabular}{|c|c|c|c|}
    \hline
    \rule{0pt}{12pt}
    node number  & 4  & 8 & 16 \\
    \hline
    \rule{0pt}{12pt}
    test accuracy(\%)  & 79.29  & 79.12 & 79.01 \\
    \rule{0pt}{12pt}
    time(mins)  & 1260  & 703 & 390 \\
    \hline
  \end{tabular}
\end{table}

\begin{figure}[htbp]
\centering
\includegraphics[scale=0.4]{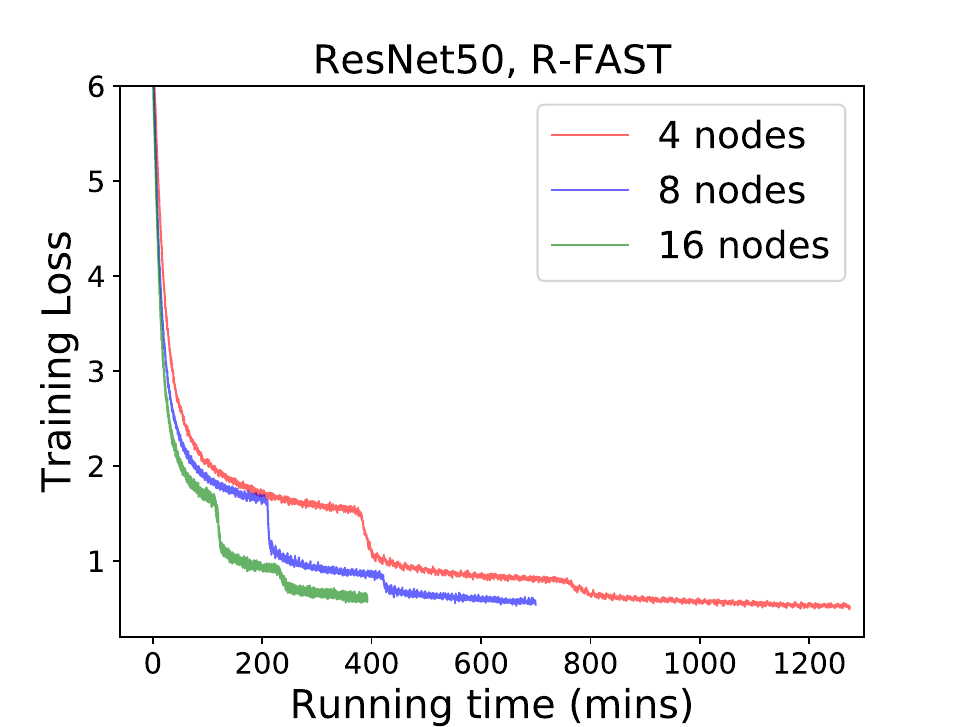}
\caption{Performance comparison in training ResNet-50 for R-FAST over 4, 8, and 16 nodes.}
\label{node_scale}
\end{figure}

\textbf{Scalability in the number of nodes.} To explore the scalability of our proposed R-FAST on different numbers of nodes for the task of large-scale image classification, we implement R-FAST over a network of ${4, 8, 16}$ nodes. We use a directed ring as the communication topology for each of the above networks and the total epoch for training is 90. 
It can be observed from Fig.~\ref{node_scale} and Table \ref{table_node_scale} that  the training speed of R-FAST increases almost linearly with the number of nodes while guaranteeing a small loss of accuracy, reflecting the good scalability of R-FAST in the number of nodes.

\section{Conclusion} 
\label{Conclusion}
In this paper, we proposed a new robust fully asynchronous method for solving large-scale distributed machine learning problems. We have provided the theoretical guarantee for the proposed R-FAST method for both strongly convex and non-convex cases.
Experiments have shown that R-FAST can run much faster than the synchronous counterparts and achieve higher testing accuracy than the existing well-known asynchronous algorithms, especially in the presence of stragglers, thanks to the introduced asynchronous mechanism and robust gradient tracking scheme.
Most importantly, R-FAST can work on two spanning-tree graphs for training, which makes it flexible in the design of network topology and is thus very suitable for scenarios where communication efficiency is a key concern and consensus protocols are difficult to design. It would be thus of great importance to explore new efficient communication topologies for message exchange.

\bibliographystyle{IEEEtran}
\bibliography{mybib}

\begin{thebibliography}{10}
\providecommand{\url}[1]{#1}
\csname url@samestyle\endcsname
\providecommand{\newblock}{\relax}
\providecommand{\bibinfo}[2]{#2}
\providecommand{\BIBentrySTDinterwordspacing}{\spaceskip=0pt\relax}
\providecommand{\BIBentryALTinterwordstretchfactor}{4}
\providecommand{\BIBentryALTinterwordspacing}{\spaceskip=\fontdimen2\font plus
\BIBentryALTinterwordstretchfactor\fontdimen3\font minus \fontdimen4\font\relax}
\providecommand{\BIBforeignlanguage}[2]{{%
\expandafter\ifx\csname l@#1\endcsname\relax
\typeout{** WARNING: IEEEtran.bst: No hyphenation pattern has been}%
\typeout{** loaded for the language `#1'. Using the pattern for}%
\typeout{** the default language instead.}%
\else
\language=\csname l@#1\endcsname
\fi
#2}}
\providecommand{\BIBdecl}{\relax}
\BIBdecl

\bibitem{lecun2015deep}
Y.~LeCun, Y.~Bengio, and G.~Hinton, ``Deep learning,'' \emph{nature}, vol. 521, no. 7553, pp. 436--444, 2015.

\bibitem{voulodimos2018deep}
A.~Voulodimos, N.~Doulamis, A.~Doulamis, and E.~Protopapadakis, ``Deep learning for computer vision: A brief review,'' \emph{Computational intelligence and neuroscience}, vol. 2018, 2018.

\bibitem{hirschberg2015advances}
J.~Hirschberg and C.~D. Manning, ``Advances in natural language processing,'' \emph{Science}, vol. 349, no. 6245, pp. 261--266, 2015.

\bibitem{muhammad2020deep}
K.~Muhammad, A.~Ullah, J.~Lloret, J.~Del~Ser, and V.~H.~C. de~Albuquerque, ``Deep learning for safe autonomous driving: Current challenges and future directions,'' \emph{IEEE Transactions on Intelligent Transportation Systems}, vol.~22, no.~7, pp. 4316--4336, 2020.

\bibitem{langer2020distributed}
M.~Langer, Z.~He, W.~Rahayu, and Y.~Xue, ``Distributed training of deep learning models: A taxonomic perspective,'' \emph{IEEE Transactions on Parallel and Distributed Systems}, vol.~31, no.~12, pp. 2802--2818, 2020.

\bibitem{nemirovski2009robust}
A.~Nemirovski, A.~Juditsky, G.~Lan, and A.~Shapiro, ``Robust stochastic approximation approach to stochastic programming,'' \emph{SIAM Journal on optimization}, vol.~19, no.~4, pp. 1574--1609, 2009.

\bibitem{moulines2011non}
E.~Moulines and F.~Bach, ``Non-asymptotic analysis of stochastic approximation algorithms for machine learning,'' \emph{Advances in neural information processing systems}, vol.~24, 2011.

\bibitem{ghadimi2013stochastic}
S.~Ghadimi and G.~Lan, ``Stochastic first-and zeroth-order methods for nonconvex stochastic programming,'' \emph{SIAM Journal on Optimization}, vol.~23, no.~4, pp. 2341--2368, 2013.

\bibitem{dean2012large}
J.~Dean, G.~Corrado, R.~Monga, K.~Chen, M.~Devin, M.~Mao, M.~Ranzato, A.~Senior, P.~Tucker, K.~Yang \emph{et~al.}, ``Large scale distributed deep networks,'' \emph{Advances in neural information processing systems}, vol.~25, 2012.

\bibitem{zinkevich2010parallelized}
M.~Zinkevich, M.~Weimer, L.~Li, and A.~Smola, ``Parallelized stochastic gradient descent,'' \emph{Advances in neural information processing systems}, vol.~23, 2010.

\bibitem{mcmahan2017communication}
B.~McMahan, E.~Moore, D.~Ramage, S.~Hampson, and B.~A. y~Arcas, ``Communication-efficient learning of deep networks from decentralized data,'' in \emph{Artificial intelligence and statistics}.\hskip 1em plus 0.5em minus 0.4em\relax PMLR, 2017, pp. 1273--1282.

\bibitem{sergeev2018horovod}
A.~Sergeev and M.~Del~Balso, ``Horovod: fast and easy distributed deep learning in tensorflow,'' \emph{arXiv preprint arXiv:1802.05799}, 2018.

\bibitem{goyal2017accurate}
P.~Goyal, P.~Doll{\'a}r, R.~Girshick, P.~Noordhuis, L.~Wesolowski, A.~Kyrola, A.~Tulloch, Y.~Jia, and K.~He, ``Accurate, large minibatch sgd: Training imagenet in 1 hour,'' \emph{arXiv preprint arXiv:1706.02677}, 2017.

\bibitem{lian2017can}
X.~Lian, C.~Zhang, H.~Zhang, C.-J. Hsieh, W.~Zhang, and J.~Liu, ``Can decentralized algorithms outperform centralized algorithms? a case study for decentralized parallel stochastic gradient descent,'' \emph{Advances in Neural Information Processing Systems}, vol.~30, 2017.

\bibitem{tang2018d}
H.~Tang, X.~Lian, M.~Yan, C.~Zhang, and J.~Liu, ``$\mathrm{D}^2$: Decentralized training over decentralized data,'' in \emph{International Conference on Machine Learning}.\hskip 1em plus 0.5em minus 0.4em\relax PMLR, 2018, pp. 4848--4856.

\bibitem{zhang2019decentralized}
J.~Zhang and K.~You, ``Decentralized stochastic gradient tracking for non-convex empirical risk minimization,'' \emph{arXiv preprint arXiv:1909.02712}, 2019.

\bibitem{xin2019distributed}
R.~Xin, A.~K. Sahu, U.~A. Khan, and S.~Kar, ``Distributed stochastic optimization with gradient tracking over strongly-connected networks,'' in \emph{2019 IEEE 58th Conference on Decision and Control (CDC)}.\hskip 1em plus 0.5em minus 0.4em\relax IEEE, 2019, pp. 8353--8358.

\bibitem{luo2020prague}
Q.~Luo, J.~He, Y.~Zhuo, and X.~Qian, ``Prague: High-performance heterogeneity-aware asynchronous decentralized training,'' in \emph{Proceedings of the Twenty-Fifth International Conference on Architectural Support for Programming Languages and Operating Systems}, 2020, pp. 401--416.

\bibitem{nadiradze2021asynchronous}
G.~Nadiradze, A.~Sabour, P.~Davies, S.~Li, and D.~Alistarh, ``Asynchronous decentralized sgd with quantized and local updates,'' \emph{Advances in Neural Information Processing Systems}, vol.~34, 2021.

\bibitem{verbraeken2020survey}
J.~Verbraeken, M.~Wolting, J.~Katzy, J.~Kloppenburg, T.~Verbelen, and J.~S. Rellermeyer, ``A survey on distributed machine learning,'' \emph{ACM Computing Surveys (CSUR)}, vol.~53, no.~2, pp. 1--33, 2020.

\bibitem{ram2009asynchronous}
S.~S. Ram, A.~Nedi{\'c}, and V.~V. Veeravalli, ``Asynchronous gossip algorithms for stochastic optimization,'' in \emph{Proceedings of the 48h IEEE Conference on Decision and Control (CDC) held jointly with 2009 28th Chinese Control Conference}.\hskip 1em plus 0.5em minus 0.4em\relax IEEE, 2009, pp. 3581--3586.

\bibitem{lian2018asynchronous}
X.~Lian, W.~Zhang, C.~Zhang, and J.~Liu, ``Asynchronous decentralized parallel stochastic gradient descent,'' in \emph{International Conference on Machine Learning}.\hskip 1em plus 0.5em minus 0.4em\relax PMLR, 2018, pp. 3043--3052.

\bibitem{assran2019stochastic}
M.~Assran, N.~Loizou, N.~Ballas, and M.~Rabbat, ``Stochastic gradient push for distributed deep learning,'' in \emph{International Conference on Machine Learning}.\hskip 1em plus 0.5em minus 0.4em\relax PMLR, 2019, pp. 344--353.

\bibitem{zhang2019fully}
J.~Zhang and K.~You, ``Fully asynchronous distributed optimization with linear convergence in directed networks,'' \emph{arXiv preprint arXiv:1901.08215}, 2019.

\bibitem{tian2020achieving}
Y.~Tian, Y.~Sun, and G.~Scutari, ``Achieving linear convergence in distributed asynchronous multiagent optimization,'' \emph{IEEE Transactions on Automatic Control}, vol.~65, no.~12, pp. 5264--5279, 2020.

\bibitem{pu2020push}
S.~Pu, W.~Shi, J.~Xu, and A.~Nedi{\'c}, ``Push--pull gradient methods for distributed optimization in networks,'' \emph{IEEE Transactions on Automatic Control}, vol.~66, no.~1, pp. 1--16, 2020.

\bibitem{kungurtsev2021decentralized}
V.~Kungurtsev, M.~Morafah, T.~Javidi, and G.~Scutari, ``Decentralized asynchronous non-convex stochastic optimization on directed graphs,'' \emph{IEEE Transactions on Control of Network Systems}, 2023.

\bibitem{recht2011hogwild}
B.~Recht, C.~Re, S.~Wright, and F.~Niu, ``Hogwild!: A lock-free approach to parallelizing stochastic gradient descent,'' \emph{Advances in neural information processing systems}, vol.~24, 2011.

\bibitem{agarwal2011distributed}
A.~Agarwal and J.~C. Duchi, ``Distributed delayed stochastic optimization,'' \emph{Advances in neural information processing systems}, vol.~24, 2011.

\bibitem{bedi2022fedbc}
A.~S. Bedi, C.~Fan, A.~Koppel, A.~K. Sahu, B.~M. Sadler, F.~Huang, and D.~Manocha, ``Fedbc: Calibrating global and local models via federated learning beyond consensus,'' \emph{arXiv preprint arXiv:2206.10815}, 2022.

\bibitem{bedi2019asynchronous}
A.~S. Bedi, A.~Koppel, and K.~Rajawat, ``Asynchronous saddle point algorithm for stochastic optimization in heterogeneous networks,'' \emph{IEEE Transactions on Signal Processing}, vol.~67, no.~7, pp. 1742--1757, 2019.

\bibitem{lian2015asynchronous}
X.~Lian, Y.~Huang, Y.~Li, and J.~Liu, ``Asynchronous parallel stochastic gradient for nonconvex optimization,'' \emph{Advances in Neural Information Processing Systems}, vol.~28, 2015.

\bibitem{zhang2015staleness}
W.~Zhang, S.~Gupta, X.~Lian, and J.~Liu, ``Staleness-aware async-sgd for distributed deep learning,'' \emph{arXiv preprint arXiv:1511.05950}, 2015.

\bibitem{zhou2022towards}
Z.~Zhou, Y.~Li, X.~Ren, and S.~Yang, ``Towards efficient and stable k-asynchronous federated learning with unbounded stale gradients on non-{IID} data,'' \emph{IEEE Transactions on Parallel and Distributed Systems}, 2022.

\bibitem{kempe2003gossip}
D.~Kempe, A.~Dobra, and J.~Gehrke, ``Gossip-based computation of aggregate information,'' in \emph{44th Annual IEEE Symposium on Foundations of Computer Science, 2003. Proceedings.}\hskip 1em plus 0.5em minus 0.4em\relax IEEE, 2003, pp. 482--491.

\bibitem{spiridonoff2020robust}
A.~Spiridonoff, A.~Olshevsky, and I.~C. Paschalidis, ``Robust asynchronous stochastic gradient-push: Asymptotically optimal and network-independent performance for strongly convex functions,'' \emph{Journal of Machine Learning Research}, vol.~21, no.~58, 2020.

\bibitem{miao2021heterogeneity}
X.~Miao, X.~Nie, Y.~Shao, Z.~Yang, J.~Jiang, L.~Ma, and B.~Cui, ``Heterogeneity-aware distributed machine learning training via partial reduce,'' in \emph{Proceedings of the 2021 International Conference on Management of Data}, 2021, pp. 2262--2270.

\bibitem{di2016next}
P.~Di~Lorenzo and G.~Scutari, ``Next: In-network nonconvex optimization,'' \emph{IEEE Transactions on Signal and Information Processing over Networks}, vol.~2, no.~2, pp. 120--136, 2016.

\bibitem{xu2015augmented}
J.~Xu, S.~Zhu, Y.~C. Soh, and L.~Xie, ``Augmented distributed gradient methods for multi-agent optimization under uncoordinated constant stepsizes,'' in \emph{2015 54th IEEE Conference on Decision and Control (CDC)}.\hskip 1em plus 0.5em minus 0.4em\relax IEEE, 2015, pp. 2055--2060.

\bibitem{pu2021distributed}
S.~Pu and A.~Nedi{\'c}, ``Distributed stochastic gradient tracking methods,'' \emph{Mathematical Programming}, vol. 187, no.~1, pp. 409--457, 2021.

\bibitem{gharesifard2010does}
B.~Gharesifard and J.~Cort{\'e}s, ``When does a digraph admit a doubly stochastic adjacency matrix?'' in \emph{Proceedings of the 2010 American Control Conference}.\hskip 1em plus 0.5em minus 0.4em\relax IEEE, 2010, pp. 2440--2445.

\bibitem{pu2020robust}
S.~Pu, ``A robust gradient tracking method for distributed optimization over directed networks,'' in \emph{2020 59th IEEE Conference on Decision and Control (CDC)}.\hskip 1em plus 0.5em minus 0.4em\relax IEEE, 2020, pp. 2335--2341.

\bibitem{cai2012average}
K.~Cai and H.~Ishii, ``Average consensus on general strongly connected digraphs,'' \emph{Automatica}, vol.~48, no.~11, pp. 2750--2761, 2012.

\bibitem{xin2018linear}
R.~Xin and U.~A. Khan, ``A linear algorithm for optimization over directed graphs with geometric convergence,'' \emph{IEEE Control Systems Letters}, vol.~2, no.~3, pp. 315--320, 2018.

\bibitem{tian2019asynchronous}
Y.~Tian, Y.~Sun, and G.~Scutari, ``Asynchronous decentralized successive convex approximation,'' \emph{arXiv preprint arXiv:1909.10144}, 2019.

\bibitem{paszke2017pytorch}
A.~Paszke, S.~Gross, S.~Chintala, and G.~Chanan, ``Pytorch: Tensors and dynamic neural networks in python with strong gpu acceleration.(2017),'' \emph{URL https://github. com/pytorch/pytorch}, 2017.

\bibitem{deng2012mnist}
L.~Deng, ``The mnist database of handwritten digit images for machine learning research [best of the web],'' \emph{IEEE signal processing magazine}, vol.~29, no.~6, pp. 141--142, 2012.

\bibitem{he2016deep}
K.~He, X.~Zhang, S.~Ren, and J.~Sun, ``Deep residual learning for image recognition,'' in \emph{Proceedings of the IEEE conference on computer vision and pattern recognition}, 2016, pp. 770--778.

\bibitem{russakovsky2015imagenet}
O.~Russakovsky, J.~Deng, H.~Su, J.~Krause, S.~Satheesh, S.~Ma, Z.~Huang, A.~Karpathy, A.~Khosla, M.~Bernstein \emph{et~al.}, ``Imagenet large scale visual recognition challenge,'' \emph{International journal of computer vision}, vol. 115, no.~3, pp. 211--252, 2015.

\end{thebibliography}

\onecolumn
\appendices
\begin{center}
\LARGE{\textbf{Supplemental Material}}
\end{center}
{\footnotesize
\tableofcontents
}

\section{Proof of Proposition~\ref{proposition 1}}
\label{appendix Proof of Proposition 1}
We first give a supporting lemma that we will use many times.
\begin{Lem}
\label{Refer_Aug_DGM}
Let $\left\{ v^k \right\} _{k=0}^{\infty}$ be non-negative sequence and $\lambda \in (0,1)$. Then there holds that 
\begin{equation}
    \left( \sum_{l=0}^k{\lambda ^{k-l}v^l} \right) ^2\leqslant \frac{1}{1-\lambda}\sum_{l=0}^k{\lambda ^{k-l}\left( v^l \right) ^2}.
\end{equation}
\end{Lem}
\begin{proof}
\begin{equation*}
\left( \sum_{l=0}^k{\lambda ^{k-l}v^l} \right) ^2=\left( \sum_{l=0}^k{\lambda ^{\frac{k-l}{2}}\left( \lambda ^{\frac{k-l}{2}}v^l \right)} \right) ^2\overset{\left( a \right)}{\leqslant}\sum_{l=0}^k{\left( \lambda ^{\frac{k-l}{2}} \right) ^2}\cdot \sum_{l=0}^k{\left( \lambda ^{\frac{k-l}{2}}v^l \right) ^2}\leqslant \frac{1}{1-\lambda}\sum_{l=0}^k{\lambda ^{k-l}\left( v^l \right) ^2},
\end{equation*}
where in (a) we have used Cauchy-Swarchz inequality.
\end{proof}

Now we restate Proposition~\ref{proposition 1} here:

\vspace{0.14cm}
\noindent \textbf{Proposition 1.} \textit{Consider the R-FAST (Algorithm~\ref{Myalgorithm_RASGT_GV}) with a constant step size $\gamma$. Suppose Assumptions \ref{Ass_weight_matrix}-\ref{Ass_bounded_var} hold. Then, we have for all $k\geqslant 0$}
\begin{equation}
\label{a}
E_{c}^{k+1}\leqslant 2C_{2}^{2}E_{c}^{0}\cdot \rho ^{2k}+\frac{4\gamma ^2C_{2}^{2}}{1-\rho}\sum_{l=0}^k{\rho ^{k-l}E_{z}^{l}}+\frac{4\gamma ^2C_{2}^{2}n}{1-\rho}\sum_{l=0}^k{\rho ^{k-l}\cdot \sigma ^2},
\end{equation}
\begin{equation}
\label{b}
E_{t}^{k+1}\leqslant 3C_{1}^{2}\left\| \bar{z}^0 \right\| ^2\cdot \rho ^{2k}+\sum_{l=0}^k{\rho ^{k-l}\left( \frac{27C_{1}^{2}C_{L}^{2}}{1-\rho}E_{c}^{l}+\frac{54\gamma ^2C_{1}^{2}C_{L}^{2}}{1-\rho}E_{z}^{l}+\frac{54\gamma ^2C_{1}^{2}C_{L}^{2}n}{1-\rho}\sigma ^2 \right)},
\end{equation}
\begin{equation}
\label{c}
E_{z}^{k}\leqslant 3E_{t}^{k}+3C_{L}^{2}nE_{c}^{k}+3L^2E_{o}^{k}.
\end{equation}
Further, assuming that $F$ is $\tau$-strongly convex and $\gamma \leqslant \frac{1}{L}$, we have for all $k\geqslant 0$
\begin{equation}
\label{d}
\begin{aligned}
 E_{o}^{k+1}\leqslant & 4\left( 1-\tau \eta ^2\gamma \right) ^{-2r}E_{o}^{0}\cdot \left( 1-\tau \eta ^2\gamma \right) ^{\frac{2r}{T}\cdot \left( k+1 \right)}
 \\
 &+\frac{4\left( 1-\tau \eta ^2\gamma \right) ^{-2r}\gamma ^2}{1-\left( 1-\tau \eta ^2\gamma \right) ^{\frac{r}{T}}}\sum_{l=0}^k{\left( 1-\tau \eta ^2\gamma \right) ^{\frac{r}{T}\left( k-l \right)}\left( C_{L}^{2}nE_{c}^{l}+E_{t}^{l}+n\sigma ^2 \right)} ,
 \end{aligned}
\end{equation}
\textit{where $C_1 \triangleq \frac{2\sqrt{2S}\left( 1+\bar{m}^{-K_1} \right)}{\rho \left( 1-\bar{m}^{K_1} \right)}$.}
\begin{proof}
Our strategy is to bound these four errors in terms of linear combinations of their past values.

\textbf{Bounding consensus error} (c.f., \eqref{a}):
Applying \eqref{iterate of h^k} recursively, we get
\begin{equation}
\label{chii}
h^{k+1} = \hat{W}^{k:0} h^0-\sum_{l=0}^k  \gamma^l \hat{W}^{k:l} e_{i^l}\left( z_{i^l}^l \right )^{\top}.
\end{equation}
With~\eqref{chii}, using \eqref{new_def_of_x_psi_0} and the fact $\left( \psi ^{t+1} \right) ^{\top}\hat{W}^t=\left( \psi ^t \right) ^{\top}
$~\cite{tian2019asynchronous}, we have
\begin{equation}
\label{def_of_x_psi_0}
x_{\psi}^{k+1}= x_{\psi}^{0}-\sum_{l=0}^k{\gamma ^l\left( \psi ^l \right) ^{\top}}e_{i^l}\left( z_{i^l}^{l} \right) ^{\top}.
\end{equation}
Using \eqref{chii},  \eqref{def_of_x_psi_0}, and Lemma~\ref{lemma_contraction_W^k}, one can obtain that
\begin{align*}
\norm{h^{k+1}-\bd{1}\CV^{k+1}} 
\leq  C_2\rho ^k\left\| {h}^0-{\mathbf{1}}x_{\psi}^{0} \right\| +C_2\sum_{l=0}^k{\rho ^{k-l}\gamma}\left\| z_{i^l}^{l} \right\|.
\end{align*}
Using $(a+b)^2 \leqslant 2a^2+2b^2$ and Lemma~\ref{Refer_Aug_DGM}, we have
\begin{equation*}
\begin{aligned}
\left\| h^{k+1}-\mathbf{1}x_{\psi}^{k+1} \right\| ^2&\leqslant 2\left( C_2\rho ^k\left\| h^0-\mathbf{1}x_{\psi}^{0} \right\| \right) ^2+2\left( \gamma C_2\sum_{l=0}^k{\rho ^{k-l}\left\| z_{i^l}^{l} \right\|} \right) ^2
\\
&\leqslant 2C_{2}^{2}\rho ^{2k}\left\| h^0-\mathbf{1}x_{\psi}^{0} \right\| ^2+\frac{2\gamma ^2C_{2}^{2}}{1-\rho}\sum_{l=0}^k{\rho ^{k-l}\left\| z_{i^l}^{l} \right\| ^2}.
\end{aligned}
\end{equation*}
Taking expectation on both sides of the above inequality, we get that
\begin{align*}
{\mathbb{E}\left[ \left\| h^{k+1}-\mathbf{1}x_{\psi}^{k+1} \right\| ^2 \right] }&\leqslant 2C_{2}^{2}\rho ^{2k}\left\| h^0-\mathbf{1}x_{\psi}^{0} \right\| ^2+\frac{2\gamma ^2C_{2}^{2}}{1-\rho}\sum_{l=0}^k{\rho ^{k-l}\mathbb{E}\left[ \left\| z_{i^l}^{l} \right\| ^2 \right]}
\\
&=2C_{2}^{2}\rho ^{2k}\left\| h^0-\mathbf{1}x_{\psi}^{0} \right\| ^2+\frac{2\gamma ^2C_{2}^{2}}{1-\rho}\sum_{l=0}^k{\rho ^{k-l}\mathbb{E}\left[ \left\| z_{i^l}^{l}-\bar{z}_{i^l}^{l}+\bar{z}_{i^l}^{l} \right\| ^2 \right]}
\\
&\leqslant 2C_{2}^{2}\rho ^{2k}\left\| h^0-\mathbf{1}x_{\psi}^{0} \right\| ^2+\frac{4\gamma ^2C_{2}^{2}}{1-\rho}\sum_{l=0}^k{\rho ^{k-l}\mathbb{E}\left[ \left\| \bar{z}_{i^l}^{l} \right\| ^2 \right]}+\frac{4\gamma ^2C_{2}^{2}}{1-\rho}\sum_{l=0}^k{\rho ^{k-l}\mathbb{E}\left[ \left\| z_{i^l}^{l}-\bar{z}_{i^l}^{l} \right\| ^2 \right]}
\\
&\leqslant 2C_{2}^{2}\rho ^{2k}\left\| h^0-\mathbf{1}x_{\psi}^{0} \right\| ^2+\frac{4\gamma ^2C_{2}^{2}}{1-\rho}\sum_{l=0}^k{\rho ^{k-l}\mathbb{E}\left[ \left\| \bar{z}_{i^l}^{l} \right\| ^2 \right]}+\frac{4\gamma ^2C_{2}^{2}n}{1-\rho}\sum_{l=0}^k{\rho ^{k-l}\cdot \sigma ^2},
\end{align*}
where we used \eqref{new_byproduct} in the last inequality.

\textbf{Bounding gradient tracking error} (c.f., \eqref{b}):
Similar with~\cite{tian2020achieving} (Proposition 19, (41b)), using 
 \eqref{iterate of h^k}, \eqref{compact_form_virtual}, \eqref{virtual_sequence} and Lemma \ref{lemma_contraction_A^k},
 we can get that
\begin{equation*}
    \left\| \bar{z}_{i^{k+1}}^{k+1}-\xi _{i^{k+1}}^{k}({\bar{z}}^{k+1})^{\top}{\mathbf{1}} \right\| \leqslant C_1\rho ^k\left\| {\bar{z}}^0 \right\| +3C_1C_L\sum_{l=0}^k{\rho ^{k-l}\left( \left\| {h}^l-{\mathbf{1}}x_{\psi}^{l} \right\| +\gamma \left\| z_{i^l}^{l} \right\| \right)}.
\end{equation*}
Using $(a+b+c)^2 \leqslant 3a^2+3b^2+3c^2$ and Lemma~\ref{Refer_Aug_DGM}, we have
\begin{align*}
&\left\| \bar{z}_{i^{k+1}}^{k+1}-\xi _{i^{k+1}}^{k}(\bar{z}^{k+1})^{\top}\mathbf{1} \right\| ^2
\\
\leqslant & 3\left( C_1\rho ^k\left\| \bar{z}^0 \right\| \right) ^2+3\left( 3C_1C_L\sum_{l=0}^k{\rho ^{k-l}\left\| h^l-\mathbf{1}x_{\psi}^{l} \right\|} \right) ^2+3\left( 3\gamma C_1C_L\sum_{l=0}^k{\rho ^{k-l}\left\| z_{i^l}^{l} \right\|} \right) ^2
\\
\leqslant & 3C_{1}^{2}\rho ^{2k}\left\| \bar{z}^0 \right\| ^2+\frac{27C_{1}^{2}C_{L}^{2}}{1-\rho}\sum_{l=0}^k{\rho ^{k-l}\left\| h^l-\mathbf{1}x_{\psi}^{l} \right\| ^2}+\frac{27\gamma ^2C_{1}^{2}C_{L}^{2}}{1-\rho}\sum_{l=0}^k{\rho ^{k-l}\left\| z_{i^l}^{l} \right\| ^2}.
\end{align*}
Taking expectation on both sides of the above inequality, we get that
\begin{align*}
&\mathbb{E}\left[ \left\| \bar{z}_{i^{k+1}}^{k+1}-\xi _{i^{k+1}}^{k}(\bar{z}^{k+1})^{\top}\mathbf{1} \right\| ^2 \right] 
\\
\leqslant & 3C_{1}^{2}\rho ^{2k}\left\| \bar{z}^0 \right\| ^2+\frac{27C_{1}^{2}C_{L}^{2}}{1-\rho}\sum_{l=0}^k{\rho ^{k-l}\mathbb{E}\left[ \left\| h^l-\mathbf{1}x_{\psi}^{l} \right\| ^2 \right]}+\frac{27\gamma ^2C_{1}^{2}C_{L}^{2}}{1-\rho}\sum_{l=0}^k{\rho ^{k-l}\mathbb{E}\left[ \left\| z_{i^l}^{l} \right\| ^2 \right]}
\\
= & 3C_{1}^{2}\rho ^{2k}\left\| \bar{z}^0 \right\| ^2+\frac{27C_{1}^{2}C_{L}^{2}}{1-\rho}\sum_{l=0}^k{\rho ^{k-l}\mathbb{E}\left[ \left\| h^l-\mathbf{1}x_{\psi}^{l} \right\| ^2 \right]}+\frac{27\gamma ^2C_{1}^{2}C_{L}^{2}}{1-\rho}\sum_{l=0}^k{\rho ^{k-l}\mathbb{E}\left[ \left\| z_{i^l}^{l}-\bar{z}_{i^l}^{l}+\bar{z}_{i^l}^{l} \right\| ^2 \right]}
\\
\leqslant & 3C_{1}^{2}\rho ^{2k}\left\| \bar{z}^0 \right\| ^2+\frac{27C_{1}^{2}C_{L}^{2}}{1-\rho}\sum_{l=0}^k{\rho ^{k-l}\mathbb{E}\left[ \left\| h^l-\mathbf{1}x_{\psi}^{l} \right\| ^2 \right]}
\\
&+\frac{54\gamma ^2C_{1}^{2}C_{L}^{2}}{1-\rho}\sum_{l=0}^k{\rho ^{k-l}\mathbb{E}\left[ \left\| \bar{z}_{i^l}^{l} \right\| ^2 \right]}+\frac{54\gamma ^2C_{1}^{2}C_{L}^{2}}{1-\rho}\sum_{l=0}^k{\rho ^{k-l}\mathbb{E}\left[ \left\| z_{i^l}^{l}-\bar{z}_{i^l}^{l} \right\| ^2 \right]}
\\
\leqslant & 3C_{1}^{2}\rho ^{2k}\left\| \bar{z}^0 \right\| ^2+\frac{27C_{1}^{2}C_{L}^{2}}{1-\rho}\sum_{l=0}^k{\rho ^{k-l}\mathbb{E}\left[ \left\| h^l-\mathbf{1}x_{\psi}^{l} \right\| ^2 \right]}
\\
&+\frac{54\gamma ^2C_{1}^{2}C_{L}^{2}}{1-\rho}\sum_{l=0}^k{\rho ^{k-l}\mathbb{E}\left[ \left\| \bar{z}_{i^l}^{l} \right\| ^2 \right]}+\frac{54\gamma ^2C_{1}^{2}C_{L}^{2}n}{1-\rho}\sum_{l=0}^k{\rho ^{k-l}\cdot \sigma ^2},
\end{align*}
where we used \eqref{new_byproduct} in the last inequality.

\textbf{Bounding gradient tracking variable} (c.f., \eqref{c}):
Using triangle inequality, we get that
\begin{align*}
\left\| \bar{z}_{i^k}^{k} \right\| &\leqslant \left\| \bar{z}_{i^k}^{k}-\xi _{i^k}^{k-1}(\bar{z}^k)^{\top}\mathbf{1} \right\| +\xi _{i^k}^{k-1}\left\| \nabla F\left( x_{\psi}^{k} \right) -(\bar{z}^k)^{\top}\mathbf{1} \right\| +\xi _{i^k}^{k-1}\left\| \nabla F\left( x_{\psi}^{k} \right) -\nabla F\left( x^{\star} \right) \right\| 
\\
&\overset{\left( a \right)}{=}\left\| \bar{z}_{i^k}^{k}-\xi _{i^k}^{k-1}(\bar{z}^k)^{\top}\mathbf{1} \right\| +\xi _{i^k}^{k-1}\left\| \sum_{i=1}^n{\left( \nabla f_i\left( x_{i}^{k} \right) -\nabla f_i\left( x_{\psi}^{k} \right) \right)} \right\| +\xi _{i^k}^{k-1}\left\| \nabla F\left( x_{\psi}^{k} \right) -\nabla F\left( x^{\star} \right) \right\| 
\\
&\overset{\left( b \right)}{\leqslant} \left\| \bar{z}_{i^k}^{k}-\xi _{i^k}^{k-1}(\bar{z}^k)^{\top}\mathbf{1} \right\| +C_L\sqrt{n}\left\| x^k-\mathbf{1}x_{\psi}^{k} \right\| +L\left\| x_{\psi}^{k}-\left( x^{\star} \right) ^{\top} \right\| 
\\
&\leqslant \left\| \bar{z}_{i^k}^{k}-\xi _{i^k}^{k-1}(\bar{z}^k)^{\top}\mathbf{1} \right\| +C_L\sqrt{n}\left\| h^k-\mathbf{1}x_{\psi}^{k} \right\| +L\left\| x_{\psi}^{k}-\left( x^{\star} \right) ^{\top} \right\| ,
\end{align*}
where we used~\eqref{auxiliary tracking_property} in $(a)$ and Assumption \ref{Ass_smoo} in $(b)$. Further, we have
\begin{align*}
\mathbb{E}\left[ \left\| \bar{z}_{i^k}^{k} \right\| ^2 \right] &\leqslant \mathbb{E}\left[ \left( \left\| \bar{z}_{i^k}^{k}-\xi _{i^k}^{k-1}(\bar{z}^k)^{\top}\mathbf{1} \right\| +C_L\sqrt{n}\left\| h^k-\mathbf{1}x_{\psi}^{k} \right\| +L\left\| x_{\psi}^{k}-\left( x^{\star} \right) ^{\top} \right\| \right) ^2 \right] 
\\
&\leqslant 3\mathbb{E}\left[ \left\| \bar{z}_{i^k}^{k}-\xi _{i^k}^{k-1}(\bar{z}^k)^{\top}\mathbf{1} \right\| ^2 \right] +3C_{L}^{2}n\mathbb{E}\left[ \left\| h^k-\mathbf{1}x_{\psi}^{k} \right\| ^2 \right] +3L^2\mathbb{E}\left[ \left\| x_{\psi}^{k}-\left( x^{\star} \right) ^{\top} \right\| ^2 \right] .
\end{align*}

\textbf{Bounding optimality gap} (c.f.,~\eqref{d}):
According to \eqref{each_iterate_of_x_psi_k}, we know that $x_\psi^{k+1}=x_\psi^k-\gamma\psi_{i^k}^k(z_{i^k}^k)^\top$. Using triangle inequality, we have

\begin{equation}
\label{optimal_gap_expectation}
    \begin{aligned}
\left\| x_{\psi}^{k+1}-\left( x^{\star} \right) ^{\top} \right\| = & \left\| x_{\psi}^{k}-\gamma \psi _{i^k}^{k}\left( z_{i^k}^{k} \right) ^{\top}-\left( x^{\star} \right) ^{\top} \right\| 
\\
\leqslant & \left\| x_{\psi}^{k}-\gamma \psi _{i^k}^{k}\xi _{i^k}^{k-1}\nabla F\left( x_{\psi}^{k} \right) -\left( x^{\star} \right) ^{\top} \right\| +\gamma \psi _{i^k}^{k}\xi _{i^k}^{k-1}\left\| \nabla F\left( x_{\psi}^{k} \right) -\mathbf{1}^{\top}\bar{z}^k \right\| 
\\
&+\gamma \psi _{i^k}^{k}\left\| \bar{z}_{i^k}^{k}-\xi _{i^k}^{k-1}\left( \bar{z}^k \right) ^{\top}\mathbf{1} \right\| +\gamma \psi _{i^k}^{k}\left\| z_{i^k}^{k}-\bar{z}_{i^k}^{k} \right\| 
\\
\overset{\left( a \right)}{\leqslant} & \underset{\triangleq \omega ^k}{\underbrace{\left( 1-\tau \psi _{i^k}^{k}\xi _{i^k}^{k-1}\gamma \right) }}\left\| x_{\psi}^{k}-\left( x^{\star} \right) ^{\top} \right\| +\gamma C_L\sqrt{n}\left\| h^k-\mathbf{1}x_{\psi}^{k} \right\|
\\
&+\gamma \left\| \bar{z}_{i^k}^{k}-\xi _{i^k}^{k-1}\left( \bar{z}^k \right) ^{\top}\mathbf{1} \right\| +\gamma \left\| z_{i^k}^{k}-\bar{z}_{i^k}^{k} \right\| 
\\
\overset{\left( b \right)}{\leqslant} & \prod_{t=0}^k{\omega ^t}\left\| x_{\psi}^{0}-\left( x^{\star} \right) ^{\top} \right\| +\gamma C_L\sqrt{n}\sum_{l=0}^k{\prod_{t=l}^{k-1}{\omega ^t}\left\| h^l-\mathbf{1}x_{\psi}^{l} \right\|}
\\
&+\gamma \sum_{l=0}^k{\prod_{t=l}^{k-1}{\omega ^t}\left\| \bar{z}_{i^l}^{l}-\xi _{i^l}^{l-1}\left( \bar{z}^l \right) ^{\top}\mathbf{1} \right\|}+\gamma \sum_{l=0}^k{\prod_{t=l}^{k-1}{\omega ^t}\left\| z_{i^l}^{l}-\bar{z}_{i^l}^{l} \right\|},
    \end{aligned}
\end{equation}
where in $(a)$ we used the $\tau$-strongly convex property of $F$, $\gamma \leqslant \frac{1}{L}$,~\eqref{auxiliary tracking_property} and Assumption~\ref{Ass_smoo}; $(b)$ follows by applying the above inequality recursively.

According to Lemma~\ref{lemma_contraction_W^k} and Lemma \ref{lemma_contraction_A^k}, we know that if $i^k \in \mathcal{R}$, $\psi_{i^k}^k \geqslant \eta $ and  $\xi_{i^k}^{k-1} \geqslant \eta $ ,
then $\omega^k \leq 1- \tau \eta^2 \gamma$. Besides, noticing that $r=\left| \mathcal{R} \right|$, we thus have for any $k\geqslant0$
\begin{equation}
\label{time_window_common_roots_activate_at_once}
\prod_{t=k}^{k+T-1}{\omega ^t}\leqslant \left( 1-\tau \eta ^2\gamma \right) ^r,
\end{equation}
and for any $s\geqslant 1$
\begin{equation}
\label{contraction_factor}
\prod_{t=k}^{k+s-1}{\omega ^t}\leqslant \left( 1-\tau \eta ^2\gamma \right) ^{\lfloor \frac{s}{T} \rfloor r}\leqslant \left( 1-\tau \eta ^2\gamma \right) ^{-r}\left( 1-\tau \eta ^2\gamma \right) ^{\frac{r}{T}s},
\end{equation}
where $(1-\tau \eta ^2\gamma )^{\frac{r}{T}}$ indicates the contraction factor.
Substituting~\eqref{contraction_factor} into~\eqref{optimal_gap_expectation} leads to
\begin{align*}
\left\| x_{\psi}^{k+1}-\left( x^{\star} \right) ^{\top} \right\| \leqslant & \left( 1-\tau \eta ^2\gamma \right) ^{-r}\left( 1-\tau \eta ^2\gamma \right) ^{\frac{r}{T}\left( k+1 \right)}\left\| x_{\psi}^{0}-\left( x^{\star} \right) ^{\top} \right\| 
\\
&+\left( 1-\tau \eta ^2\gamma \right) ^{-r}\gamma C_L\sqrt{n}\sum_{l=0}^k{\left( 1-\tau \eta ^2\gamma \right) ^{\frac{r}{T}\left( k-l \right)}\left\| h^l-1x_{\psi}^{l} \right\|}
\\
&+\left( 1-\tau \eta ^2\gamma \right) ^{-r}\gamma \sum_{l=0}^k{\left( 1-\tau \eta ^2\gamma \right) ^{\frac{r}{T}\left( k-l \right)}\left\| \bar{z}_{i^l}^{l}-\xi _{i^l}^{l-1}\left( \bar{z}^l \right) ^{\top}\mathbf{1} \right\|}
\\
&+\left( 1-\tau \eta ^2\gamma \right) ^{-r}\gamma \sum_{l=0}^k{\left( 1-\tau \eta ^2\gamma \right) ^{\frac{r}{T}\left( k-l \right)}\left\| z_{i^l}^{l}-\bar{z}_{i^l}^{l} \right\|}.
\end{align*}
Using $(a+b+c+d)^2 \leqslant 4a^2+4b^2+4c^2+4d^2$ and Lemma~\ref{Refer_Aug_DGM}, we have
\begin{align*}
\left\| x_{\psi}^{k+1}-\left( x^{\star} \right) ^{\top} \right\| ^2\leqslant & 4\left( 1-\tau \eta ^2\gamma \right) ^{-2r}\left( 1-\tau \eta ^2\gamma \right) ^{\frac{2r}{T}\left( k+1 \right)}\left\| x_{\psi}^{0}-\left( x^{\star} \right) ^{\top} \right\| ^2
\\
&+\frac{4\left( 1-\tau \eta ^2\gamma \right) ^{-2r}\gamma ^2C_{L}^{2}n}{1-\left( 1-\tau \eta ^2\gamma \right) ^{\frac{r}{T}}}\sum_{l=0}^k{\left( 1-\tau \eta ^2\gamma \right) ^{\frac{r}{T}\left( k-l \right)}\left\| h^l-1x_{\psi}^{l} \right\| ^2}
\\
&+\frac{4\left( 1-\tau \eta ^2\gamma \right) ^{-2r}\gamma ^2}{1-\left( 1-\tau \eta ^2\gamma \right) ^{\frac{r}{T}}}\sum_{l=0}^k{\left( 1-\tau \eta ^2\gamma \right) ^{\frac{r}{T}\left( k-l \right)}\left\| \bar{z}_{i^l}^{l}-\xi _{i^l}^{l-1}\left( \bar{z}^l \right) ^{\top}\mathbf{1} \right\| ^2}
\\
&+\frac{4\left( 1-\tau \eta ^2\gamma \right) ^{-2r}\gamma ^2}{1-\left( 1-\tau \eta ^2\gamma \right) ^{\frac{r}{T}}}\sum_{l=0}^k{\left( 1-\tau \eta ^2\gamma \right) ^{\frac{r}{T}\left( k-l \right)}\left\| z_{i^l}^{l}-\bar{z}_{i^l}^{l} \right\| ^2}.
\end{align*}
Taking expectation on both sides of the above inequality, we get
\begin{equation}
\label{response_eq_2}
\begin{aligned}
\mathbb{E}\left[ \left\| x_{\psi}^{k+1}-\left( x^{\star} \right) ^{\top} \right\| ^2 \right] \leqslant & 4\left( 1-\tau \eta ^2\gamma \right) ^{-2r}\left( 1-\tau \eta ^2\gamma \right) ^{\frac{2r}{T}\left( k+1 \right)}\left\| x_{\psi}^{0}-\left( x^{\star} \right) ^{\top} \right\| ^2
\\
&+\frac{4\left( 1-\tau \eta ^2\gamma \right) ^{-2r}\gamma ^2C_{L}^{2}n}{1-\left( 1-\tau \eta ^2\gamma \right) ^{\frac{r}{T}}}\sum_{l=0}^k{\left( 1-\tau \eta ^2\gamma \right) ^{\frac{r}{T}\left( k-l \right)}\mathbb{E}\left[ \left\| h^l-1x_{\psi}^{l} \right\| ^2 \right]}
\\
&+\frac{4\left( 1-\tau \eta ^2\gamma \right) ^{-2r}\gamma ^2}{1-\left( 1-\tau \eta ^2\gamma \right) ^{\frac{r}{T}}}\sum_{l=0}^k{\left( 1-\tau \eta ^2\gamma \right) ^{\frac{r}{T}\left( k-l \right)}\mathbb{E}\left[ \left\| \bar{z}_{i^l}^{l}-\xi _{i^l}^{l-1}\left( \bar{z}^l \right) ^{\top}\mathbf{1} \right\| ^2 \right]}
\\
&+\frac{4\left( 1-\tau \eta ^2\gamma \right) ^{-2r}\gamma ^2}{1-\left( 1-\tau \eta ^2\gamma \right) ^{\frac{r}{T}}}\sum_{l=0}^k{\left( 1-\tau \eta ^2\gamma \right) ^{\frac{r}{T}\left( k-l \right)}\mathbb{E}\left[ \left\| z_{i^k}^{k}-\bar{z}_{i^k}^{k} \right\| ^2 \right]}
\\
\leqslant & 4\left( 1-\tau \eta ^2\gamma \right) ^{-2r}\left( 1-\tau \eta ^2\gamma \right) ^{\frac{2r}{T}\left( k+1 \right)}\left\| x_{\psi}^{0}-\left( x^{\star} \right) ^{\top} \right\| ^2
\\
&+\frac{4\left( 1-\tau \eta ^2\gamma \right) ^{-2r}\gamma ^2C_{L}^{2}n}{1-\left( 1-\tau \eta ^2\gamma \right) ^{\frac{r}{T}}}\sum_{l=0}^k{\left( 1-\tau \eta ^2\gamma \right) ^{\frac{r}{T}\left( k-l \right)}\mathbb{E}\left[ \left\| h^l-1x_{\psi}^{l} \right\| ^2 \right]}
\\
&+\frac{4\left( 1-\tau \eta ^2\gamma \right) ^{-2r}\gamma ^2}{1-\left( 1-\tau \eta ^2\gamma \right) ^{\frac{r}{T}}}\sum_{l=0}^k{\left( 1-\tau \eta ^2\gamma \right) ^{\frac{r}{T}\left( k-l \right)}\mathbb{E}\left[ \left\| \bar{z}_{i^l}^{l}-\xi _{i^l}^{l-1}\left( \bar{z}^l \right) ^{\top}\mathbf{1} \right\| ^2 \right]}
\\
&+\frac{4\left( 1-\tau \eta ^2\gamma \right) ^{-2r}\gamma ^2n}{1-\left( 1-\tau \eta ^2\gamma \right) ^{\frac{r}{T}}}\sum_{l=0}^k{\left( 1-\tau \eta ^2\gamma \right) ^{\frac{r}{T}\left( k-l \right)}\sigma ^2},
\end{aligned}
\end{equation}
where we used \eqref{new_byproduct} in the last inequality.

The proof of Proposition~\ref{proposition 1} is completed.
\end{proof}

\section{Proof of Lemma~\ref{lemma8}}
\label{appendix_proof_of_lemma8}
We restate Lemma~\ref{lemma8} here:

\vspace{0.14cm}
\noindent \textbf{Lemma 6.}
\textit{Suppose Assumptions \ref{Ass_weight_matrix}-\ref{Ass_bounded_var} hold. Let the constant step size $\gamma$ satisfy}
\begin{equation}
    \gamma \,\,< \frac{1}{\sqrt{3\varrho _cC_{L}^{2}n+3\varrho _t}}.
\end{equation}
\textit{Then, for all $k\geqslant 0$, we have}
\begin{equation}
    \begin{aligned}
\sum_{l=0}^k{\mathbb{E}\left\| h^l-\mathbf{1}x_{\psi}^{l} \right\| ^2}\leqslant & \frac{3\varrho _c\gamma ^2}{1-3\left( \varrho _cC_{L}^{2}n+\varrho _t \right) \gamma ^2}\sum_{l=0}^k{\mathbb{E}\left\| \nabla F\left( x_{\psi}^{l} \right) \right\| ^2}
\\
&+\frac{\left( 1-3\varrho _t\gamma ^2 \right) c_c+3c_t\varrho _c\gamma ^2}{1-3\left( \varrho _cC_{L}^{2}n+\varrho _t \right) \gamma ^2}+\frac{\varrho _cn\gamma ^2\left( k+1 \right)}{1-3\left( \varrho _cC_{L}^{2}n+\varrho _t \right) \gamma ^2}\sigma ^2
,
    \end{aligned}
\end{equation}
\textit{and}
\begin{equation}
    \begin{aligned}
\sum_{l=0}^k{\mathbb{E}\left\| \bar{z}_{i^l}^{l}-\xi _{i^l}^{l-1}(\bar{z}^l)^{\top}\mathbf{1} \right\| ^2}\leqslant & \frac{\varrho _tn\gamma ^2\left( k+1 \right)}{1-3\left( \varrho _cC_{L}^{2}n+\varrho _t \right) \gamma ^2}\sigma ^2
+\frac{3\varrho _t\gamma ^2}{1-3\left( \varrho _cC_{L}^{2}n+\varrho _t \right) \gamma ^2}\sum_{l=0}^k{\mathbb{E}\left\| \nabla F\left( x_{\psi}^{l} \right) \right\| ^2}
\\
&+\frac{\left( 1-3\varrho _cC_{L}^{2}n\gamma ^2 \right) c_t+3c_c\varrho _tC_{L}^{2}n\gamma ^2}{1-3\left( \varrho _cC_{L}^{2}n+\varrho _t \right) \gamma ^2},
    \end{aligned}
\end{equation}
\textit{where $c_c$, $\varrho _c$, $c_t$ and $\varrho _t$ are constants given as below:}
\begin{equation}
    c_c\triangleq\left( 1+\frac{2C_{2}^{2}}{1-\rho ^2} \right) \left\| h^0-1x_{\psi}^{0} \right\| ^2, \qquad \varrho _c\triangleq\frac{4C_{2}^{2}}{\left( 1-\rho \right) ^2},
\end{equation}

\begin{equation}
\begin{aligned}
&c_t\triangleq \frac{3C_{1}^{2}\left\| \bar{z}^0 \right\| ^2}{1-\rho ^2}+\frac{27C_{1}^{2}C_{L}^{2}\left( 2C_{2}^{2}+1-\rho ^2 \right)}{\left( 1-\rho \right) ^4}\left\| h^0-\mathbf{1}x_{\psi}^{0} \right\| ^2+\left\| \bar{z}_{i^0}^{0}-\xi _{i^0}^{-1}(\bar{z}^0)^{\top}\mathbf{1} \right\| ^2
,  
\\
&\varrho _t\triangleq \frac{54C_{1}^{2}C_{L}^{2}\left[ 4C_{2}^{2}+\left( 1-\rho \right) ^2 \right]}{\left( 1-\rho \right) ^4}.
\end{aligned}
\end{equation}

\begin{proof}
According to~\eqref{a}, we know that for $l \geqslant 1$:
\begin{equation}
\mathbb{E}\left[ \left\| h^l-\mathbf{1}x_{\psi}^{l} \right\| ^2 \right] \leqslant 2C_{2}^{2}\rho ^{2\left( l-1 \right)}\left\| h^0-\mathbf{1}x_{\psi}^{0} \right\| ^2+\frac{4\gamma ^2C_{2}^{2}}{1-\rho}\sum_{t=0}^{l-1}{\rho ^{l-1-t}\mathbb{E}\left[ \left\| \bar{z}_{i^t}^{t} \right\| ^2 \right]}+\frac{4\gamma ^2C_{2}^{2}n\sigma ^2}{\left( 1-\rho \right) ^2}.
\end{equation}

Then, for $k \geqslant 0$, we have
\begin{equation}
\label{eq_60}
\begin{aligned}
&\sum_{l=0}^k{\mathbb{E}\left[ \left\| h^l-\mathbf{1}x_{\psi}^{l} \right\| ^2 \right]}
\\
= & \left\| h^0-\mathbf{1}x_{\psi}^{0} \right\| ^2+\sum_{l=1}^k{\mathbb{E}\left[ \left\| h^l-\mathbf{1}x_{\psi}^{l} \right\| ^2 \right]}
\\
\leqslant & \left\| h^0-\mathbf{1}x_{\psi}^{0} \right\| ^2+2C_{2}^{2}\left\| h^0-\mathbf{1}x_{\psi}^{0} \right\| ^2\sum_{l=1}^k{\rho ^{2\left( l-1 \right)}}+\frac{4\gamma ^2C_{2}^{2}}{1-\rho}\sum_{l=1}^k{\sum_{t=0}^{l-1}{\rho ^{l-1-t}\mathbb{E}\left[ \left\| \bar{z}_{i^t}^{t} \right\| ^2 \right]}}+\frac{4\gamma ^2C_{2}^{2}n\sigma ^2}{\left( 1-\rho \right) ^2}\left( k+1 \right) 
\\
\leqslant & \left\| h^0-\mathbf{1}x_{\psi}^{0} \right\| ^2+\frac{2C_{2}^{2}\left\| h^0-\mathbf{1}x_{\psi}^{0} \right\| ^2}{1-\rho ^2}+\frac{4\gamma ^2C_{2}^{2}}{1-\rho}\sum_{t=0}^{k-1}{\sum_{l=t+1}^k{\rho ^{l-1-t}\mathbb{E}\left[ \left\| \bar{z}_{i^t}^{t} \right\| ^2 \right]}}+\frac{4\gamma ^2C_{2}^{2}n\sigma ^2}{\left( 1-\rho \right) ^2}\left( k+1 \right) 
\\
\leqslant & \left( 1+\frac{2C_{2}^{2}}{1-\rho ^2} \right) \left\| h^0-\mathbf{1}x_{\psi}^{0} \right\| ^2+\frac{4\gamma ^2C_{2}^{2}}{\left( 1-\rho \right) ^2}\sum_{t=0}^{k-1}{\mathbb{E}\left[ \left\| \bar{z}_{i^t}^{t} \right\| ^2 \right]}+\frac{4\gamma ^2C_{2}^{2}n\sigma ^2}{\left( 1-\rho \right) ^2}\left( k+1 \right) 
\\
\leqslant & \underset{c_c}{\underbrace{\left( 1+\frac{2C_{2}^{2}}{1-\rho ^2} \right) \left\| h^0-\mathbf{1}x_{\psi}^{0} \right\| ^2}}+\underset{\varrho _c}{\underbrace{\frac{4C_{2}^{2}}{\left( 1-\rho \right) ^2}}}\cdot \gamma ^2\sum_{l=0}^k{\mathbb{E}\left[ \left\| \bar{z}_{i^l}^{l} \right\| ^2 \right]}+\underset{\varrho _c}{\underbrace{\frac{4C_{2}^{2}}{\left( 1-\rho \right) ^2}}}\gamma ^2n\sigma ^2\left( k+1 \right) .
\end{aligned}
\end{equation}

According to~\eqref{b}, we know that for $l \geqslant 1$:
\begin{equation}
\begin{aligned}
\mathbb{E}\left[ \left\| \bar{z}_{i^l}^{l}-\xi _{i^l}^{l-1}(\bar{z}^l)^{\top}\mathbf{1} \right\| ^2 \right] \leqslant & 3C_{1}^{2}\rho ^{2\left( l-1 \right)}\left\| \bar{z}^0 \right\| ^2+\frac{27C_{1}^{2}C_{L}^{2}}{1-\rho}\sum_{t=0}^{l-1}{\rho ^{l-1-t}\mathbb{E}\left[ \left\| h^t-\mathbf{1}x_{\psi}^{t} \right\| ^2 \right]}
\\
&+\frac{54\gamma ^2C_{1}^{2}C_{L}^{2}}{1-\rho}\sum_{t=0}^{l-1}{\rho ^{l-1-t}\mathbb{E}\left[ \left\| \bar{z}_{i^t}^{t} \right\| ^2 \right]}+\frac{54\gamma ^2C_{1}^{2}C_{L}^{2}n\sigma ^2}{\left( 1-\rho \right) ^2}.
\end{aligned}
\end{equation}

Then, for $k \geqslant 0$, we have
\begin{equation}
\label{wanqi_yang}
\begin{aligned}
& \sum_{l=0}^k{\mathbb{E}\left[ \left\| \bar{z}_{i^l}^{l}-\xi _{i^l}^{l-1}(\bar{z}^l)^{\top}\mathbf{1} \right\| ^2 \right]}
\\
= &\left\| \bar{z}_{i^0}^{0}-\xi _{i^0}^{-1}(\bar{z}^0)^{\top}\mathbf{1} \right\| ^2+\sum_{l=1}^k{\mathbb{E}\left[ \left\| \bar{z}_{i^l}^{l}-\xi _{i^l}^{l-1}(\bar{z}^l)^{\top}\mathbf{1} \right\| ^2 \right]}
\\
\leqslant & \left\| \bar{z}_{i^0}^{0}-\xi _{i^0}^{-1}(\bar{z}^0)^{\top}\mathbf{1} \right\| ^2+3C_{1}^{2}\left\| \bar{z}^0 \right\| ^2\sum_{l=1}^k{\rho ^{2\left( l-1 \right)}}+\frac{27C_{1}^{2}C_{L}^{2}}{1-\rho}\sum_{l=1}^k{\sum_{t=0}^{l-1}{\rho ^{l-1-t}\cdot \mathbb{E}\left[ \left\| h^t-\mathbf{1}x_{\psi}^{t} \right\| ^2 \right]}}
\\
& +\frac{54\gamma ^2C_{1}^{2}C_{L}^{2}}{1-\rho}\sum_{l=1}^k{\sum_{t=0}^{l-1}{\rho ^{l-1-t}\cdot \mathbb{E}\left[ \left\| \bar{z}_{i^t}^{t} \right\| ^2 \right]}}+\frac{54\gamma ^2C_{1}^{2}C_{L}^{2}n\sigma ^2}{\left( 1-\rho \right) ^2}\left( k+1 \right) .
\end{aligned}
\end{equation}

Using the fact that
\begin{equation*}
\sum_{l=1}^k{\sum_{t=0}^{l-1}{\rho ^{l-1-t}\cdot \mathbb{E}\left[ \left\| h^t-\mathbf{1}x_{\psi}^{t} \right\| ^2 \right]}}=\sum_{t=0}^{k-1}{\sum_{l=t+1}^k{\rho ^{l-1-t}\cdot \mathbb{E}\left[ \left\| h^t-\mathbf{1}x_{\psi}^{t} \right\| ^2 \right]}}\leqslant \frac{1}{1-\rho}\sum_{t=0}^{k-1}{\mathbb{E}\left[ \left\| h^t-\mathbf{1}x_{\psi}^{t} \right\| ^2 \right]}
\end{equation*}
and \begin{equation*}
\sum_{l=1}^k{\sum_{t=0}^{l-1}{\rho ^{l-1-t}\cdot \mathbb{E}\left[ \left\| \bar{z}_{i^t}^{t} \right\| ^2 \right]}}=\sum_{t=0}^{k-1}{\sum_{l=t+1}^k{\rho ^{l-1-t}\cdot \mathbb{E}\left[ \left\| \bar{z}_{i^t}^{t} \right\| ^2 \right]}}\leqslant \frac{1}{1-\rho}\sum_{t=0}^{k-1}{\mathbb{E}\left[ \left\| \bar{z}_{i^t}^{t} \right\| ^2 \right]},
\end{equation*}
\eqref{wanqi_yang} becomes
\begin{equation}
\label{eq_62}
\begin{aligned}
& \sum_{l=0}^k{\mathbb{E}\left[ \left\| \bar{z}_{i^l}^{l}-\xi _{i^l}^{l-1}(\bar{z}^l)^{\top}\mathbf{1} \right\| ^2 \right]}
\\
\leqslant & \left\| \bar{z}_{i^0}^{0}-\xi _{i^0}^{-1}(\bar{z}^0)^{\top}\mathbf{1} \right\| ^2+\frac{3C_{1}^{2}\left\| \bar{z}^0 \right\| ^2}{1-\rho ^2}+\frac{27C_{1}^{2}C_{L}^{2}}{\left( 1-\rho \right) ^2}\sum_{t=0}^{k-1}{\mathbb{E}\left[ \left\| h^t-\mathbf{1}x_{\psi}^{t} \right\| ^2 \right]}
\\
& +\frac{54\gamma ^2C_{1}^{2}C_{L}^{2}}{\left( 1-\rho \right) ^2}\sum_{t=0}^{k-1}{\mathbb{E}\left[ \left\| \bar{z}_{i^t}^{t} \right\| ^2 \right]}+\frac{54\gamma ^2C_{1}^{2}C_{L}^{2}n\sigma ^2}{\left( 1-\rho \right) ^2}\left( k+1 \right) 
\\
\leqslant & \left\| \bar{z}_{i^0}^{0}-\xi _{i^0}^{-1}(\bar{z}^0)^{\top}\mathbf{1} \right\| ^2+\frac{3C_{1}^{2}\left\| \bar{z}^0 \right\| ^2}{1-\rho ^2}+\frac{27C_{1}^{2}C_{L}^{2}}{\left( 1-\rho \right) ^2}\sum_{l=0}^k{\mathbb{E}\left[ \left\| h^l-\mathbf{1}x_{\psi}^{l} \right\| ^2 \right]}
\\
& +\frac{54\gamma ^2C_{1}^{2}C_{L}^{2}}{\left( 1-\rho \right) ^2}\sum_{l=0}^k{\mathbb{E}\left[ \left\| \bar{z}_{i^l}^{l} \right\| ^2 \right]}+\frac{54\gamma ^2C_{1}^{2}C_{L}^{2}n\sigma ^2}{\left( 1-\rho \right) ^2}\left( k+1 \right) 
\\
\overset{\left( a \right)}{\leqslant} & \left\| \bar{z}_{i^0}^{0}-\xi _{i^0}^{-1}(\bar{z}^0)^{\top}\mathbf{1} \right\| ^2+\frac{3C_{1}^{2}\left\| \bar{z}^0 \right\| ^2}{1-\rho ^2}+\frac{54\gamma ^2C_{1}^{2}C_{L}^{2}}{\left( 1-\rho \right) ^2}\sum_{l=0}^k{\mathbb{E}\left[ \left\| \bar{z}_{i^l}^{l} \right\| ^2 \right]}+\frac{54\gamma ^2C_{1}^{2}C_{L}^{2}n\sigma ^2}{\left( 1-\rho \right) ^2}\left( k+1 \right) 
\\
& +\frac{27C_{1}^{2}C_{L}^{2}}{\left( 1-\rho \right) ^2}\left\{ \left( 1+\frac{2C_{2}^{2}}{1-\rho ^2} \right) \left\| h^0-\mathbf{1}x_{\psi}^{0} \right\| ^2+\frac{4\gamma ^2C_{2}^{2}}{\left( 1-\rho \right) ^2}\sum_{l=0}^k{\mathbb{E}\left[ \left\| \bar{z}_{i^l}^{l} \right\| ^2 \right]}+\frac{4\gamma ^2C_{2}^{2}n\sigma ^2}{\left( 1-\rho \right) ^2}\left( k+1 \right) \right\} 
\\
= & \underset{c_t}{\underbrace{\frac{3C_{1}^{2}\left\| \bar{z}^0 \right\| ^2}{1-\rho ^2}+\frac{27C_{1}^{2}C_{L}^{2}\left( 2C_{2}^{2}+1-\rho ^2 \right)}{\left( 1-\rho \right) ^4}\left\| h^0-\mathbf{1}x_{\psi}^{0} \right\| ^2+\left\| \bar{z}_{i^0}^{0}-\xi _{i^0}^{-1}(\bar{z}^0)^{\top}\mathbf{1} \right\| ^2}}
\\
& +\underset{\varrho _t}{\underbrace{\frac{54C_{1}^{2}C_{L}^{2}\left[ 2C_{2}^{2}+\left( 1-\rho \right) ^2 \right]}{\left( 1-\rho \right) ^4}}}\cdot \gamma ^2\sum_{l=0}^k{\mathbb{E}\left[ \left\| \bar{z}_{i^l}^{l} \right\| ^2 \right]}+\underset{\varrho _t}{\underbrace{\frac{54C_{1}^{2}C_{L}^{2}\left[ 2C_{2}^{2}+\left( 1-\rho \right) ^2 \right]}{\left( 1-\rho \right) ^4}}}\cdot \gamma ^2n\sigma ^2\left( k+1 \right) .
\end{aligned}
\end{equation}
where in $(a)$ we have used~\eqref{eq_60}.

Further, for~\eqref{eq_60} and~\eqref{eq_62}, using $\norm{a+b+c}^2 \leqslant 3\norm{a}^2+3\norm{b}^2+3\norm{c}^2$, we have
\begin{align*}
&\sum_{l=0}^k{\mathbb{E}\left[ \left\| h^l-\mathbf{1}x_{\psi}^{l} \right\| ^2 \right]}
\\
\leqslant & c_c+3\varrho _c\gamma ^2\sum_{l=0}^k{\mathbb{E}\left[ \left\| \bar{z}_{i^l}^{l}-\xi _{i^l}^{l-1}\left( \bar{z}^l \right) ^{\top}\mathbf{1} \right\| ^2 \right]}+3\varrho _c\gamma ^2\sum_{l=0}^k{\left( \xi _{i^l}^{l-1} \right) ^2\mathbb{E}\left[ \left\| \nabla F\left( x_{\psi}^{l} \right) \right\| ^2 \right]}
\\
&+3\varrho _c\gamma ^2\sum_{l=0}^k{\left( \xi _{i^l}^{l-1} \right) ^2\mathbb{E}\left[ \left\| \nabla F\left( x_{\psi}^{l} \right) -\left( \bar{z}^l \right) ^{\top}\mathbf{1} \right\| ^2 \right]}+\varrho _c\gamma ^2\sum_{l=0}^k{n\sigma ^2}
\\
\overset{\left( a \right)}{\leqslant} & c_c+3\varrho _c\gamma ^2\sum_{l=0}^k{\mathbb{E}\left[ \left\| \bar{z}_{i^l}^{l}-\xi _{i^l}^{l-1}\left( \bar{z}^l \right) ^{\top}\mathbf{1} \right\| ^2 \right]}+3\varrho _c\gamma ^2\sum_{l=0}^k{\mathbb{E}\left[ \left\| \sum_{i=1}^n{\left( \nabla f_i\left( x_{i}^{k} \right) -\nabla f_i\left( x_{\psi}^{k} \right) \right)} \right\| ^2 \right]}
\\
&+\varrho _c\gamma ^2\sum_{l=0}^k{\left( 3\mathbb{E}\left[ \left\| \nabla F\left( x_{\psi}^{l} \right) \right\| ^2 \right] +n\sigma ^2 \right)}
\\
\overset{\left( b \right)}{\leqslant} & c_c+3\varrho _c\gamma ^2\sum_{l=0}^k{\mathbb{E}\left[ \left\| \bar{z}_{i^l}^{l}-\xi _{i^l}^{l-1}\left( \bar{z}^l \right) ^{\top}\mathbf{1} \right\| ^2 \right]}+3\varrho _cC_{L}^{2}n\gamma ^2\sum_{l=0}^k{\mathbb{E}\left[ \left\| h^l-\mathbf{1}x_{\psi}^{l} \right\| ^2 \right]}
\\
&+\varrho _c\gamma ^2\sum_{l=0}^k{\left( 3\mathbb{E}\left[ \left\| \nabla F\left( x_{\psi}^{l} \right) \right\| ^2 \right] +n\sigma ^2 \right)},
\end{align*}

and
\begin{align*}
&\sum_{l=0}^k{\mathbb{E}\left[ \left\| \bar{z}_{i^l}^{l}-\xi _{i^l}^{l-1}\left( \bar{z}^l \right) ^{\top}\mathbf{1} \right\| ^2 \right]}
\\
\leqslant & c_t+3\varrho _t\gamma ^2\sum_{l=0}^k{\mathbb{E}\left[ \left\| \bar{z}_{i^l}^{l}-\xi _{i^l}^{l-1}\left( \bar{z}^l \right) ^{\top}\mathbf{1} \right\| ^2 \right]}+3\varrho _t\gamma ^2\sum_{l=0}^k{\left( \xi _{i^l}^{l-1} \right) ^2\mathbb{E}\left[ \left\| \nabla F\left( x_{\psi}^{l} \right) \right\| ^2 \right]}
\\
&+3\varrho _t\gamma ^2\sum_{l=0}^k{\left( \xi _{i^l}^{l-1} \right) ^2\mathbb{E}\left[ \left\| \nabla F\left( x_{\psi}^{l} \right) -\left( \bar{z}^l \right) ^{\top}\mathbf{1} \right\| ^2 \right]}+\varrho _t\gamma ^2\sum_{l=0}^k{n\sigma ^2}
\\
\overset{\left( c \right)}{\leqslant} & c_t+3\varrho _t\gamma ^2\sum_{l=0}^k{\mathbb{E}\left[ \left\| \bar{z}_{i^l}^{l}-\xi _{i^l}^{l-1}\left( \bar{z}^l \right) ^{\top}\mathbf{1} \right\| ^2 \right]}+3\varrho _t\gamma ^2\sum_{l=0}^k{\mathbb{E}\left[ \left\| \sum_{i=1}^n{\left( \nabla f_i\left( x_{i}^{l} \right) -\nabla f_i\left( x_{\psi}^{l} \right) \right)} \right\| ^2 \right]}
\\
&+\varrho _t\gamma ^2\sum_{l=0}^k{\left( 3\mathbb{E}\left[ \left\| \nabla F\left( x_{\psi}^{l} \right) \right\| ^2 \right] +n\sigma ^2 \right)}
\\
\overset{\left( d \right)}{\leqslant} & c_t+3\varrho _t\gamma ^2\sum_{l=0}^k{\mathbb{E}\left[ \left\| \bar{z}_{i^l}^{l}-\xi _{i^l}^{l-1}\left( \bar{z}^l \right) ^{\top}\mathbf{1} \right\| ^2 \right]}+3\varrho _tC_{L}^{2}n\gamma ^2\sum_{l=0}^k{\mathbb{E}\left[ \left\| h^l-\mathbf{1}x_{\psi}^{l} \right\| ^2 \right]}
\\
&+\varrho _t\gamma ^2\sum_{l=0}^k{\left( 3\mathbb{E}\left[ \left\| \nabla F\left( x_{\psi}^{l} \right) \right\| ^2 \right] +n\sigma ^2 \right)}
,
\end{align*}
where we used \eqref{auxiliary tracking_property} in $(a)$ and $(c)$, and used Assumption \ref{Ass_smoo} in $(b)$ and $(d)$. Moreover, if we treat $\sum_{l=0}^k{\mathbb{E}\left[ \left\| {h}^l-\mathbf{1}x_{\psi}^{l} \right\| ^2 \right]}$ and $\sum_{l=0}^k{\mathbb{E}\left[ \left\| \bar{z}_{i^l}^{l}-\xi _{i^l}^{l-1} ({\bar{z}}^l)^{\top}\mathbf{1} \right\| ^2 \right]}$ as new variables $x$ and $y$ respectively, then the above two interconnected inequalities can be rewritten by
\begin{align*}
    \left\{ \begin{array}{c}
	x\leqslant c_c+3\varrho _c\gamma ^2\cdot y+3\varrho _cC_{L}^{2}n\gamma ^2\cdot x+\varrho _c\gamma ^2\sum_{l=0}^{k}{\left( 3\mathbb{E}\left[ \left\| \nabla F\left( x_{\psi}^{l} \right) \right\| ^2 \right] +n\sigma ^2 \right)}\\
	y\leqslant c_t+3\varrho _t\gamma ^2\cdot y+3\varrho _tC_{L}^{2}n\gamma ^2\cdot x+\varrho _t\gamma ^2\sum_{l=0}^{k}{\left( 3\mathbb{E}\left[ \left\| \nabla F\left( x_{\psi}^{l} \right) \right\| ^2 \right] +n\sigma ^2 \right)}\\
\end{array} \right. .
\end{align*}

Using simple calculation, we can easily know that when $\gamma$ satisfies~\eqref{the_first_gamma_condition}, i.e., $\gamma < \frac{1}{\sqrt{3\varrho _cC_{L}^{2}n+3\varrho _t}}$, the intersection of 2 lines is exactly the upper bound of $x$ and $y$, and the intersection is
\begin{align*}
x=&\frac{3\varrho _c\gamma ^2}{1-3\left( \varrho _cC_{L}^{2}n+\varrho _t \right) \gamma ^2}\sum_{l=0}^k{\mathbb{E}\left[ \left\| \nabla F\left( x_{\psi}^{l} \right) \right\| ^2 \right]}+\frac{\left( 1-3\varrho _t\gamma ^2 \right) c_c+3c_t\varrho _c\gamma ^2}{1-3\left( \varrho _cC_{L}^{2}n+\varrho _t \right) \gamma ^2}
+\frac{\varrho _cn\gamma ^2\left( k+1 \right)}{1-3\left( \varrho _cC_{L}^{2}n+\varrho _t \right) \gamma ^2}\sigma ^2 ,
\\
y=&\frac{3\varrho _t\gamma ^2}{1-3\left( \varrho _cC_{L}^{2}n+\varrho _t \right) \gamma ^2}\sum_{l=0}^k{\mathbb{E}\left\| \nabla F\left( x_{\psi}^{l} \right) \right\| ^2}+\frac{\left( 1-3\varrho _cC_{L}^{2}n\gamma ^2 \right) c_t+3c_c\varrho _tC_{L}^{2}n\gamma ^2}{1-3\left( \varrho _cC_{L}^{2}n+\varrho _t \right) \gamma ^2}
+\frac{\varrho _tn\gamma ^2\left( k+1 \right)}{1-3\left( \varrho _cC_{L}^{2}n+\varrho _t \right) \gamma ^2}\sigma ^2
.
\end{align*}

The proof of Lemma \ref{lemma8} is thus completed.
\end{proof}

\section{Proof of Lemma~\ref{new_lemma7}}
\label{appecdix_proof_of_new_lemma_7}
To prove Lemma~\ref{new_lemma7}, we resort to two supporting lemmas and give them first.

The first supporting lemma as follows provides the lower and upper bound of $\mathbb{E} \left\| \nabla F\left( x_{\psi}^{k} \right) \right\| ^2 $ within a time window of length $T$, in term of its value at the starting point in this window, up to certain errors.
\begin{Lem}
\label{lemma9}
Suppose Assumptions \ref{Ass_weight_matrix}-\ref{Ass_bounded_var} hold. For $\forall k \geqslant0$ and $t\in \left[ 1,T-1 \right] $, $\mathbb{E} \left\| \nabla F\left( x_{\psi}^{kT+t} \right) \right\| ^2  
$ can be lower bounded by
\begin{small}
\begin{equation}
    \label{lower_bound_of_F_psi}
    \begin{aligned}
\mathbb{E}\left[ \left\| \nabla F\left( x_{\psi}^{kT+t} \right) \right\| ^2 \right] \geqslant & \frac{1}{2}\mathbb{E}\left[ \left\| \nabla F\left( x_{\psi}^{kT} \right) \right\| ^2 \right] -4\gamma ^2L^2T\sum_{t=0}^{T-1}{\mathbb{E}\left[ \left\| \bar{z}_{i^{kT+t}}^{kT+t}-\xi _{i^{kT+t}}^{kT+t-1} ({\bar{z}}^{kT+t})^{\top}\mathbf{1} \right\| ^2 \right]}
\\
&-4\gamma ^2L^2T\sum_{t=0}^{T-1}{\left( \psi _{i^{kT+t}}^{kT+t} \right) ^2\left( \xi _{i^{kT+t}}^{kT+t-1} \right) ^2\mathbb{E}\left[ \left\| \mathbf{1}^{\top}{\bar{z}}^{kT+t} \right\| ^2 \right]}-2\gamma ^2L^2T^2n\sigma ^2,
    \end{aligned}
\end{equation}
\end{small}
and can be upper bounded by 
\begin{equation}
    \label{upper_bound_of_F_psi}
    \begin{aligned}
\mathbb{E}\left[ \left\| \nabla F\left( x_{\psi}^{kT+t} \right) \right\| ^2 \right] \leqslant & 2\mathbb{E}\left[ \left\| \nabla F\left( x_{\psi}^{kT} \right) \right\| ^2 \right] +8\gamma ^2L^2T\sum_{t=0}^{T-1}{\mathbb{E}\left[ \left\| \bar{z}_{i^{kT+t}}^{kT+t}-\xi _{i^{kT+t}}^{kT+t-1} ({\bar{z}}^{kT+t})^{\top}\mathbf{1} \right\| ^2 \right]}
\\
&+8\gamma ^2L^2T\sum_{t=0}^{T-1}{\left( \psi _{i^{kT+t}}^{kT+t} \right) ^2\left( \xi _{i^{kT+t}}^{kT+t-1} \right) ^2\mathbb{E}\left[ \left\| \mathbf{1}^{\top}{\bar{z}}^{kT+t} \right\| ^2 \right]}+4\gamma ^2L^2T^2n\sigma ^2.
    \end{aligned}
\end{equation}
\end{Lem}
\begin{proof}
Applying \eqref{each_iterate_of_x_psi_k} recursively, 
the update of $x_{\psi}^{kT+t}$ can be written as
\begin{equation}
\label{yangyang}
x_{\psi}^{kT+t}=x_{\psi}^{kT}-\gamma \sum_{j=0}^{t-1}{\psi _{i^{kT+j}}^{kT+j}(z_{i^{kT+j}}^{kT+j})^\top} .
\end{equation}

Using the smoothness of $\,\,F\left( \cdot \right) $, we have 
\begin{equation}
\label{zhuzhu}
\begin{aligned}
& \mathbb{E}\left[ \left\| \nabla F\left( x_{\psi}^{kT+t} \right) -\nabla F\left( x_{\psi}^{kT} \right) \right\| ^2 \right] 
\\
\leqslant & L^2\mathbb{E}\left[ \left\| x_{\psi}^{kT+t}-x_{\psi}^{kT} \right\| ^2 \right] 
\\
\overset{\eqref{yangyang}}{=} & \gamma ^2L^2\mathbb{E}\left[ \left\| \sum_{j=0}^{t-1}{\psi _{i^{kT+j}}^{kT+j}z_{i^{kT+j}}^{kT+j}} \right\| ^2 \right] 
\\
\leqslant & \gamma ^2L^2\left( T-1 \right) \sum_{j=0}^{T-2}{\left( \psi _{i^{kT+j}}^{kT+j} \right) ^2\mathbb{E}\left[ \left\| z_{i^{kT+j}}^{kT+j} \right\| ^2 \right]}
\\
\leqslant & \gamma ^2L^2T\sum_{t=0}^{T-1}{\left( \psi _{i^{kT+t}}^{kT+t} \right) ^2\mathbb{E}\left[ \left\| z_{i^{kT+t}}^{kT+t} \right\| ^2 \right]}
\\
= & \gamma ^2L^2T\sum_{t=0}^{T-1}{\left( \psi _{i^{kT+t}}^{kT+t} \right) ^2\mathbb{E}\left[ \left\| z_{i^{kT+t}}^{kT+t}-\bar{z}_{i^{kT+t}}^{kT+t}+\bar{z}_{i^{kT+t}}^{kT+t} \right\| ^2 \right]}
\\
\leqslant & 2\gamma ^2L^2T\sum_{t=0}^{T-1}{\left( \psi _{i^{kT+t}}^{kT+t} \right) ^2\mathbb{E}\left[ \left\| \bar{z}_{i^{kT+t}}^{kT+t} \right\| ^2 \right]}+2\gamma ^2L^2Tn\sum_{t=0}^{T-1}{\left( \psi _{i^{kT+t}}^{kT+t} \right) ^2\sigma ^2}
\\
\leqslant & 4\gamma ^2L^2T\sum_{t=0}^{T-1}{\mathbb{E}\left[ \left\| \bar{z}_{i^{kT+t}}^{kT+t}-\xi _{i^{kT+t}}^{kT+t-1}(\bar{z}^{kT+t})^{\top}\mathbf{1} \right\| ^2 \right]}
\\
& +4\gamma ^2L^2T\sum_{t=0}^{T-1}{\left( \psi _{i^{kT+t}}^{kT+t} \right) ^2\left( \xi _{i^{kT+t}}^{kT+t-1} \right) ^2\mathbb{E}\left[ \left\| \mathbf{1}^{\top}\bar{z}^{kT+t} \right\| ^2 \right]}+2\gamma ^2L^2T^2n\sigma ^2 .
\end{aligned}
\end{equation}

Using $\left\| a+b \right\| ^2\leqslant 2\left\| a \right\| ^2+2\left\| b \right\| ^2$, we have
\begin{equation*}
\begin{aligned}
&\mathbb{E}\left[ \left\| \nabla F\left( x_{\psi}^{kT} \right) \right\| ^2 \right] 
\\
= &\mathbb{E}\left[ \left\| \nabla F\left( x_{\psi}^{kT} \right) -\nabla F\left( x_{\psi}^{kT+t} \right) +\nabla F\left( x_{\psi}^{kT+t} \right) \right\| ^2 \right] 
\\
\leqslant & 2\mathbb{E}\left[ \left\| \nabla F\left( x_{\psi}^{kT} \right) -\nabla F\left( x_{\psi}^{kT+t} \right) \right\| ^2 \right] +2\mathbb{E}\left[ \left\| \nabla F\left( x_{\psi}^{kT+t} \right) \right\| ^2 \right] 
\\
\overset{\eqref{zhuzhu}}{\leqslant} & 8\gamma ^2L^2T\sum_{t=0}^{T-1}{\mathbb{E}\left[ \left\| \bar{z}_{i^{kT+t}}^{kT+t}-\xi _{i^{kT+t}}^{kT+t-1}({\bar{z}}^{kT+t})^{\top}\mathbf{1} \right\| ^2 \right]}
\\
&+8\gamma ^2L^2T\sum_{t=0}^{T-1}{\left( \psi _{i^{kT+t}}^{kT+t} \right) ^2\left( \xi _{i^{kT+t}}^{kT+t-1} \right) ^2\mathbb{E}\left[ \left\| {\mathbf{1}}^{\top}{\bar{z}}^{kT+t} \right\| ^2 \right]}+4\gamma ^2L^2T^2n\sigma ^2+2\mathbb{E}\left[ \left\| \nabla F\left( x_{\psi}^{kT+t} \right) \right\| ^2 \right] .
\end{aligned}
\end{equation*}

By rearranging the terms of the above inequality we can obtain the lower bound of $\mathbb{E}\left[ \left\| \nabla F\left( x_{\psi}^{kT+t} \right) \right\| ^2 \right]$ as
\begin{equation*}
\begin{aligned}
\mathbb{E}\left[ \left\| \nabla F\left( x_{\psi}^{kT+t} \right) \right\| ^2 \right] \geqslant & \frac{1}{2}\mathbb{E}\left[ \left\| \nabla F\left( x_{\psi}^{kT} \right) \right\| ^2 \right] -4\gamma ^2L^2T\sum_{t=0}^{T-1}{\mathbb{E}\left[ \left\| \bar{z}_{i^{kT+t}}^{kT+t}-\xi _{i^{kT+t}}^{kT+t-1} ({\bar{z}}^{kT+t})^{\top}\mathbf{1} \right\| ^2 \right]}
\\
&-4\gamma ^2L^2T\sum_{t=0}^{T-1}{\left( \psi _{i^{kT+t}}^{kT+t} \right) ^2\left( \xi _{i^{kT+t}}^{kT+t-1} \right) ^2\mathbb{E}\left[ \left\| {\mathbf{1}}^{\top}{\bar{z}}^{kT+t} \right\| ^2 \right]}-2\gamma ^2L^2T^2n\sigma ^2 .
\end{aligned}
\end{equation*}

Similarly, the upper bound of $\mathbb{E}\left[ \left\| \nabla F\left( x_{\psi}^{kT+t} \right) \right\| ^2 \right]$ can be obtained by
\begin{equation*}
\begin{aligned}
\mathbb{E}\left[ \left\| \nabla F\left( x_{\psi}^{kT+t} \right) \right\| ^2 \right] = & \mathbb{E}\left[ \left\| \nabla F\left( x_{\psi}^{kT+t} \right) -\nabla F\left( x_{\psi}^{kT} \right) +\nabla F\left( x_{\psi}^{kT} \right) \right\| ^2 \right] 
\\
\leqslant & 2\mathbb{E}\left[ \left\| \nabla F\left( x_{\psi}^{kT+t} \right) -\nabla F\left( x_{\psi}^{kT} \right) \right\| ^2 \right] +2\mathbb{E}\left[ \left\| \nabla F\left( x_{\psi}^{kT} \right) \right\| ^2 \right] 
\\
\overset{\eqref{zhuzhu}}{\leqslant} & 2\mathbb{E}\left[ \left\| \nabla F\left( x_{\psi}^{kT} \right) \right\| ^2 \right] +8\gamma ^2L^2T\sum_{t=0}^{T-1}{\mathbb{E}\left[ \left\| \bar{z}_{i^{kT+t}}^{kT+t}-\xi _{i^{kT+t}}^{kT+t-1}({\bar{z}}^{kT+t})^{\top}\mathbf{1}\right\| ^2 \right]}
\\
& +8\gamma ^2L^2T\sum_{t=0}^{T-1}{\left( \psi _{i^{kT+t}}^{kT+t} \right) ^2\left( \xi _{i^{kT+t}}^{kT+t-1} \right) ^2\mathbb{E}\left[ \left\| {\mathbf{1}}^{\top}{\bar{z}}^{kT+t} \right\| ^2 \right]}+4\gamma ^2L^2T^2n\sigma ^2 .
\end{aligned}
\end{equation*}

The proof of Lemma \ref{lemma9} is completed.
\end{proof}

The second supporting lemma as follows is obtained by applying the descent lemma recursively from $k=0$ to a given total iteration.
\begin{Lem}
\label{lemma10}
Suppose Assumptions \ref{Ass_weight_matrix}-\ref{Ass_bounded_var} hold. For the given total iteration of $\bar{k}T$ for $\bar{k} \geqslant 1$, we have
\begin{equation}
    \label{running_sum_descent_lemma}
    \begin{aligned}
&\frac{\gamma}{2}\sum_{k=0}^{\bar{k}-1}{\sum_{t=0}^{T-1}{\psi _{i^{kT+t}}^{kT+t}\xi _{i^{kT+t}}^{kT+t-1}\mathbb{E}\left[ \left\| \nabla F\left( x_{\psi}^{kT+t} \right) \right\| ^2 \right]}}-\frac{\gamma ^2}{2}\sum_{k=0}^{\bar{k}T-1}{\mathbb{E}\left[ \left\| \nabla F\left( x_{\psi}^{k} \right) \right\| ^2 \right]}
\\
\leqslant & F\left( x_{\psi}^{0} \right) -F^{\star}+\frac{\gamma C_{L}^{2}n}{2}\sum_{k=0}^{\bar{k}T-1}{\mathbb{E}\left[ \left\| h^k-\mathbf{1}x_{\psi}^{k} \right\| ^2 \right]}+\left( \frac{1}{2}+2L\gamma ^2 \right) \sum_{k=0}^{\bar{k}T-1}{\mathbb{E}\left[ \left\| \bar{z}_{i^k}^{k}-\xi _{i^k}^{k-1}(\bar{z}^k)^{\top}\mathbf{1} \right\| ^2 \right]}
\\
&+Ln\bar{k}T\gamma ^2\sigma ^2-\sum_{k=0}^{\bar{k}T-1}{\left( \frac{\gamma \psi _{i^k}^{k}\xi _{i^k}^{k-1}}{2}-2L\gamma ^2\left( \psi _{i^k}^{k} \right) ^2\left( \xi _{i^k}^{k-1} \right) ^2 \right) \mathbb{E}\left[ \left\| \mathbf{1}^{\top}\bar{z}^k \right\| ^2 \right]} .
    \end{aligned}
\end{equation}
\end{Lem}
\begin{proof}
With~\eqref{each_iterate_of_x_psi_k}, 
applying the descent lemma to $F$ at $x_\psi^k$ and $x_\psi^{k+1}$, gives that
\begin{equation}
\label{descent_lemma_state_1}
    \begin{aligned}
&\mathbb{E}\left[ F\left( x_{\psi}^{k+1} \right) \right] 
\\
\leqslant & \mathbb{E}\left[ F\left( x_{\psi}^{k} \right) \right] +\gamma \psi _{i^k}^{k}\mathbb{E}\left[ \left< \nabla F\left( x_{\psi}^{k} \right) ,-\left( z_{i^k}^{k} \right) ^{\top} \right> \right] +\frac{L\gamma ^2\left( \psi _{i^k}^{k} \right) ^2}{2}\mathbb{E}\left[ \left\| z_{i^k}^{k} \right\| ^2 \right] 
\\
\leqslant & \mathbb{E}\left[ F\left( x_{\psi}^{k} \right) \right] -\gamma \psi _{i^k}^{k}\mathbb{E}\left[ \left< \nabla F\left( x_{\psi}^{k} \right) ,\left( \bar{z}_{i^k}^{k} \right) ^{\top} \right> \right] +\gamma \psi _{i^k}^{k}\mathbb{E}\left[ \left< \nabla F\left( x_{\psi}^{k} \right) ,\left( \bar{z}_{i^k}^{k} \right) ^{\top}-\left( z_{i^k}^{k} \right) ^{\top} \right> \right] 
\\
&+L\gamma ^2\left( \psi _{i^k}^{k} \right) ^2\mathbb{E}\left[ \left\| \bar{z}_{i^k}^{k} \right\| ^2 \right] +L\gamma ^2\left( \psi _{i^k}^{k} \right) ^2\mathbb{E}\left[ \left\| z_{i^k}^{k}-\bar{z}_{i^k}^{k} \right\| ^2 \right] 
\\
\overset{\left( a \right)}{\leqslant} & \mathbb{E}\left[ F\left( x_{\psi}^{k} \right) \right] \underset{T_1}{\underbrace{-\gamma \psi _{i^k}^{k}\mathbb{E}\left[ \left< \nabla F\left( x_{\psi}^{k} \right) ,\left( \bar{z}_{i^k}^{k} \right) ^{\top}-\xi _{i^k}^{k-1}\mathbf{1}^{\top}\bar{z}^k \right> \right] }}\underset{T_2}{\underbrace{-\gamma \psi _{i^k}^{k}\xi _{i^k}^{k-1}\mathbb{E}\left[ \left< \nabla F\left( x_{\psi}^{k} \right) ,\mathbf{1}^{\top}\bar{z}^k \right> \right] }}
\\
&+2L\gamma ^2\left( \psi _{i^k}^{k} \right) ^2\mathbb{E}\left[ \left\| \bar{z}_{i^k}^{k}-\xi _{i^k}^{k-1}\left( \bar{z}^k \right) ^{\top}\mathbf{1} \right\| ^2 \right] +2L\gamma ^2\left( \psi _{i^k}^{k} \right) ^2\left( \xi _{i^k}^{k-1} \right) ^2\mathbb{E}\left[ \left\| \mathbf{1}^{\top}\bar{z}^k \right\| ^2 \right] 
\\
&+\gamma \psi _{i^k}^{k}\mathbb{E}\left[ \left< \nabla F\left( x_{\psi}^{k} \right) ,\left( \bar{z}_{i^k}^{k} \right) ^{\top}-\left( z_{i^k}^{k} \right) ^{\top} \right> \right] +Ln\gamma ^2\sigma ^2  ,
    \end{aligned}
\end{equation}
where in $(a)$ we used Lemma \ref{variance of the tracking and auxiliary}.
Using Cauchy's inequality, we can bound $T_1$ by
\begin{equation}
\label{value_of_T_1}
    T_1\leqslant \frac{\gamma ^2\psi _{i^k}^{k}}{2}\mathbb{E}\left[ \left\| \nabla F\left( x_{\psi}^{k} \right) \right\| ^2 \right] +\frac{\psi _{i^k}^{k}}{2}\mathbb{E}\left[ \left\| \bar{z}_{i^k}^{k}-\xi _{i^k}^{k-1}\left( \bar{z}^k \right) ^{\top}\mathbf{1} \right\| ^2 \right] .
\end{equation}

Using $\left< a,b \right> =\frac{1}{2}\left( \left\| a \right\| ^2+\left\| b \right\| ^2-\left\| a-b \right\| ^2 \right) 
$, we can bound $T_2$ by
\begin{equation}
\label{value_of_T_2}
\begin{aligned}
T_2=&-\frac{\gamma \psi _{i^k}^{k}\xi _{i^k}^{k-1}}{2}\left( \mathbb{E}\left[ \left\| \nabla F\left( x_{\psi}^{k} \right) \right\| ^2 \right] +\mathbb{E}\left[ \left\| \mathbf{1}^{\top}\bar{z}^k \right\| ^2 \right] -\mathbb{E}\left[ \left\| \nabla F\left( x_{\psi}^{k} \right) -\mathbf{1}^{\top}\bar{z}^k \right\| ^2 \right] \right) 
\\
=&-\frac{\gamma \psi _{i^k}^{k}\xi _{i^k}^{k-1}}{2}\mathbb{E}\left[ \left\| \nabla F\left( x_{\psi}^{k} \right) \right\| ^2 \right] -\frac{\gamma \psi _{i^k}^{k}\xi _{i^k}^{k-1}}{2}\mathbb{E}\left[ \left\| \mathbf{1}^{\top}\bar{z}^k \right\| ^2 \right] +\frac{\gamma \psi _{i^k}^{k}\xi _{i^k}^{k-1}}{2}\mathbb{E}\left[ \left\| \nabla F\left( x_{\psi}^{k} \right) -\mathbf{1}^{\top}\bar{z}^k \right\| ^2 \right] 
\\
\leqslant &-\frac{\gamma \psi _{i^k}^{k}\xi _{i^k}^{k-1}}{2}\mathbb{E}\left[ \left\| \nabla F\left( x_{\psi}^{k} \right) \right\| ^2 \right] -\frac{\gamma \psi _{i^k}^{k}\xi _{i^k}^{k-1}}{2}\mathbb{E}\left[ \left\| \mathbf{1}^{\top}\bar{z}^k \right\| ^2 \right] +\frac{\gamma \psi _{i^k}^{k}\xi _{i^k}^{k-1}}{2}C_{L}^{2}n\mathbb{E}\left[ \left\| h^k-\mathbf{1}x_{\psi}^{k} \right\| ^2 \right] .
\end{aligned}
\end{equation}

Substituting \eqref{value_of_T_1} and \eqref{value_of_T_2} into \eqref{descent_lemma_state_1}, yields that
\begin{align*}
&\mathbb{E}\left[ F\left( x_{\psi}^{k+1} \right) \right] 
\\
\leqslant & \mathbb{E}\left[ F\left( x_{\psi}^{k} \right) \right] -\frac{\gamma \psi _{i^k}^{k}\xi _{i^k}^{k-1}}{2}\mathbb{E}\left[ \left\| \nabla F\left( x_{\psi}^{k} \right) \right\| ^2 \right] +\frac{\gamma ^2\psi _{i^k}^{k}}{2}\mathbb{E}\left[ \left\| \nabla F\left( x_{\psi}^{k} \right) \right\| ^2 \right] 
\\
&+\frac{\gamma \psi _{i^k}^{k}\xi _{i^k}^{k-1}}{2}C_{L}^{2}n\mathbb{E}\left[ \left\| h^k-\mathbf{1}x_{\psi}^{k} \right\| ^2 \right] +\left( \frac{\psi _{i^k}^{k}}{2}+2L\gamma ^2\left( \psi _{i^k}^{k} \right) ^2 \right) \mathbb{E}\left[ \left\| \bar{z}_{i^k}^{k}-\xi _{i^k}^{k-1}\left( \bar{z}^k \right) ^{\top}\mathbf{1} \right\| ^2 \right] 
\\
&-\left( \frac{\gamma \psi _{i^k}^{k}\xi _{i^k}^{k-1}}{2}-2L\gamma ^2\left( \psi _{i^k}^{k} \right) ^2\left( \xi _{i^k}^{k-1} \right) ^2 \right) \mathbb{E}\left[ \left\| \mathbf{1}^{\top}\bar{z}^k \right\| ^2 \right] +\gamma \psi _{i^k}^{k}\mathbb{E}\left[ \left< \nabla F\left( x_{\psi}^{k} \right) ,\left( \bar{z}_{i^k}^{k} \right) ^{\top}-\left( z_{i^k}^{k} \right) ^{\top} \right> \right]
+Ln\gamma ^2\sigma ^2
\\
\leqslant & \mathbb{E}\left[ F\left( x_{\psi}^{k} \right) \right] -\frac{\gamma \psi _{i^k}^{k}\xi _{i^k}^{k-1}}{2}\mathbb{E}\left[ \left\| \nabla F\left( x_{\psi}^{k} \right) \right\| ^2 \right] +\frac{\gamma ^2}{2}\mathbb{E}\left[ \left\| \nabla F\left( x_{\psi}^{k} \right) \right\| ^2 \right] +\frac{\gamma}{2}C_{L}^{2}n\mathbb{E}\left[ \left\| h^k-\mathbf{1}x_{\psi}^{k} \right\| ^2 \right] 
\\
&+\left( \frac{1}{2}+2L\gamma ^2 \right) \mathbb{E}\left[ \left\| \bar{z}_{i^k}^{k}-\xi _{i^k}^{k-1}\left( \bar{z}^k \right) ^{\top}\mathbf{1} \right\| ^2 \right] -\left( \frac{\gamma \psi _{i^k}^{k}\xi _{i^k}^{k-1}}{2}-2L\gamma ^2\left( \psi _{i^k}^{k} \right) ^2\left( \xi _{i^k}^{k-1} \right) ^2 \right) \mathbb{E}\left[ \left\| \mathbf{1}^{\top}\bar{z}^k \right\| ^2 \right] 
\\
&+\gamma \psi _{i^k}^{k}\mathbb{E}\left[ \left< \nabla F\left( x_{\psi}^{k} \right) ,(\bar{z}_{i^k}^{k})^{\top}-(z_{i^k}^{k})^{\top} \right> \right] +Ln\gamma ^2\sigma ^2.
\end{align*}

Summing the above inequality over $k$ from $0$ to $\bar{k}T-1$, and taking full expectations we get
\begin{equation*}
\begin{aligned}
&\frac{\gamma}{2}\sum_{k=0}^{\bar{k}-1}{\sum_{t=0}^{T-1}{\psi _{i^{kT+t}}^{kT+t}\xi _{i^{kT+t}}^{kT+t-1}\mathbb{E}\left[ \left\| \nabla F\left( x_{\psi}^{kT+t} \right) \right\| ^2 \right]}}-\frac{\gamma ^2}{2}\sum_{k=0}^{\bar{k}T-1}{\mathbb{E}\left[ \left\| \nabla F\left( x_{\psi}^{k} \right) \right\| ^2 \right]}
\\
\leqslant & F\left( x_{\psi}^{0} \right) -F^{\star}+\frac{\gamma C_{L}^{2}n}{2}\sum_{k=0}^{\bar{k}T-1}{\mathbb{E}\left[ \left\| h^k-1x_{\psi}^{k} \right\| ^2 \right]}+\left( \frac{1}{2}+2L\gamma ^2 \right) \sum_{k=0}^{\bar{k}T-1}{\mathbb{E}\left[ \left\| \bar{z}_{i^k}^{k}-\xi _{i^k}^{k-1}\left( \bar{z}^k \right) ^{\top}\mathbf{1} \right\| ^2 \right]}
\\
&+Ln\bar{k}T\gamma ^2\sigma ^2-\sum_{k=0}^{\bar{k}T-1}{\left( \frac{\gamma \psi _{i^k}^{k}\xi _{i^k}^{k-1}}{2}-2L\gamma ^2\left( \psi _{i^k}^{k} \right) ^2\left( \xi _{i^k}^{k-1} \right) ^2 \right) \mathbb{E}\left[ \left\| \mathbf{1}^{\top}\bar{z}^k \right\| ^2 \right]},
\end{aligned}
\end{equation*}
which completes the proof of Lemma~\ref{lemma10}.
\end{proof}

With the above two supporting lemmas, we are ready to prove Lemma~\ref{new_lemma7}.
According to~\eqref{lower_bound_of_F_psi} in Lemma~\ref{lemma9}, 
the lower bound of $\sum_{t=0}^{T-1}{\psi _{i^{kT+t}}^{kT+t}\xi _{i^{kT+t}}^{kT+t-1}\mathbb{E}\left[ \left\| \nabla F\left( x_{\psi}^{kT+t} \right) \right\| ^2 \right]}$
 can be obtained by
\begin{equation}
\label{zhu_yang}
\begin{aligned}
&\sum_{t=0}^{T-1}{\psi _{i^{kT+t}}^{kT+t}\xi _{i^{kT+t}}^{kT+t-1}\mathbb{E}\left[ \left\| \nabla F\left( x_{\psi}^{kT+t} \right) \right\| ^2 \right]}
\\
= & \psi _{i^{kT}}^{kT}\xi _{i^{kT}}^{kT-1}\mathbb{E}\left[ \left\| \nabla F\left( x_{\psi}^{kT} \right) \right\| ^2 \right] +\sum_{t=1}^{T-1}{\psi _{i^{kT+t}}^{kT+t}\xi _{i^{kT+t}}^{kT+t-1}\mathbb{E}\left[ \left\| \nabla F\left( x_{\psi}^{kT+t} \right) \right\| ^2 \right]}
\\
\geqslant & \frac{\psi _{i^{kT}}^{kT}\xi _{i^{kT}}^{kT-1}}{2}\mathbb{E}\left[ \left\| \nabla F\left( x_{\psi}^{kT} \right) \right\| ^2 \right] +\sum_{t=1}^{T-1}{\frac{\psi _{i^{kT+t}}^{kT+t}\xi _{i^{kT+t}}^{kT+t-1}}{2}\mathbb{E}\left[ \left\| \nabla F\left( x_{\psi}^{kT} \right) \right\| ^2 \right]}
-2\gamma ^2L^2T^2n\sigma ^2\sum_{t=1}^{T-1}{\psi _{i^{kT+t}}^{kT+t}\xi _{i^{kT+t}}^{kT+t-1}}
\\
&-\left( 4\gamma ^2L^2T\sum_{t=0}^{T-1}{\mathbb{E}\left[ \left\| \bar{z}_{i^{kT+t}}^{kT+t}-\xi _{i^{kT+t}}^{kT+t-1}(\bar{z}^{kT+t})^{\top}\mathbf{1} \right\| ^2 \right]} \right) \sum_{t=1}^{T-1}{\psi _{i^{kT+t}}^{kT+t}\xi _{i^{kT+t}}^{kT+t-1}}
\\
&-\left( 4\gamma ^2L^2T\sum_{t=0}^{T-1}{\left( \psi _{i^{kT+t}}^{kT+t} \right) ^2\left( \xi _{i^{kT+t}}^{kT+t-1} \right) ^2\mathbb{E}\left[ \left\| \mathbf{1}^{\top}\bar{z}^{kT+t} \right\| ^2 \right]} \right) \sum_{t=1}^{T-1}{\psi _{i^{kT+t}}^{kT+t}\xi _{i^{kT+t}}^{kT+t-1}}
\\
\overset{\left( a \right)}{\geqslant} & \frac{\mathbb{E}\left[ \left\| \nabla F\left( x_{\psi}^{kT} \right) \right\| ^2 \right]}{2}\sum_{t=0}^{T-1}{\psi _{i^{kT+t}}^{kT+t}\xi _{i^{kT+t}}^{kT+t-1}}-\left( 4\gamma ^2L^2T\sum_{t=0}^{T-1}{\mathbb{E}\left[ \left\| \bar{z}_{i^{kT+t}}^{kT+t}-\xi _{i^{kT+t}}^{kT+t-1}(\bar{z}^{kT+t})^{\top}\mathbf{1} \right\| ^2 \right]} \right) \left( T-1 \right) 
\\
&-\left( 4\gamma ^2L^2T\sum_{t=0}^{T-1}{\left( \psi _{i^{kT+t}}^{kT+t} \right) ^2\left( \xi _{i^{kT+t}}^{kT+t-1} \right) ^2\mathbb{E}\left[ \left\| \mathbf{1}^{\top}\bar{z}^{kT+t} \right\| ^2 \right]} \right) \left( T-1 \right) -2\gamma ^2L^2T^2n\sigma ^2\left( T-1 \right) 
\\
\overset{\left( b \right)}{\geqslant} & \frac{r\eta ^2\mathbb{E}\left[ \left\| \nabla F\left( x_{\psi}^{kT} \right) \right\| ^2 \right]}{2}-4\gamma ^2L^2T^2\sum_{t=0}^{T-1}{\mathbb{E}\left[ \left\| \bar{z}_{i^{kT+t}}^{kT+t}-\xi _{i^{kT+t}}^{kT+t-1}(\bar{z}^{kT+t})^{\top}\mathbf{1} \right\| ^2 \right]}
\\
&-4\gamma ^2L^2T^2\sum_{t=0}^{T-1}{\left( \psi _{i^{kT+t}}^{kT+t} \right) ^2\left( \xi _{i^{kT+t}}^{kT+t-1} \right) ^2\mathbb{E}\left[ \left\| \mathbf{1}^{\top}\bar{z}^{kT+t} \right\| ^2 \right]}-2\gamma ^2L^2T^3n\sigma ^2
,
\end{aligned}
\end{equation}
where we used 
\begin{equation}
\sum_{t=1}^{T-1}{\psi _{i^{kT+t}}^{kT+t}\xi _{i^{kT+t}}^{kT+t-1}}\leqslant T-1
\end{equation}
and
\begin{equation}
\label{non_convx_roots_descent}
\sum_{t=0}^{T-1}{\psi _{i^{kT+t}}^{kT+t}\xi _{i^{kT+t}}^{kT+t-1}}\geqslant r\eta ^2
\end{equation}
 in $(a)$ and $(b)$ respectively, according to Assumption \ref{Ass_graph} and \ref{Ass_asyn}, Lemma \ref{lemma_contraction_W^k} and \ref{lemma_contraction_A^k}.

Further, with simple calculation, we can lower bound $\sum_{k=0}^{\bar{k}-1}{\sum_{t=0}^{T-1}{\psi _{i^{kT+t}}^{kT+t}\xi _{i^{kT+t}}^{kT+t-1}\mathbb{E}\left[ \left\| \nabla F\left( x_{\psi}^{kT+t} \right) \right\| ^2 \right]}}
$ by
\begin{equation}
\label{yang_zhu}
\begin{aligned}
&\sum_{k=0}^{\bar{k}-1}{\sum_{t=0}^{T-1}{\psi _{i^{kT+t}}^{kT+t}\xi _{i^{kT+t}}^{kT+t-1}\mathbb{E}\left[ \left\| \nabla F\left( x_{\psi}^{kT+t} \right) \right\| ^2 \right]}}
\\
\overset{\eqref{zhu_yang}}{\geqslant} & \sum_{k=0}^{\bar{k}-1}{\frac{r\eta ^2\mathbb{E}\left[ \left\| \nabla F\left( x_{\psi}^{kT} \right) \right\| ^2 \right]}{2}}-4\gamma ^2L^2T^2\sum_{k=0}^{\bar{k}-1}{\sum_{t=0}^{T-1}{\mathbb{E}\left[ \left\| \bar{z}_{i^{kT+t}}^{kT+t}-\xi _{i^{kT+t}}^{kT+t-1}(\bar{z}^{kT+t})^{\top}\mathbf{1} \right\| ^2 \right]}}
\\
&-\sum_{k=0}^{\bar{k}-1}{\sum_{t=0}^{T-1}{4\gamma ^2L^2T^2\left( \psi _{i^{kT+t}}^{kT+t} \right) ^2\left( \xi _{i^{kT+t}}^{kT+t-1} \right) ^2\mathbb{E}\left[ \left\| \mathbf{1}^{\top}\bar{z}^{kT+t} \right\| ^2 \right]}}-2L^2\bar{k}T^3n\gamma ^2\sigma ^2
\\
=&\frac{r\eta ^2}{2}\sum_{k=0}^{\bar{k}-1}{\mathbb{E}\left[ \left\| \nabla F\left( x_{\psi}^{kT} \right) \right\| ^2 \right]}-4\gamma ^2L^2T^2\sum_{k=0}^{\bar{k}T-1}{\mathbb{E}\left[ \left\| \bar{z}_{i^k}^{k}-\xi _{i^k}^{k-1}(\bar{z}^k)^{\top}\mathbf{1} \right\| ^2 \right]}
\\
&-\sum_{k=0}^{\bar{k}T-1}{4\gamma ^2L^2T^2\left( \psi _{i^k}^{k} \right) ^2\left( \xi _{i^k}^{k-1} \right) ^2\mathbb{E}\left[ \left\| \mathbf{1}^{\top}\bar{z}^k \right\| ^2 \right]}-2L^2\bar{k}T^3n\gamma ^2\sigma ^2 .
\end{aligned}
\end{equation}

Substituting the above~\eqref{yang_zhu} into~\eqref{running_sum_descent_lemma}, gives that
\begin{equation}
\label{number_1_inequality}
\begin{aligned}
&\frac{\gamma r\eta ^2}{4}\sum_{k=0}^{\bar{k}-1}{\mathbb{E}\left[ \left\| \nabla F\left( x_{\psi}^{kT} \right) \right\| ^2 \right]}-\frac{\gamma ^2}{2}\sum_{k=0}^{\bar{k}T-1}{\mathbb{E}\left[ \left\| \nabla F\left( x_{\psi}^{k} \right) \right\| ^2 \right]}
\\
\leqslant & F\left( x_{\psi}^{0} \right) -F^{\star}+\frac{\gamma C_{L}^{2}n}{2}\sum_{k=0}^{\bar{k}T-1}{\mathbb{E}\left[ \left\| h^k-\mathbf{1}x_{\psi}^{k} \right\| ^2 \right]}
\\
&+\left( \frac{1}{2}+2L\gamma ^2+2L^2T^2\gamma ^3 \right) \sum_{k=0}^{\bar{k}T-1}{\mathbb{E}\left[ \left\| \bar{z}_{i^k}^{k}-\xi _{i^k}^{k-1}(\bar{z}^k)^{\top}\mathbf{1} \right\| ^2 \right]}+Ln\bar{k}T\gamma ^2\sigma ^2+L^2\bar{k}T^3n\gamma ^3\sigma ^2
\\
&-\sum_{k=0}^{\bar{k}T-1}{\left( \frac{\gamma \psi _{i^k}^{k}\xi _{i^k}^{k-1}}{2}-2L\gamma ^2\left( \psi _{i^k}^{k} \right) ^2\left( \xi _{i^k}^{k-1} \right) ^2-2\gamma ^3L^2T^2\left( \psi _{i^k}^{k} \right) ^2\left( \xi _{i^k}^{k-1} \right) ^2 \right) \mathbb{E}\left[ \left\| \mathbf{1}^{\top}\bar{z}^k \right\| ^2 \right]} .
\end{aligned}
\end{equation}

Summing (\ref{upper_bound_of_F_psi}) over $k$ from $0$ to $\bar{k}-1$ and over $t$ from $0$ to $T-1$, we have
\begin{equation*}
\begin{aligned}
&\sum_{k=0}^{\bar{k}-1}{\sum_{t=0}^{T-1}{\mathbb{E}\left[ \left\| \nabla F\left( x_{\psi}^{kT+t} \right) \right\| ^2 \right]}}
\\
\leqslant & 2\sum_{k=0}^{\bar{k}-1}{\sum_{t=0}^{T-1}{\mathbb{E}\left[ \left\| \nabla F\left( x_{\psi}^{kT} \right) \right\| ^2 \right]}}+8\gamma ^2L^2T\sum_{k=0}^{\bar{k}-1}{\sum_{t=0}^{T-1}{\left( \sum_{t=0}^{T-1}{\mathbb{E}\left[ \left\| \bar{z}_{i^{kT+t}}^{kT+t}-\xi _{i^{kT+t}}^{kT+t-1}(\bar{z}^{kT+t})^{\top}\mathbf{1} \right\| ^2 \right]} \right)}}
\\
&+\sum_{k=0}^{\bar{k}-1}{\sum_{t=0}^{T-1}{\left( \sum_{t=0}^{T-1}{8\gamma ^2L^2T\left( \psi _{i^{kT+t}}^{kT+t} \right) ^2\left( \xi _{i^{kT+t}}^{kT+t-1} \right) ^2\mathbb{E}\left[ \left\| \mathbf{1}^{\top}\bar{z}^{kT+t} \right\| ^2 \right]} \right)}}+4L^2\bar{k}T^3n\gamma ^2\sigma ^2,
\end{aligned}
\end{equation*}
i.e.,
\begin{equation}
    \label{number_2_inequality}
    \begin{aligned}
\sum_{k=0}^{\bar{k}T-1}{\mathbb{E}\left[ \left\| \nabla F\left( x_{\psi}^{k} \right) \right\| ^2 \right]}\leqslant & 2T\sum_{k=0}^{\bar{k}-1}{\mathbb{E}\left[ \left\| \nabla F\left( x_{\psi}^{kT} \right) \right\| ^2 \right]}+8\gamma ^2L^2T^2\sum_{k=0}^{\bar{k}T-1}{\mathbb{E}\left[ \left\| \bar{z}_{i^k}^{k}-\xi _{i^k}^{k-1}(\bar{z}^k)^{\top}\mathbf{1} \right\| ^2 \right]}
\\
&+\sum_{k=0}^{\bar{k}T-1}{8\gamma ^2L^2T^2\left( \psi _{i^k}^{k} \right) ^2\left( \xi _{i^k}^{k-1} \right) ^2\mathbb{E}\left[ \left\| \mathbf{1}^{\top}\bar{z}^k \right\| ^2 \right]}+4L^2\bar{k}T^3n\gamma ^2\sigma ^2 .
    \end{aligned}
\end{equation}

Computing~\eqref{number_1_inequality}$+$ $\frac{\gamma r\eta ^2}{8T}\times $\eqref{number_2_inequality}, gives that
\begin{equation}
\label{new_combine_of_two_ineq}
\begin{aligned}
&\left( \frac{\gamma r\eta ^2}{8T}-\frac{\gamma ^2}{2} \right) \sum_{k=0}^{\bar{k}T-1}{\mathbb{E}\left[ \left\| \nabla F\left( x_{\psi}^{k} \right) \right\| ^2 \right]}
\\
\leqslant & F\left( x_{\psi}^{0} \right) -F^{\star}+\frac{\gamma C_{L}^{2}n}{2}\sum_{k=0}^{\bar{k}T-1}{\mathbb{E}\left[ \left\| h^k-\mathbf{1}x_{\psi}^{k} \right\| ^2 \right]}
\\
&+\left( \frac{1}{2}+2L\gamma ^2+\left( 2T^2+r\eta ^2T \right) L^2\gamma ^3 \right) \sum_{k=0}^{\bar{k}T-1}{\mathbb{E}\left[ \left\| \bar{z}_{i^k}^{k}-\xi _{i^k}^{k-1}(\bar{z}^k)^{\top}\mathbf{1} \right\| ^2 \right]}
\\
&+Ln\bar{k}T\gamma ^2\sigma ^2+L^2\bar{k}T^3n\gamma ^3\sigma ^2+\frac{1}{2}r\eta ^2L^2\bar{k}T^2n\gamma ^3\sigma ^2
\\
&-\sum_{k=0}^{\bar{k}T-1}{\left( \frac{\gamma \psi _{i^k}^{k}\xi _{i^k}^{k-1}}{2}-2L\gamma ^2\left( \psi _{i^k}^{k} \right) ^2\left( \xi _{i^k}^{k-1} \right) ^2-\left( 2T^2+r\eta ^2T \right) L^2\gamma ^3\left( \psi _{i^k}^{k} \right) ^2\left( \xi _{i^k}^{k-1} \right) ^2 \right) \mathbb{E}\left[ \left\| \mathbf{1}^{\top}\bar{z}^k \right\| ^2 \right]}.
\end{aligned}
\end{equation}

According to the condition in~\eqref{step_size_condition_1} satisfied by the constant step size, i.e., $\gamma \leqslant \min \left\{ \frac{2}{\left( 2T^2+r\eta ^2T \right) L},\frac{1}{8L} \right\} $, we have 
\begin{equation}
\label{yang_w_1}
\frac{1}{2}+2L\gamma ^2+\left( 2T^2+r\eta ^2T \right) L^2\gamma ^3\leqslant \frac{1+\gamma}{2}
\end{equation}
and 
\begin{equation}
\label{yang_w_2}
\frac{\gamma \psi _{i^k}^{k}\xi _{i^k}^{k-1}}{2}-2L\gamma ^2\left( \psi _{i^k}^{k} \right) ^2\left( \xi _{i^k}^{k-1} \right) ^2-\left( 2T^2+r\eta ^2T \right) L^2\gamma ^3\left( \psi _{i^k}^{k} \right) ^2\left( \xi _{i^k}^{k-1} \right) ^2\geqslant 0 .
\end{equation}

Substituting~\eqref{yang_w_1} and~\eqref{yang_w_2} into~\eqref{new_combine_of_two_ineq}, we obtain
\begin{equation*}
\begin{aligned}
\left( \frac{\gamma r\eta ^2}{8T}-\frac{\gamma ^2}{2} \right) \sum_{k=0}^{\bar{k}T-1}{\mathbb{E}\left[ \left\| \nabla F\left( x_{\psi}^{k} \right) \right\| ^2 \right]}
\leqslant & F\left( x_{\psi}^{0} \right) -F^{\star}+\frac{1+\gamma}{2}\sum_{k=0}^{\bar{k}T-1}{\mathbb{E}\left[ \left\| \bar{z}_{i^k}^{k}-\xi _{i^k}^{k-1}(\bar{z}^k)^{\top}\mathbf{1} \right\| ^2 \right]}+\frac{1}{2}r\eta ^2L^2\bar{k}T^2n\gamma ^3\sigma ^2
\\
&+\frac{\gamma C_{L}^{2}n}{2}\sum_{k=0}^{\bar{k}T-1}{\mathbb{E}\left[ \left\| h^k-\mathbf{1}x_{\psi}^{k} \right\| ^2 \right]}+Ln\bar{k}T\gamma ^2\sigma ^2+L^2\bar{k}T^3n\gamma ^3\sigma ^2,
\end{aligned}
\end{equation*}
which completes the proof of Lemma~\ref{new_lemma7}.

\section{Proof of Theorem~\ref{Thm_sublinear_conv}}
\label{appendix_proof_thm_non_convex}
Considering \eqref{complex_descent_lemma} in Lemma~\ref{new_lemma7}, we first provide an upper bound for
$\sum_{k=0}^{\bar{k}T-1}{\mathbb{E}\left\| h^k-\mathbf{1}x_{\psi}^{k} \right\| ^2}\,\,
$ and $\sum_{k=0}^{\bar{k}T-1}{\mathbb{E}\left\| \bar{z}_{i^k}^{k}-\xi _{i^k}^{k-1}(\bar{z}^k)^{\top}\mathbf{1} \right\| ^2}
$ in term of $\sum_{k=0}^{\bar{k}T-1}{\mathbb{E}\left\| \nabla F\left( x_{\psi}^{k} \right) \right\| ^2}
$, which can be easily obtained by invoking~\eqref{upper_bound_running_sum_consensus_error} and~\eqref{upper_bound_running_sum_tracking_error} in Lemma~\ref{lemma8}. Substituting these obtained bounds into~\eqref{complex_descent_lemma} and rearranging the terms yields
\begin{equation}
\label{qingwang_xinduo}
\begin{aligned}
& \left( \frac{r\eta ^2}{8}-\frac{T\gamma}{2}-\frac{3\varrho _tT+3\left( \varrho _t+\varrho _cC_{L}^{2}n \right) T\gamma}{2\left( 1-3\left( \varrho _cC_{L}^{2}n+\varrho _t \right) \gamma ^2 \right)}\gamma \right) \frac{1}{\bar{k}T}\cdot \sum_{k=0}^{\bar{k}T-1}{\mathbb{E}\left\| \nabla F\left( x_{\psi}^{k} \right) \right\| ^2}
\\
\leqslant & \frac{1}{\bar{k}}\left\{ \frac{F\left( x_{\psi}^{0} \right) -F^{\star}}{\gamma}+\frac{C_{L}^{2}n}{2}\cdot \frac{c_c+3c_t\varrho _c\gamma ^2}{1-3\left( \varrho _cC_{L}^{2}n+\varrho _t \right) \gamma ^2}+\frac{1+\gamma}{2\gamma}\cdot \frac{c_t+3c_c\varrho _tC_{L}^{2}n\gamma ^2}{1-3\left( \varrho _cC_{L}^{2}n+\varrho _t \right) \gamma ^2} \right\} 
\\
& +\frac{\left( 1+\gamma \right) \varrho _tnT\gamma}{2\left( 1-3\left( \varrho _cC_{L}^{2}n+\varrho _t \right) \gamma ^2 \right)}\sigma ^2+\frac{1}{2}r\eta ^2L^2T^2n\gamma ^2\sigma ^2+\frac{C_{L}^{2}n}{2}\cdot \frac{\varrho _cnT\gamma ^2}{1-3\left( \varrho _cC_{L}^{2}n+\varrho _t \right) \gamma ^2}\sigma ^2
+LnT\gamma \sigma ^2+L^2T^3n\gamma ^2\sigma ^2.
\end{aligned}
\end{equation}

If $\gamma$ further satisfies
\begin{small}
\begin{equation}
\label{step_size_condition_2}
\,\,\gamma \,\,\leqslant \min \left\{ \frac{1}{\sqrt{6\left( \varrho _cC_{L}^{2}n+\varrho _t \right)}},\frac{1+6\varrho _t}{6\left( \varrho _t+\varrho _cC_{L}^{2}n \right)},\frac{r\eta ^2}{16\left( 1+6\varrho _t \right) T} \right\} ,
\end{equation}
\end{small}
we can further obtain from~\eqref{qingwang_xinduo} that
\begin{equation}
\label{nips}
\begin{aligned}
& \frac{1}{\bar{k}T}\sum_{k=0}^{\bar{k}T-1}{\mathbb{E}\left\| \nabla F\left( x_{\psi}^{k} \right) \right\| ^2}
\\
\leqslant & \frac{16\left( F\left( x_{\psi}^{0} \right) -F^{\star}+c_t \right)}{r\eta ^2\gamma \bar{k}}+\frac{16\left( C_{L}^{2}nc_c+c_t \right) +48c_c\varrho _tC_{L}^{2}n\gamma +48\left( c_t\varrho _c+c_c\varrho _t \right) C_{L}^{2}n\gamma ^2}{r\eta ^2\bar{k}}
\\
& +\frac{16}{r\eta ^2}\left( \varrho _tnT+\varrho _cC_{L}^{2}n^2T+L^2T^3n+\frac{1}{2}r\eta ^2L^2T^2n \right) \gamma ^2\sigma ^2+\frac{16}{r\eta ^2}\left( \varrho _tnT+LnT \right) \gamma \sigma ^2.
\end{aligned}
\end{equation}

By now, the constant step size $\gamma$ needs to satisfy~\eqref{the_first_gamma_condition},~\eqref{step_size_condition_1} and~\eqref{step_size_condition_2}, i.e., 
\begin{equation}
\label{total_condion_of_stepsize}
\gamma \,\,\leqslant \underset{\triangleq \bar{\gamma}}{\underbrace{\min \left\{ \frac{1}{\sqrt{6\left( \varrho _cC_{L}^{2}n+\varrho _t \right)}},\frac{2}{\left( 2T^2+r\eta ^2T \right) L},\frac{1}{8L},\frac{1+6\varrho _t}{6\left( \varrho _t+\varrho _cC_{L}^{2}n \right)},\frac{r\eta ^2}{16\left( 1+6\varrho _t \right) T} \right\} }}
.
\end{equation}

Next, we move to bound $\frac{1}{\bar{k}T}\sum_{k=0}^{\bar{k}T-1}{\mathbb{E}\left[ M_F\left( {x}^k \right) \right]}$.
To this end, we first establish a relationship between $\mathbb{E}\left[ M_F\left( x^k \right) \right]$ with $\mathbb{E}\left\| \nabla F\left( x_{\psi}^{k} \right) \right\| ^2$ and $\mathbb{E}\left\| h^k-\mathbf{1}x_{\psi}^{k} \right\| ^2$ as follows:
\begin{equation}
\label{lemma7}
\begin{aligned}
\mathbb{E}\left[ M_F\left( x^k \right) \right] \leqslant & 2\mathbb{E}\left\| \nabla F\left( x_{\psi}^{k} \right) \right\| ^2+2\mathbb{E}\left\| \nabla F\left( \bar{x}^k \right) -\nabla F\left( x_{\psi}^{k} \right) \right\| ^2+2\mathbb{E}\left\| x^k-\mathbf{1}x_{\psi}^{k} \right\| ^2+2\mathbb{E}\left\| \mathbf{1}x_{\psi}^{k}-\mathbf{1}\left( \bar{x}^k \right) ^{\top} \right\| ^2
\\
 \overset{\left( a \right)}{\leqslant} & 2\mathbb{E}\left\| \nabla F\left( x_{\psi}^{k} \right) \right\| ^2+\left( \frac{2L^2}{n}+4 \right) \mathbb{E}\left\| x^k-\mathbf{1}x_{\psi}^{k} \right\| ^2
\\
\leqslant & 2\mathbb{E}\left\| \nabla F\left( x_{\psi}^{k} \right) \right\| ^2+\left( \frac{2L^2}{n}+4 \right) \mathbb{E}\left\| h^k-\mathbf{1}x_{\psi}^{k} \right\| ^2,
\end{aligned}
\end{equation}
where in $(a)$ we used $\mathbb{E}\left\| \nabla F\left( \bar{x}^k \right) -\nabla F\left( x_{\psi}^{k} \right) \right\| ^2\leqslant L^2\mathbb{E}\left\| \left( \bar{x}^k \right) ^{\top}-x_{\psi}^{k} \right\| ^2=\frac{L^2}{n}\mathbb{E}\left\| \frac{\mathbf{1}\mathbf{1}^{\top}}{n}\left( \mathbf{1}x_{\psi}^{k}-x^k \right) \right\| ^2$, $\left\| \frac{\mathbf{1}\mathbf{1}^{\top}}{n} \right\| _2=1$ and $\mathbb{E}\left\| \mathbf{1}x_{\psi}^{k}-\mathbf{1}\left( \bar{x}^k \right) ^{\top} \right\| ^2=\mathbb{E}\left\| \frac{\mathbf{1}\mathbf{1}^{\top}}{n}\left( \mathbf{1}x_{\psi}^{k}-x^k \right) \right\| ^2$.

Combining~\eqref{lemma7} and \eqref{upper_bound_running_sum_consensus_error} in Lemma \ref{lemma8}, we have
\begin{equation*}
\begin{aligned}
& \frac{1}{\bar{k}T}\sum_{k=0}^{\bar{k}T-1}{\mathbb{E}\left[ M_F\left( x^k \right) \right]}
\\
\leqslant & \frac{2}{\bar{k}T}\sum_{k=0}^{\bar{k}T-1}{\mathbb{E}\left\| \nabla F\left( x_{\psi}^{k} \right) \right\| ^2}+\left( \frac{2L^2}{n}+4 \right) \frac{1}{\bar{k}T}\cdot 
\\
& \left\{ \frac{3\varrho _c\gamma ^2}{1-3\left( \varrho _cC_{L}^{2}n+\varrho _t \right) \gamma ^2}\sum_{k=0}^{\bar{k}T-1}{\mathbb{E}\left\| \nabla F\left( x_{\psi}^{k} \right) \right\| ^2}+\frac{\left( 1-3\varrho _t\gamma ^2 \right) c_c+3c_t\varrho _c\gamma ^2}{1-3\left( \varrho _cC_{L}^{2}n+\varrho _t \right) \gamma ^2}+\frac{\varrho _cn\bar{k}T\gamma ^2}{1-3\left( \varrho _cC_{L}^{2}n+\varrho _t \right) \gamma ^2}\sigma ^2 \right\} .
\end{aligned}
\end{equation*}

Substituting $\gamma \,\,\leqslant \frac{1}{\sqrt{6\left( \varrho _cC_{L}^{2}n+\varrho _t \right)}}$ in \eqref{total_condion_of_stepsize}, into the above inequality, we further have 
\begin{equation*}
\begin{aligned}
\frac{1}{\bar{k}T}\sum_{k=0}^{\bar{k}T-1}{\mathbb{E}\left[ M_F\left( x^k \right) \right]}\leqslant & \frac{2}{\bar{k}T}\sum_{k=0}^{\bar{k}T-1}{\mathbb{E}\left\| \nabla F\left( x_{\psi}^{k} \right) \right\| ^2}+\frac{12\left( L^2+2n \right) \varrho _c\gamma ^2}{n}\frac{1}{\bar{k}T}\sum_{k=0}^{\bar{k}T-1}{\mathbb{E}\left\| \nabla F\left( x_{\psi}^{k} \right) \right\| ^2}
\\
& +\frac{4\left( L^2+2n \right) \left( c_c+3c_t\varrho _c\gamma ^2 \right)}{n\bar{k}T}+4\left( L^2+2n \right) \varrho _c\gamma ^2\sigma ^2.
\end{aligned}
\end{equation*}

Substituting the upper bound of $\frac{1}{\bar{k}T} \sum_{k=0}^{\bar{k}T-1}{\mathbb{E}\left\| \nabla F\left( x_{\psi}^{k} \right) \right\| ^2}$ (c.f., \eqref{nips}) into the above inequality, and plugging $\gamma \leqslant \frac{1}{8L}$ in \eqref{total_condion_of_stepsize}, yields that
\begin{equation}
\label{no_total_K}
\frac{1}{\bar{k}T}\sum_{k=0}^{\bar{k}T-1}{\mathbb{E}\left[ M_F\left( x^k \right) \right]}\leqslant \frac{32\left( F\left( x_{\psi}^{0} \right) -F^{\star}+c_t \right)}{r\eta ^2\gamma \bar{k}}
+\frac{E_1}{\bar{k}}+\frac{32T\left( n\varrho _t+nL \right)}{\eta ^2}\gamma \sigma ^2+E_2\gamma ^2\sigma ^2 ,
\end{equation}
where the constants
\begin{equation}
\label{def_of_E1}
\begin{aligned}
E_1\triangleq & \frac{32\left( C_{L}^{2}nc_c+c_t \right)}{r\eta ^2}+\frac{4\left( L^2+2n \right) c_c}{nT}+\frac{12c_c\varrho _tC_{L}^{2}n}{r\eta ^2L}+\frac{24\left( F\left( x_{\psi}^{0} \right) -F^{\star}+c_t \right) \left( L^2+2n \right) \varrho _c}{nr\eta ^2L}
\\
&+\frac{3\left( c_t\varrho _c+c_c\varrho _t \right) C_{L}^{2}n}{2r\eta ^2L^2}+\frac{3\left( C_{L}^{2}nc_c+c_t \right) \left( L^2+2n \right) \varrho _c}{nr\eta ^2L^2}+\frac{3\left( L^2+2n \right) c_t\varrho _c}{16nTL^2}
\\
&+\frac{9\left( L^2+2n \right) c_c\varrho _c\varrho _tC_{L}^{2}}{8r\eta ^2L^3}\,\,+\frac{9\left( L^2+2n \right) \left( c_t\varrho _c+c_c\varrho _t \right) \varrho _cC_{L}^{2}}{64r\eta ^2L^4},
\end{aligned}
\end{equation}
and
\begin{equation}
\label{def_of_E2}
\begin{aligned}
E_2\triangleq & \frac{3\left( L^2+2n \right) \varrho _c}{r\eta ^2L^2}\left( \varrho _tT+\varrho _cC_{L}^{2}nT+L^2T^3+\frac{1}{2}r\eta ^2L^2T^2 \right) +4\left( L^2+2n \right) \varrho _c
\\
&+\frac{24}{r\eta ^2L}\left( L^2+2n \right) \varrho _c\left( \varrho _tT+LT \right) +\frac{32}{r\eta ^2}\left( \varrho _tnT+\varrho _cC_{L}^{2}n^2T+L^2T^3n+\frac{1}{2}r\eta ^2L^2T^2n \right) .
\end{aligned}
\end{equation}

Let $K=\bar{k}T$, \eqref{no_total_K} becomes
\begin{equation*}
\frac{1}{K}\sum_{k=0}^{K-1}{\mathbb{E}\left[ M_F\left( x^k \right) \right]}\leqslant \frac{32T\left( F\left( x_{\psi}^{0} \right) -F^{\star}+c_t \right)}{\eta^2 \gamma Kr}
+\frac{TE_1}{K}+\frac{32T\left( n\varrho _t+nL \right)}{\eta^2}\gamma \sigma ^2+E_2\gamma ^2\sigma ^2,
\end{equation*}
which completes the proof of Theorem~\ref{Thm_sublinear_conv}.

\section{Analysis of consensus scheme on augmented system}
\label{appendix_augmented_graph_of_W}
\begin{figure}[h]
\centering
\includegraphics[scale=0.5]{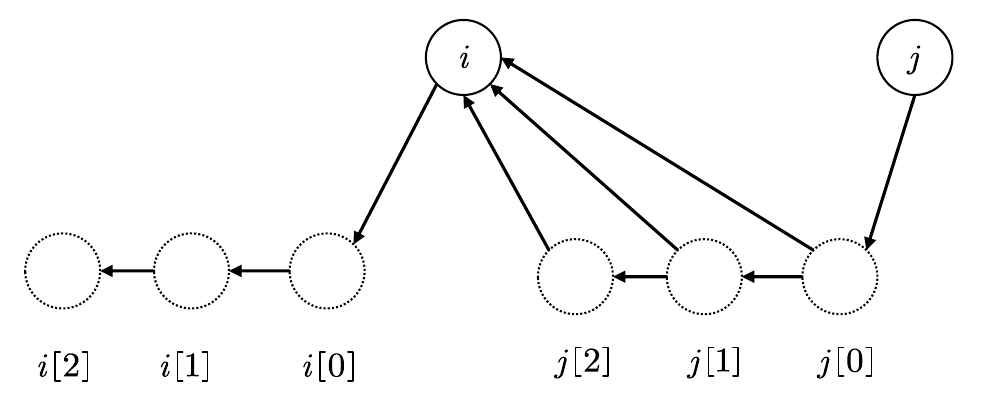}
\caption{Illustration of constructing an augmented graph when the maximum delay $D=2$. The above shows that three non-computing \emph{virtual} nodes  are added for each node.}
\label{fig_W}
\end{figure}
The results in this section and Appendix~\ref{appendix_augmented_graph_of_A} are similar to that of~\cite{tian2020achieving}, although we are working with more general underlying topology. We include them for completeness.

It follows from Algorithm \ref{Myalgorithm_RASGT_GV} that the recursions of $v_{i}^{k}$ and ${x}_{i}^{k}$, $i\in \mathcal{V}$, can be described as follows:
\begin{subequations}
\begin{align} 
& v_{i^k}^{k+1}=x_{i^k}^k -\gamma^k z_{i^k}^k, \\
& {x}_{i^k}^{k+1}=w_{i^ki^k}{v}_{i^k}^{k+1}+\sum_{j\in \mathcal{N}_{i^k}^{\text{in}}(W)}w_{i^kj}v_{j}^{k-d_{v,j}^k},\\
& v_{j}^{k+1} = v_{j}^{k}, \, x_{j}^{k+1} = x_{j}^{k}, \quad \forall j \in \mathcal{V} \setminus \{i^k\}.
\end{align}
\end{subequations}
We add $D+1$ virtual nodes for each node $i$, denoted by $i[0], i[1],...,i[D]$ (see Fig.~\ref{fig_W}) and used for storing delayed information $v_i^{k}, v_i^{k-1},...,v_i^{k-D}$. Any virtual node $i[d], d=D,D-1,...1$ can only receive information from virtual node $i[d-1]$; $i[0]$ can only receive information from the real node $i$ or doesn't change the value. We give an example  to illustrate the dynamics: at iteration $k$, any virtual node $i[d]$ for $i\in \mathcal{V}$ and $d=D,D-1,...,1$ receives the information from $i[d-1]$ and replaces its own value with the received one ; 
virtual node $i^k[0]$ replaces its own value with the information $x_{i^k}^k-\gamma^k z_{i^k}^k$  received from the real node $i^k$;
virtual nodes $j[0]$ for $j\ne i^k$ keep the value unchanged; the value of  real node $i^k$ becomes a weighted average of the $x_{i^k}^k-\gamma^kz_{i^k}^k$ and $v_j^{k-d_{v,j}^k}$ received from the virtual nodes $j[d_{v,j}^k]$ for $j\in \mathcal{N}_{i^k}^{\mathrm{in}}\left( W \right)$; and the other real nodes keep the value unchanged. 

Define ${v}^k \triangleq [{v}_1^k, \cdots, {v}_n^k]^{\top}\in \mathbb{R}^{n\times p}$. Construct the $(D+2)n \times p$ dimensional concatenated variables of augmented system as   
\begin{equation}
h^{k} \triangleq [{{x}^k}; {{v}^k}; {{v}^{k-1}}; \cdots; {{v}^{k-D}}]\in \mathbb{R}^{\left( D+2 \right) n\times p},
\end{equation}
and the augmented matrix $\hat{W}^k \in \mathbb{R}^{\left( D+2 \right) n\times \left( D+2 \right) n}$, defined as
\begin{equation}
\label{def_of_hat_W}
\hat{W}_{rm}^k  \triangleq 
\left\{
\begin{aligned}
& w_{i^k i^k}  ,  && \text{if }  r=m = i^k ; \\
& w_{i^k j},        && \text{if }  r = i^k,\, m= j+ (d_{v,j}^{k}+1) n  ; \\
& 1 ,  && \text{if }  r = m \in \{1,2,\ldots,2n\} \setminus \{i^k, i^k+n\};  \\
& 1 ,  && \text{if }  r \in  \{2n+1,2n+2,\ldots, (D+2)n\} \\
&      &&  \cup \{i^k+n\} \text{ and } m = r-n;\\
& 0,   && \text{otherwise}.
\end{aligned}
\right.
\end{equation}
The update of the augmented system can be rewritten in a compact form as
\begin{align}
{h}^{k+1} = \hat{W}^k({h}^{k} - \gamma^k e_{i^k} \left( z_{i^k}^k \right )^{\top}).
\end{align}

\section{Analysis of gradient tracking scheme on augmented system}
\label{appendix_augmented_graph_of_A}
\begin{figure}[h]
\centering
\includegraphics[scale=0.5]{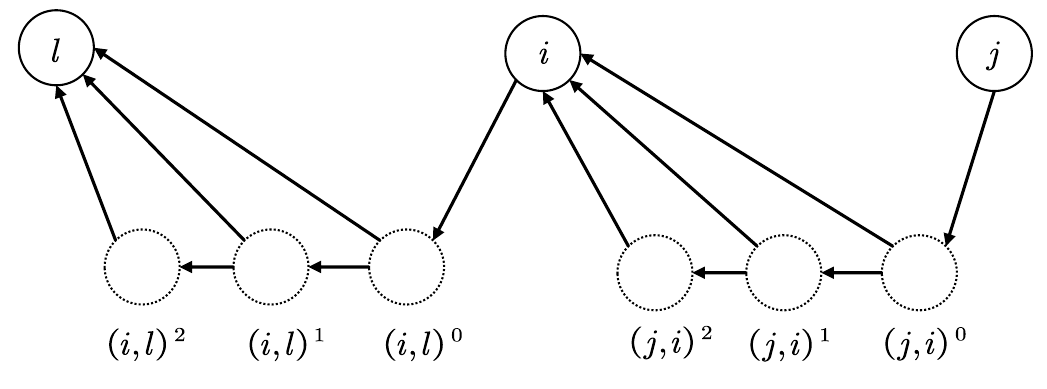}
\caption{When the maximum delay $D=2$,  three noncomputing node  are added for each edges $(j,i)\in \mathcal{E}(A)$ and $(i,l)\in \mathcal{E}(A)$.}
\label{fig:original}
\end{figure}

In what follows, we call the real nodes in $\mathcal{G}(A)$  {\it computing nodes} and call the virtual nodes {\it noncomputing nodes}. 
Each computing node $j\in \mathcal{V}$ can only send information to the noncomputing nodes $(j,i)^0$, for $i\in \mathcal{N}^{\text{out}}_j(A)$; each noncomputing node  $(j,i)^d$ can either  send information to the next noncomputing node $(j,i)^{d+1}$, or to the computing node  $i$; see Fig.~\ref{fig:original}
 .

 For further description, we define  $\mathcal{T}_i \triangleq \left\{ k  \, \big\vert \, i^{k} =i   , \, k\in \mathbb{N}_0\right\}$ and it denotes the set of global iteration counter at which the computing node $i\in \mathcal{V}$ wakes up;  let  $\mathcal{T}_i^{k} \triangleq \left\{ t\in \mathcal{T}_i \, \big\vert \, t \leq k  \right\}$,
it  follows from (S.2 c) and (S.4) in Algorithm \ref{Myalgorithm_RASGT_GV}  that  \begin{equation}\label{eq:rho_z}
 \hspace{-0.5cm}\rho_{ij}^{k}  =  \!\sum_{t \in \mathcal{T}^{k-1}_j} a_{ij}z^{t + 1/2}_{j}  \,\,\,\text{and}\,\,\,
 \tilde{\rho}^{k}_{ij} =  \!\rho_{ij}^{k-1-d_{\rho,j}^{k-1}},\quad   (j,i)\in \mathcal{E}(A).
\end{equation} 
Remember that each computing node  $i$ stores $z_i^0$ at $k=0$, and the values of the noncomputing nodes are initialized to $0.$ 
At the beginning of iteration $k$, every computing node  $i$  will store   $z_i^k$ whereas  every noncomputing node $(j,i)^d$ for $0\leq d \leq D-1$, stores  the mass $a_{ij}z_j$  (if any)   generated by $j$ for $i$ at iteration  $k-d-1$ (thus $k-d-1 \in \mathcal{T}_j^{k-1}$), i.e., $a_{ij}z_j^{k-(d+1)+1/2}$ (cf. (S.2 c) in Algorithm \ref{Myalgorithm_RASGT_GV}), and not been used by  $i$ yet; otherwise it stores $0$. 
Formally, we have
\begin{equation}
\label{eq:virtual_info_d}
z^k_{(j,i)^d} \triangleq  a_{ij}z_j^{t + 1/2 }  \cdot 1{\left[ t = k-d-1 \in \mathcal{T}_j^{k-1} ~\& ~t+1>k-1-d_{\rho ,j}^{k-1} \right].} 
\end{equation}

The virtual node $(j,i)^D$ cumulates  all the masses $a_{ij}z_j^{k-(d+1)+1/2}$ with   $d \geq D$, not  received by $i$ yet: 
\begin{equation}\label{eq:virtual_info}
\begin{aligned}
z^k_{(j,i)^D} \triangleq \sum_{t \in  \mathcal{T}_j^{k-D-1},\, t +1 > k-1-d_{\rho ,j}^{k-1}} a_{ij}z_j^{t + 1/2}.
\end{aligned}
\end{equation}

Next we write the update of the $z$-variables of both the computing and noncomputing nodes, absorbing the $(\rho, \tilde{\rho})$-variables using \eqref{eq:rho_z}-\eqref{eq:virtual_info}.

The update of augmented system can be can be divided into two steps:sum-step and push-step.
 In the sum-step, the update of the $z$-variables of the computing nodes can be written as:

 \begin{subequations} \label{eq:pf_sum}
 \begin{equation}
     z_{i^k}^{k+\frac{1}{2}}=z_{i^k}^{k}+\sum_{j\in \mathcal{N}_{i^k}^{\mathrm{in}}(A)}{\left( \rho _{i^kj}^{k-d_{\rho,j}^k}-\tilde{\rho}_{i^kj}^{k} \right)}+\epsilon ^k\overset{\eqref{eq:rho_z}-\eqref{eq:virtual_info}}{=}z_{i^k}^{k}+\sum_{j\in \mathcal{N}_{i^k}^{\mathrm{in}}(A)}{\sum_{d=d_{\rho,j}^k}^D{z_{(j,i^k)^d}^{k}}}+\epsilon ^k; \label{eq:pf_sum_1}
 \end{equation}
 \begin{equation}
     z^{k+\frac{1}{2}}_{j} = z^{k }_{j},  \quad  j \in \mathcal{V}\setminus \{i^k\} .\label{eq:pf_sum_2}
 \end{equation}
 i.e., node $i^k$ builds the update $z_{i^k}^{k} \!\!\to\!\! z_{i^k}^{k+\frac{1}{2}}$ based upon the masses transmitted by the noncomputing nodes    
 $(j,i^k)^{d_{\rho,j}^k},(j,i^k)^{d_{\rho,j}^k+1},\ldots,$ $(j,i^k)^D$ (cf. \eqref{eq:pf_sum_1}).  And the other computing nodes keep their masses unchanged (cf. \eqref{eq:pf_sum_2}).  The  updates of the noncomputing nodes is set to
 \begin{align}
 & z^{k+\frac{1}{2}}_{(j,i^k)^d} \triangleq 0,  \quad   d=d_{\rho,j}^k,\ldots,D,\quad   j \in \mathcal{N}_{i^k}^{\text{in}}(A);\label{eq:pf_sum_3}\smallskip \\
& z^{k+\frac{1}{2}}_{(j^\prime,i)^\tau} \triangleq z^{k}_{(j^\prime,i)^\tau}, \quad   \text{for all the other }   (j^\prime,i)^\tau\in \widehat{\mathcal{V}}.\label{eq:pf_sum_4}
\end{align} 
\end{subequations}

The  noncomputing nodes in  \eqref{eq:pf_sum_3}    set  their    variables to zero (as they transferred their  masses to   $i^k$) while  the  other  noncomputing nodes  keep their   variables unchanged (cf.~\eqref{eq:pf_sum_4}).

 In the push-step, the update of the $z$-variables of the computing nodes are as follows:
 \begin{subequations}\label{eq:pf_push}
\begin{align}
  &z_{i^k}^{k+1} = a_{i^k i^k}\,z_{i^k}^{k+\frac{1}{2}}; \label{eq:pf_push_1}\\
    & z^{k+1}_{j} = z^{k+ \frac{1}{2}}_{j}, \qquad \text{ for } j \in \mathcal{V}\setminus \{i^k\}. \label{eq:pf_push_2}
\end{align}
i.e., node $i^k$ keeps the portion  $a_{i^k i^k} {z}_{i^k}^{k+ \frac{1}{2}}$  of the new generated mass (cf. \eqref{eq:pf_push_1}) whereas  the other computing nodes  do not change their variables (cf. \eqref{eq:pf_push_2}).  The  noncomputing nodes update as:   \begin{align}
& z^{k+1}_{(i^k,\ell)^{0}} \triangleq  a_{\ell i^k}\, z_{i^k}^{k+1/2}, \quad   \ell \in \mathcal{N}_{i^k}^{\text{out}}(A); \label{eq:pf_push_5}\\
& z^{k+1}_{(i,j)^{0}} \triangleq 0, \quad  (i,j) \in \mathcal{E}(A),\quad   i \neq i^k; \label{eq:pf_push_6}\\
& z^{k+1}_{(i,j)^{d}} \triangleq z^{k+\frac{1}{2}}_{(i,j)^{d-1}},  \quad d= 1 ,\ldots, D-1, \quad   (i,j) \in \mathcal{E}(A); \label{eq:pf_push_4} \\
& z^{k+1}_{(i,j)^{D}} \triangleq  z^{k+\frac{1}{2}}_{(i,j)^{D}} + z^{k+\frac{1}{2}}_{(i,j)^{D-1}}, \quad   (i,j) \in \mathcal{E}(A). \label{eq:pf_push_3} 
 \end{align}
 \end{subequations}
 i.e., the computing node $i^k$ pushes its masses  $a_{\ell i^k} {z}_{i^k}^{k+\frac{1}{2}}$    to the noncomputing nodes $(i^k, \ell)^0$, with   $\ell \in \mathcal{N}_{i^k}^{\text{out}}(A)$ (cf. \eqref{eq:pf_push_5}). The other noncomputing nodes $(i,j)^{0}$,   $i \neq i^k$ set their variables to zero (cf. \eqref{eq:pf_push_6}). The noncomputing nodes $(i,j)^d$ for $0\leq d \leq D-1$, transfers  their mass to the next  noncomputing node $(i,j)^{d+1}$ (cf. \eqref{eq:pf_push_3}, \eqref{eq:pf_push_4}).

Remember that the $S \times p
$ dimensional concatenated tracking variables $\hat{z}^k$ is defined in \eqref{def_hat_z^k}.
The transition matrix $S^k$ of the sum step is defined as

\[
\eC_{hm}^k  \triangleq 
\left\{
\begin{aligned}
& 1 ,  && \text{if }  m \in \{ (j, i^k)^{d} \mid  d_{\rho,j}^k \leq d \leq D \} \\
&        && \text{and }  h = i^k; \\
& 1 ,  && \text{if }  m \in \widehat{\mathcal{V}} \setminus \{ (j, i^k)^{d} \mid  d_{\rho,j}^k \leq d \leq D \} \\
&        && \text{and }  h = m; \\
& 0,   && \text{otherwise}.
\end{aligned}
\right.
\]

Therefore, the sum-step can be written in compact form as 
\begin{equation}\label{eq:sum_matrix}
\hat{z}^{k+\frac{1}{2}} = S^k \hat{z}^k + e_{i^k} (\epsilon^k)^\top.
\end{equation}

Define the transition matrix $P^k$ of the push step as

\begin{equation}
\label{def_of_P_k}
    P_{hm}^{k}\triangleq \left\{ \begin{aligned}
	&a_{ji^k},\quad \text{if } m=i^k\,\,\text{and } h=(j,i^k)^0,j\in \mathcal{N}_{i^k}^{\mathrm{out}}(A);\\
	&a_{i^ki^k},\quad \text{if } m=h=i^k;\\
	&1,\qquad  \text{if } m=h\in \mathcal{V}\setminus i^k;\\
	&1,\qquad \text{if } m=(i,j)^d,\,h=(i,j)^{d+1},(i,j)\in \mathcal{E}(A),\,0\le d\le D-1;\\
	&1,\qquad \text{if } m=h=(i,j)^D,\,(i,j)\in \mathcal{E}(A);\\
	&0,\qquad \mathrm{otherwise}\\
\end{aligned} \right. 
\end{equation}
then, the push-step can be written as 
\begin{align}\label{eq:push_matrix}
\hat{z}^{k+1} = P^k \hat{z}^{k+\frac{1}{2}}.
\end{align}

Combing \eqref{eq:sum_matrix} and \eqref{eq:push_matrix}, yields 
\begin{align}
\label{eq:tracking_dynamics-z}
\hat{z}^{k+1}=\hat{A}^k \hat{z}^k + P^k e_{i^k} (\epsilon^k)^\top,
\end{align}
where
\begin{equation}
\label{def_of_hat_A_k}
\hat{A}^k=P^k S^k,
\end{equation}
 with initialization $z^{0}_i  \in \mathbb{R}^p$ for $i \in  \mathcal{V}$  and $z^{0}_i = 0$ for $i \in  \widehat{\mathcal{V}} \setminus \mathcal{V}$.

\section{Used Weight Matrices and Topology Design}
\label{weghit_matrix_used}
In this section, we present the specific weight matrices used in the experiments in the main text and provide some other possible topologies to illustrate the flexibility of topology design.
\\
\\
\textbf{Weight matrices used in experiments.} The corresponding row (resp. column) stochastic weight matrix $W$ (resp. $A$) satisfying Assumption \ref{Ass_weight_matrix} can be easily designed by knowing the number of in-neighbors (resp. out-neighbors).

\begin{figure}[htbp]
    \centering
    {
        \begin{minipage}[t]{0.75\textwidth}
        \centering          
         \includegraphics[width=\textwidth]{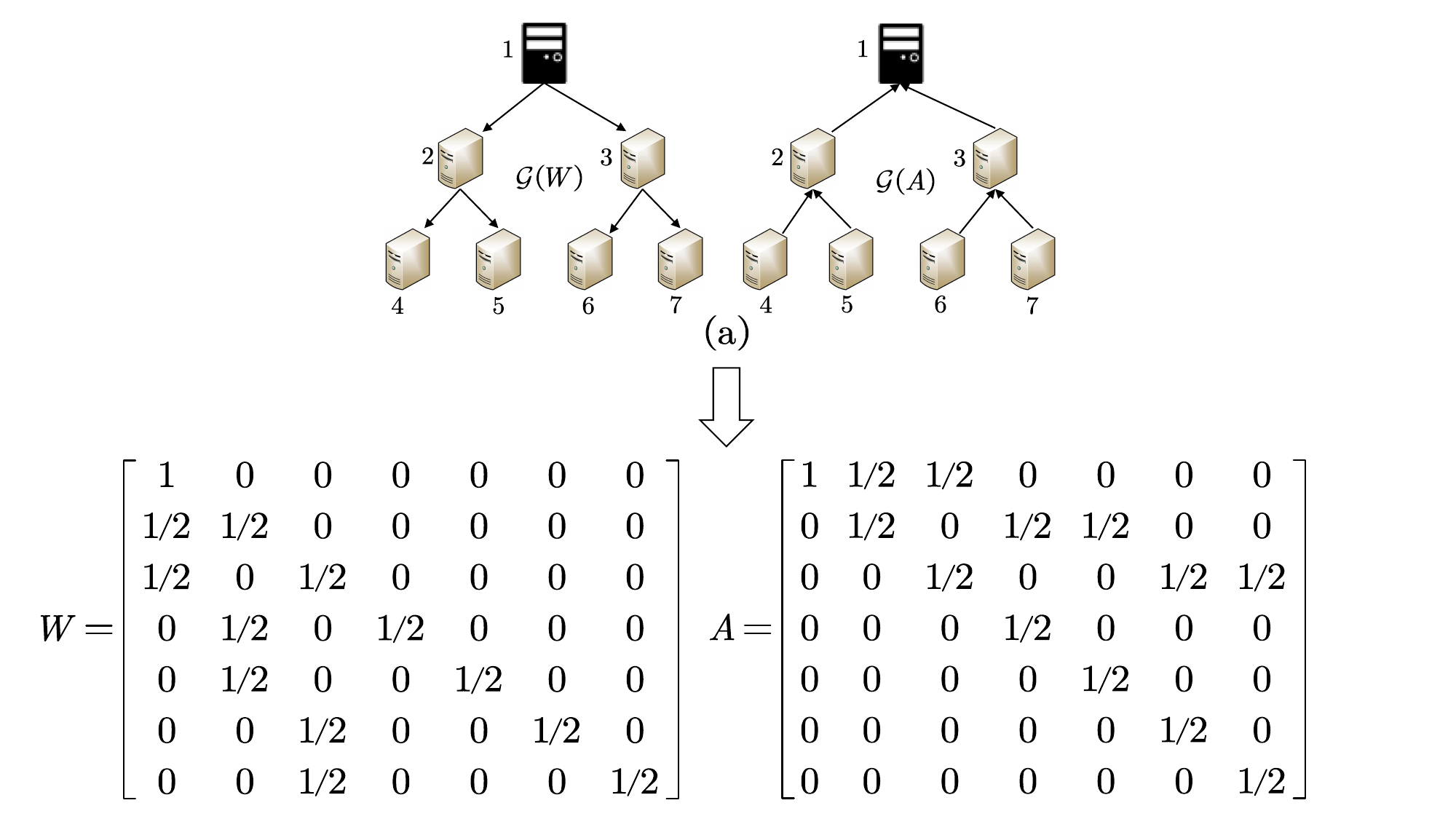}
        \end{minipage}%
    }
    \caption{Weight matrices for binary tree graph. }
    \label{matrix_tree}
\end{figure}

For binary tree graph, the corresponding weight matrices are designed as Fig.~\ref{matrix_tree}.

\begin{figure}[htbp]
    \centering
    {
        \begin{minipage}[t]{0.75\textwidth}
        \centering          
         \includegraphics[width=\textwidth]{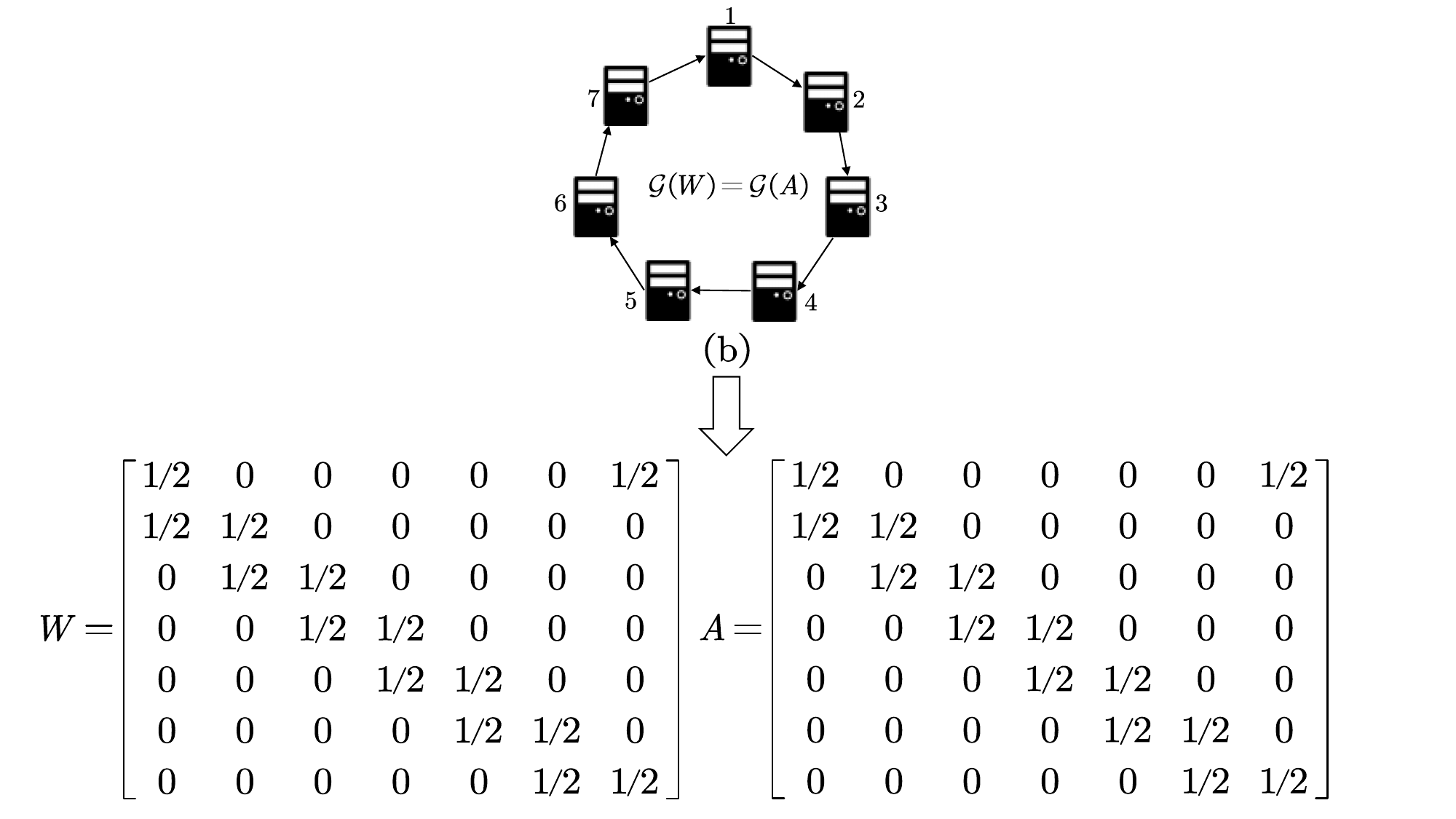}
        \end{minipage}%
    }
    \caption{Weight matrices for directed ring graph. }
    \label{matrix_ring}
\end{figure}

For directed ring graph, the corresponding weight matrices are designed as Fig.~\ref{matrix_ring}.

\begin{figure}[htbp]
    \centering
    {
        \begin{minipage}[t]{0.75\textwidth}
        \centering          
         \includegraphics[width=\textwidth]{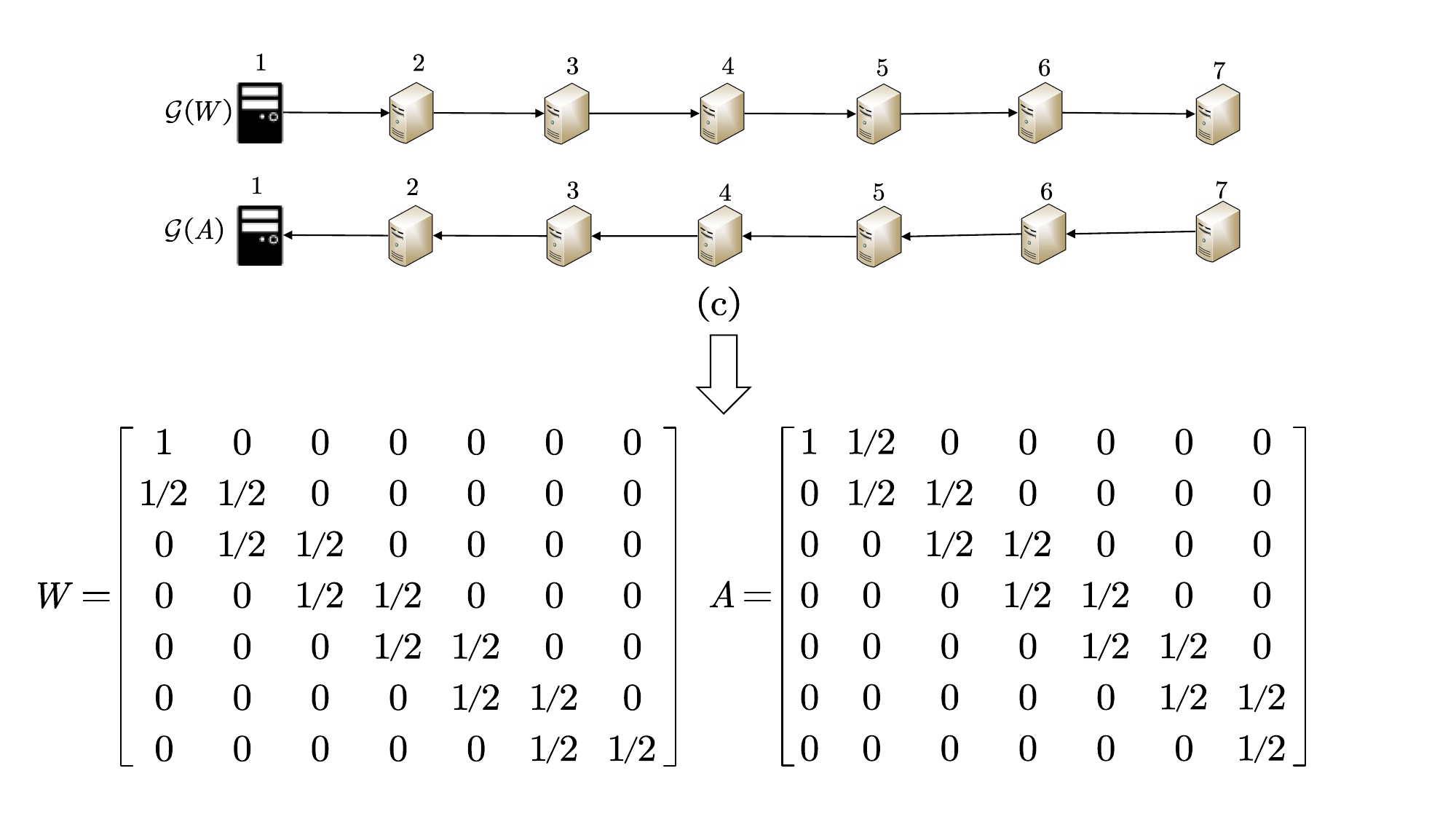}
        \end{minipage}%
    }
    \caption{Weight matrices for line graph. }
    \label{matrix_line}
\end{figure}

For line graph, the corresponding weight matrices are designed as Fig.~\ref{matrix_line}.

\begin{figure}[htbp]
    \centering
    {
        \begin{minipage}[t]{0.75\textwidth}
        \centering          
         \includegraphics[width=\textwidth]{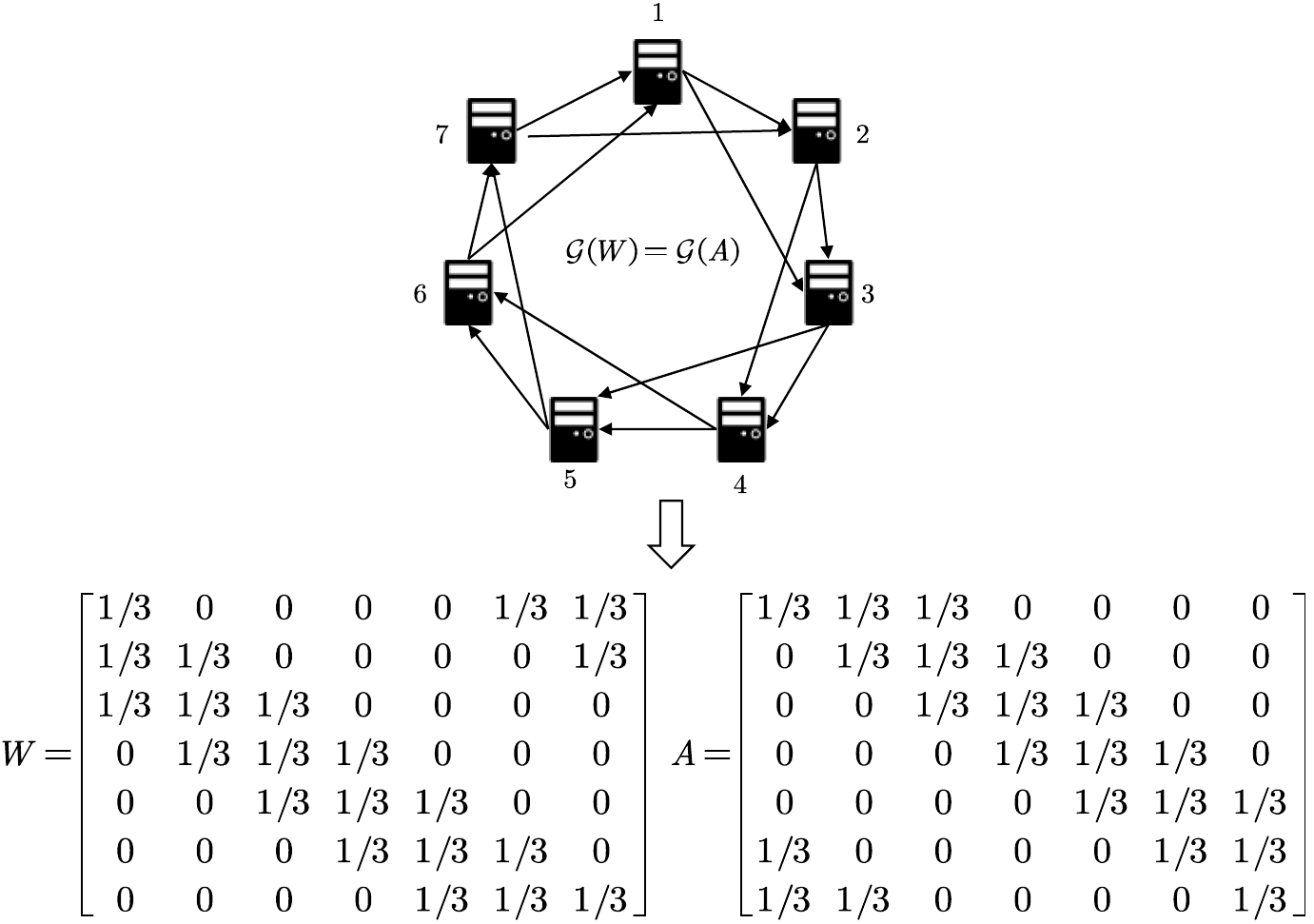}
        \end{minipage}%
    }
    \caption{Weight matrices for exponential graph. }
    \label{matrix_exponential}
\end{figure}

For exponential graph, the corresponding weight matrices are designed as Fig.~\ref{matrix_exponential}.

\begin{figure}[htbp]
    \centering
    {
        \begin{minipage}[t]{0.75\textwidth}
        \centering          
         \includegraphics[width=\textwidth]{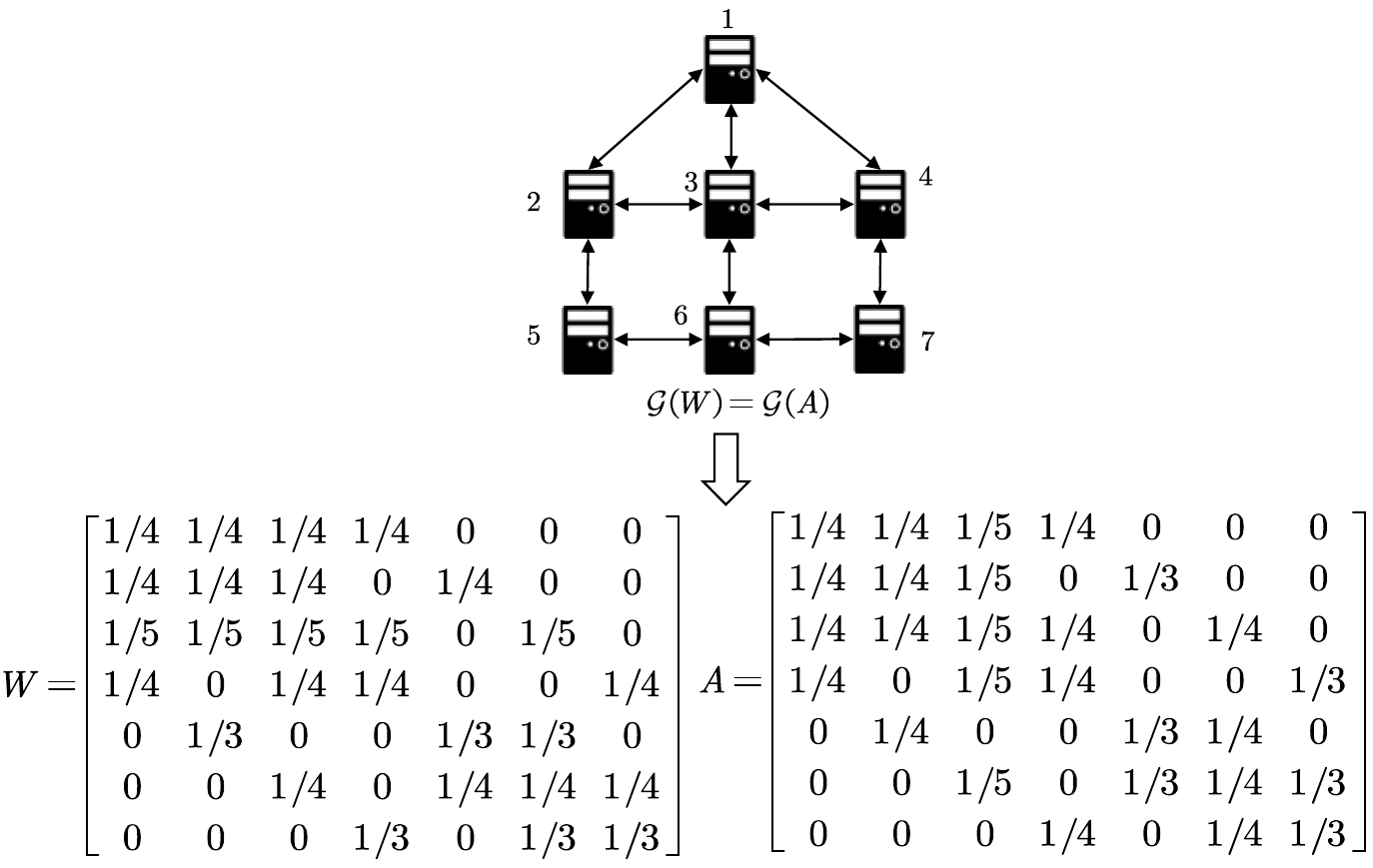}
        \end{minipage}%
    }
    \caption{Weight matrices for mesh graph. }
    \label{matrix_mesh}
\end{figure}

For mesh graph, the corresponding weight matrices are designed as Fig.~\ref{matrix_mesh}.

\begin{figure}[htbp]
    \centering
    {
        \begin{minipage}[t]{0.9\textwidth}
        \centering          
         \includegraphics[width=\textwidth]{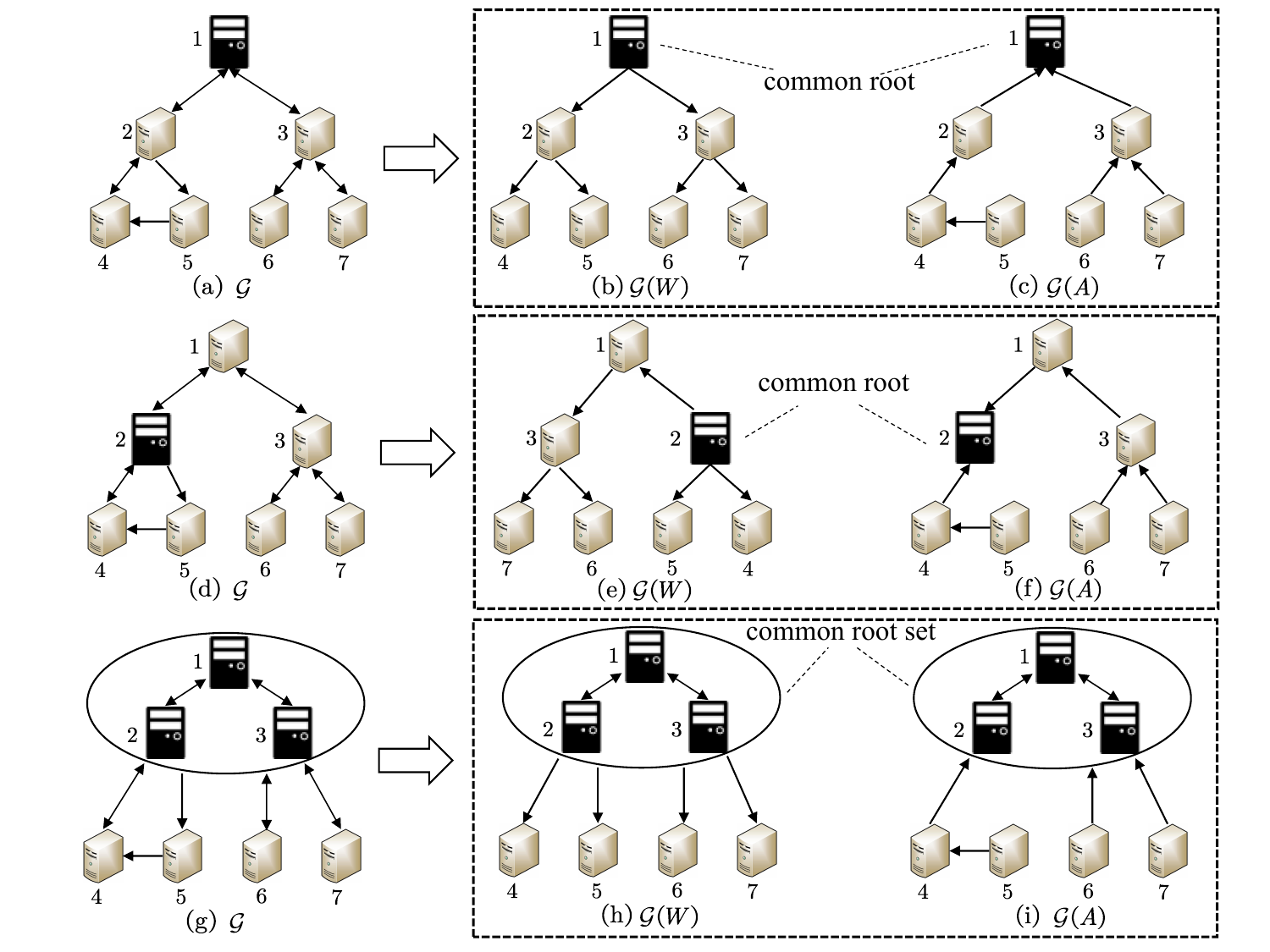}
        \end{minipage}%
    }
    \caption{Illustration of the simplicity and flexibility of topology design. The left sub-figures (a), (d) and (g) denote the original topology and the middle ((b),(e), (h)) and right subfigures ((c),(f), (i)) represent the two non-strongly-connected sub-graphs}
    \label{Fig_topology_design}
\end{figure}

\noindent
Furthermore, we report several other possible graphs to illustrate the simplicity and flexibility of our proposed R-FAST algorithm for communication topology design.

\textbf{Flexibility of communication topology design.}
We present three different strategies of topology design based on an \textit{unbalanced} strongly connected graph $\mathcal{G}$ in Fig.~\ref{Fig_topology_design}, where each row of subfigures indicates a way to splitting the original topology $\mathcal{G}$ into two non-strongly-connected sub-graphs $\mathcal{G}(W)$ and $\mathcal{G}(A)$ with different sets of common root nodes satisfying Assumption~\ref{Ass_graph}. It follows that the minimum assumption of the proposed R-FAST algorithm imposed on the topology allows us to freely design the topologies of sub-graphs and the corresponding weight matrices of R-FAST. For instance, we can see from the bottom row of subfigures that nodes $\{ 1,2,3\}$ are chosen as common roots of sub-graphs $\mathcal{G}(W)$ and $\mathcal{G}(A)$, which resembles the group of servers in the Pramater-Server structure~\cite{mcmahan2017communication} that have higher computing power and communication bandwidth while other nodes serve as workers (clients) to provide training data and calculated gradient vectors. As such, our proposed R-FAST algorithm enjoys great flexibility of communication topology design to account for ad-hoc requirement in real training environments.

\end{document}